\documentclass[longauth]{aaEC}
\usepackage{graphicx}
\usepackage{natbib}
\usepackage{scalerel}
\usepackage{comment}
\usepackage{amsmath} 
\usepackage{amssymb}
\usepackage{mathtools}
\usepackage{microtype}
\usepackage{siunitx}
\usepackage{multirow}
\usepackage[table]{xcolor}
\definecolor{CiteRed}{RGB}{110, 0, 0}
\graphicspath{{fig/}}
\newcommand{\sersic}{S\'ersic }

\usepackage{txfonts}
\usepackage[pdfencoding=auto,psdextra]{hyperref}
\hypersetup{
    colorlinks=true,
    linkcolor=blue, 
    filecolor=magenta,      
    urlcolor=blue, 
    citecolor=blue, 
    pdftitle={Euclid Quick Data Release (Q1). Searching for giant gravitational arcs in galaxy clusters with mask region-based convolutional neural networks},
    pdfkeywords={Gravitational lensing: strong -- Galaxies: clusters: general -- Techniques: image processing},
    pdfauthor={Lorenzo Bazzanini},
    pdfcreator={\LaTeX}
}
\makeatletter
\renewcommand*\aa@pageof{, page \thepage{} of \pageref*{LastPage}}
\makeatother

\usepackage[utf8]{inputenc}

\newcommand{\linenumbers}[0]{}

\usepackage{euclid}

\begin{document}
    \title{Euclid Quick Data Release (Q1)}
    \subtitle{Searching for giant gravitational arcs in galaxy clusters with mask region-based convolutional neural networks}   
    
\newcommand{\orcid}[1]{} 
\author{Euclid Collaboration: L.~Bazzanini\orcid{0000-0003-0727-0137}\thanks{\email{bzzlnz[at]unife[dot]it}}\inst{\ref{aff1},\ref{aff2}}
\and G.~Angora\orcid{0000-0002-0316-6562}\inst{\ref{aff3},\ref{aff1}}
\and P.~Bergamini\orcid{0000-0003-1383-9414}\inst{\ref{aff4},\ref{aff2}}
\and M.~Meneghetti\orcid{0000-0003-1225-7084}\inst{\ref{aff2},\ref{aff5}}
\and P.~Rosati\orcid{0000-0002-6813-0632}\inst{\ref{aff1},\ref{aff2}}
\and A.~Acebron\orcid{0000-0003-3108-9039}\inst{\ref{aff6},\ref{aff7}}
\and C.~Grillo\orcid{0000-0002-5926-7143}\inst{\ref{aff4},\ref{aff7}}
\and M.~Lombardi\orcid{0000-0002-3336-4965}\inst{\ref{aff4},\ref{aff2}}
\and R.~Ratta\orcid{0009-0009-4651-1248}\inst{\ref{aff1}}
\and M.~Fogliardi\orcid{0009-0006-4964-5311}\inst{\ref{aff1}}
\and G.~Di~Rosa\orcid{0009-0001-9416-0923}\inst{\ref{aff1}}
\and D.~Abriola\orcid{0009-0005-4230-3266}\inst{\ref{aff4}}
\and M.~D'Addona\orcid{0000-0003-3445-0483}\inst{\ref{aff3},\ref{aff8}}
\and G.~Granata\orcid{0000-0002-9512-3788}\inst{\ref{aff9},\ref{aff1}}
\and L.~Leuzzi\orcid{0009-0006-4479-7017}\inst{\ref{aff2}}
\and A.~Mercurio\orcid{0000-0001-9261-7849}\inst{\ref{aff3},\ref{aff8},\ref{aff10}}
\and J.~M.~Palencia\orcid{0000-0003-0942-817X}\inst{\ref{aff6}}
\and S.~Schuldt\orcid{0000-0003-2497-6334}\inst{\ref{aff4},\ref{aff7}}
\and E.~Vanzella\orcid{0000-0002-5057-135X}\inst{\ref{aff2}}
\and C.~Tortora\orcid{0000-0001-7958-6531}\inst{\ref{aff3}}
\and B.~Altieri\orcid{0000-0003-3936-0284}\inst{\ref{aff11}}
\and S.~Andreon\orcid{0000-0002-2041-8784}\inst{\ref{aff12}}
\and N.~Auricchio\orcid{0000-0003-4444-8651}\inst{\ref{aff2}}
\and C.~Baccigalupi\orcid{0000-0002-8211-1630}\inst{\ref{aff13},\ref{aff14},\ref{aff15},\ref{aff16}}
\and M.~Baldi\orcid{0000-0003-4145-1943}\inst{\ref{aff17},\ref{aff2},\ref{aff5}}
\and A.~Balestra\orcid{0000-0002-6967-261X}\inst{\ref{aff18}}
\and S.~Bardelli\orcid{0000-0002-8900-0298}\inst{\ref{aff2}}
\and P.~Battaglia\orcid{0000-0002-7337-5909}\inst{\ref{aff2}}
\and A.~Biviano\orcid{0000-0002-0857-0732}\inst{\ref{aff14},\ref{aff13}}
\and E.~Branchini\orcid{0000-0002-0808-6908}\inst{\ref{aff19},\ref{aff20},\ref{aff12}}
\and M.~Brescia\orcid{0000-0001-9506-5680}\inst{\ref{aff21},\ref{aff3}}
\and S.~Camera\orcid{0000-0003-3399-3574}\inst{\ref{aff22},\ref{aff23},\ref{aff24}}
\and G.~Ca\~nas-Herrera\orcid{0000-0003-2796-2149}\inst{\ref{aff25},\ref{aff26}}
\and V.~Capobianco\orcid{0000-0002-3309-7692}\inst{\ref{aff24}}
\and C.~Carbone\orcid{0000-0003-0125-3563}\inst{\ref{aff7}}
\and J.~Carretero\orcid{0000-0002-3130-0204}\inst{\ref{aff27},\ref{aff28}}
\and M.~Castellano\orcid{0000-0001-9875-8263}\inst{\ref{aff29}}
\and G.~Castignani\orcid{0000-0001-6831-0687}\inst{\ref{aff2}}
\and S.~Cavuoti\orcid{0000-0002-3787-4196}\inst{\ref{aff3},\ref{aff30}}
\and A.~Cimatti\inst{\ref{aff31}}
\and C.~Colodro-Conde\inst{\ref{aff32}}
\and G.~Congedo\orcid{0000-0003-2508-0046}\inst{\ref{aff33}}
\and L.~Conversi\orcid{0000-0002-6710-8476}\inst{\ref{aff34},\ref{aff11}}
\and Y.~Copin\orcid{0000-0002-5317-7518}\inst{\ref{aff35}}
\and A.~Costille\inst{\ref{aff36}}
\and F.~Courbin\orcid{0000-0003-0758-6510}\inst{\ref{aff37},\ref{aff38},\ref{aff39}}
\and H.~M.~Courtois\orcid{0000-0003-0509-1776}\inst{\ref{aff40}}
\and M.~Cropper\orcid{0000-0003-4571-9468}\inst{\ref{aff41}}
\and A.~Da~Silva\orcid{0000-0002-6385-1609}\inst{\ref{aff42},\ref{aff43}}
\and H.~Degaudenzi\orcid{0000-0002-5887-6799}\inst{\ref{aff44}}
\and G.~De~Lucia\orcid{0000-0002-6220-9104}\inst{\ref{aff14}}
\and H.~Dole\orcid{0000-0002-9767-3839}\inst{\ref{aff45}}
\and F.~Dubath\orcid{0000-0002-6533-2810}\inst{\ref{aff44}}
\and C.~A.~J.~Duncan\orcid{0009-0003-3573-0791}\inst{\ref{aff33}}
\and X.~Dupac\inst{\ref{aff11}}
\and S.~Dusini\orcid{0000-0002-1128-0664}\inst{\ref{aff46}}
\and S.~Escoffier\orcid{0000-0002-2847-7498}\inst{\ref{aff47}}
\and M.~Fabricius\orcid{0000-0002-7025-6058}\inst{\ref{aff48},\ref{aff49}}
\and M.~Farina\orcid{0000-0002-3089-7846}\inst{\ref{aff50}}
\and R.~Farinelli\inst{\ref{aff2}}
\and F.~Faustini\orcid{0000-0001-6274-5145}\inst{\ref{aff29},\ref{aff51}}
\and S.~Ferriol\inst{\ref{aff35}}
\and F.~Finelli\orcid{0000-0002-6694-3269}\inst{\ref{aff2},\ref{aff52}}
\and M.~Frailis\orcid{0000-0002-7400-2135}\inst{\ref{aff14}}
\and E.~Franceschi\orcid{0000-0002-0585-6591}\inst{\ref{aff2}}
\and M.~Fumana\orcid{0000-0001-6787-5950}\inst{\ref{aff7}}
\and S.~Galeotta\orcid{0000-0002-3748-5115}\inst{\ref{aff14}}
\and K.~George\orcid{0000-0002-1734-8455}\inst{\ref{aff53}}
\and W.~Gillard\orcid{0000-0003-4744-9748}\inst{\ref{aff47}}
\and B.~Gillis\orcid{0000-0002-4478-1270}\inst{\ref{aff33}}
\and C.~Giocoli\orcid{0000-0002-9590-7961}\inst{\ref{aff2},\ref{aff5}}
\and J.~Gracia-Carpio\inst{\ref{aff48}}
\and A.~Grazian\orcid{0000-0002-5688-0663}\inst{\ref{aff18}}
\and F.~Grupp\inst{\ref{aff48},\ref{aff49}}
\and L.~Guzzo\orcid{0000-0001-8264-5192}\inst{\ref{aff4},\ref{aff12},\ref{aff54}}
\and S.~V.~H.~Haugan\orcid{0000-0001-9648-7260}\inst{\ref{aff55}}
\and J.~Hoar\inst{\ref{aff11}}
\and W.~Holmes\inst{\ref{aff56}}
\and I.~M.~Hook\orcid{0000-0002-2960-978X}\inst{\ref{aff57}}
\and F.~Hormuth\inst{\ref{aff58}}
\and A.~Hornstrup\orcid{0000-0002-3363-0936}\inst{\ref{aff59},\ref{aff60}}
\and K.~Jahnke\orcid{0000-0003-3804-2137}\inst{\ref{aff61}}
\and M.~Jhabvala\inst{\ref{aff62}}
\and B.~Joachimi\orcid{0000-0001-7494-1303}\inst{\ref{aff63}}
\and E.~Keih\"anen\orcid{0000-0003-1804-7715}\inst{\ref{aff64}}
\and S.~Kermiche\orcid{0000-0002-0302-5735}\inst{\ref{aff47}}
\and A.~Kiessling\orcid{0000-0002-2590-1273}\inst{\ref{aff56}}
\and M.~Kilbinger\orcid{0000-0001-9513-7138}\inst{\ref{aff65}}
\and B.~Kubik\orcid{0009-0006-5823-4880}\inst{\ref{aff35}}
\and M.~Kunz\orcid{0000-0002-3052-7394}\inst{\ref{aff66}}
\and H.~Kurki-Suonio\orcid{0000-0002-4618-3063}\inst{\ref{aff67},\ref{aff68}}
\and R.~Laureijs\inst{\ref{aff69}}
\and A.~M.~C.~Le~Brun\orcid{0000-0002-0936-4594}\inst{\ref{aff70}}
\and D.~Le~Mignant\orcid{0000-0002-5339-5515}\inst{\ref{aff36}}
\and S.~Ligori\orcid{0000-0003-4172-4606}\inst{\ref{aff24}}
\and P.~B.~Lilje\orcid{0000-0003-4324-7794}\inst{\ref{aff55}}
\and V.~Lindholm\orcid{0000-0003-2317-5471}\inst{\ref{aff67},\ref{aff68}}
\and I.~Lloro\orcid{0000-0001-5966-1434}\inst{\ref{aff71}}
\and G.~Mainetti\orcid{0000-0003-2384-2377}\inst{\ref{aff72}}
\and D.~Maino\inst{\ref{aff4},\ref{aff7},\ref{aff54}}
\and E.~Maiorano\orcid{0000-0003-2593-4355}\inst{\ref{aff2}}
\and O.~Mansutti\orcid{0000-0001-5758-4658}\inst{\ref{aff14}}
\and O.~Marggraf\orcid{0000-0001-7242-3852}\inst{\ref{aff73}}
\and M.~Martinelli\orcid{0000-0002-6943-7732}\inst{\ref{aff29},\ref{aff74}}
\and N.~Martinet\orcid{0000-0003-2786-7790}\inst{\ref{aff36}}
\and F.~Marulli\orcid{0000-0002-8850-0303}\inst{\ref{aff75},\ref{aff2},\ref{aff5}}
\and R.~J.~Massey\orcid{0000-0002-6085-3780}\inst{\ref{aff76}}
\and E.~Medinaceli\orcid{0000-0002-4040-7783}\inst{\ref{aff2}}
\and S.~Mei\orcid{0000-0002-2849-559X}\inst{\ref{aff77},\ref{aff78}}
\and M.~Melchior\inst{\ref{aff79}}
\and Y.~Mellier\thanks{Deceased}\inst{\ref{aff80},\ref{aff81}}
\and E.~Merlin\orcid{0000-0001-6870-8900}\inst{\ref{aff29}}
\and G.~Meylan\inst{\ref{aff82}}
\and A.~Mora\orcid{0000-0002-1922-8529}\inst{\ref{aff83}}
\and M.~Moresco\orcid{0000-0002-7616-7136}\inst{\ref{aff75},\ref{aff2}}
\and L.~Moscardini\orcid{0000-0002-3473-6716}\inst{\ref{aff75},\ref{aff2},\ref{aff5}}
\and C.~Neissner\orcid{0000-0001-8524-4968}\inst{\ref{aff84},\ref{aff28}}
\and S.-M.~Niemi\orcid{0009-0005-0247-0086}\inst{\ref{aff25}}
\and C.~Padilla\orcid{0000-0001-7951-0166}\inst{\ref{aff84}}
\and S.~Paltani\orcid{0000-0002-8108-9179}\inst{\ref{aff44}}
\and F.~Pasian\orcid{0000-0002-4869-3227}\inst{\ref{aff14}}
\and K.~Pedersen\inst{\ref{aff85}}
\and W.~J.~Percival\orcid{0000-0002-0644-5727}\inst{\ref{aff86},\ref{aff87},\ref{aff88}}
\and V.~Pettorino\orcid{0000-0002-4203-9320}\inst{\ref{aff25}}
\and S.~Pires\orcid{0000-0002-0249-2104}\inst{\ref{aff65}}
\and G.~Polenta\orcid{0000-0003-4067-9196}\inst{\ref{aff51}}
\and M.~Poncet\inst{\ref{aff89}}
\and L.~A.~Popa\inst{\ref{aff90}}
\and L.~Pozzetti\orcid{0000-0001-7085-0412}\inst{\ref{aff2}}
\and F.~Raison\orcid{0000-0002-7819-6918}\inst{\ref{aff48}}
\and A.~Renzi\orcid{0000-0001-9856-1970}\inst{\ref{aff91},\ref{aff46}}
\and J.~Rhodes\orcid{0000-0002-4485-8549}\inst{\ref{aff56}}
\and G.~Riccio\inst{\ref{aff3}}
\and E.~Romelli\orcid{0000-0003-3069-9222}\inst{\ref{aff14}}
\and M.~Roncarelli\orcid{0000-0001-9587-7822}\inst{\ref{aff2}}
\and R.~Saglia\orcid{0000-0003-0378-7032}\inst{\ref{aff49},\ref{aff48}}
\and Z.~Sakr\orcid{0000-0002-4823-3757}\inst{\ref{aff92},\ref{aff93},\ref{aff94}}
\and A.~G.~S\'anchez\orcid{0000-0003-1198-831X}\inst{\ref{aff48}}
\and D.~Sapone\orcid{0000-0001-7089-4503}\inst{\ref{aff95}}
\and B.~Sartoris\orcid{0000-0003-1337-5269}\inst{\ref{aff49},\ref{aff14}}
\and P.~Schneider\orcid{0000-0001-8561-2679}\inst{\ref{aff73}}
\and T.~Schrabback\orcid{0000-0002-6987-7834}\inst{\ref{aff96}}
\and A.~Secroun\orcid{0000-0003-0505-3710}\inst{\ref{aff47}}
\and G.~Seidel\orcid{0000-0003-2907-353X}\inst{\ref{aff61}}
\and S.~Serrano\orcid{0000-0002-0211-2861}\inst{\ref{aff97},\ref{aff98},\ref{aff99}}
\and P.~Simon\inst{\ref{aff73}}
\and C.~Sirignano\orcid{0000-0002-0995-7146}\inst{\ref{aff91},\ref{aff46}}
\and G.~Sirri\orcid{0000-0003-2626-2853}\inst{\ref{aff5}}
\and L.~Stanco\orcid{0000-0002-9706-5104}\inst{\ref{aff46}}
\and J.~Steinwagner\orcid{0000-0001-7443-1047}\inst{\ref{aff48}}
\and P.~Tallada-Cresp\'{i}\orcid{0000-0002-1336-8328}\inst{\ref{aff27},\ref{aff28}}
\and A.~N.~Taylor\inst{\ref{aff33}}
\and I.~Tereno\orcid{0000-0002-4537-6218}\inst{\ref{aff42},\ref{aff100}}
\and N.~Tessore\orcid{0000-0002-9696-7931}\inst{\ref{aff63},\ref{aff41}}
\and S.~Toft\orcid{0000-0003-3631-7176}\inst{\ref{aff101},\ref{aff102}}
\and R.~Toledo-Moreo\orcid{0000-0002-2997-4859}\inst{\ref{aff103}}
\and F.~Torradeflot\orcid{0000-0003-1160-1517}\inst{\ref{aff28},\ref{aff27}}
\and I.~Tutusaus\orcid{0000-0002-3199-0399}\inst{\ref{aff99},\ref{aff97},\ref{aff93}}
\and E.~A.~Valentijn\inst{\ref{aff69}}
\and L.~Valenziano\orcid{0000-0002-1170-0104}\inst{\ref{aff2},\ref{aff52}}
\and J.~Valiviita\orcid{0000-0001-6225-3693}\inst{\ref{aff67},\ref{aff68}}
\and T.~Vassallo\orcid{0000-0001-6512-6358}\inst{\ref{aff14}}
\and G.~Verdoes~Kleijn\orcid{0000-0001-5803-2580}\inst{\ref{aff69}}
\and A.~Veropalumbo\orcid{0000-0003-2387-1194}\inst{\ref{aff12},\ref{aff20},\ref{aff19}}
\and Y.~Wang\orcid{0000-0002-4749-2984}\inst{\ref{aff104}}
\and J.~Weller\orcid{0000-0002-8282-2010}\inst{\ref{aff49},\ref{aff48}}
\and A.~Zacchei\orcid{0000-0003-0396-1192}\inst{\ref{aff14},\ref{aff13}}
\and G.~Zamorani\orcid{0000-0002-2318-301X}\inst{\ref{aff2}}
\and E.~Zucca\orcid{0000-0002-5845-8132}\inst{\ref{aff2}}
\and M.~Ballardini\orcid{0000-0003-4481-3559}\inst{\ref{aff1},\ref{aff105},\ref{aff2}}
\and M.~Bolzonella\orcid{0000-0003-3278-4607}\inst{\ref{aff2}}
\and E.~Bozzo\orcid{0000-0002-8201-1525}\inst{\ref{aff44}}
\and C.~Burigana\orcid{0000-0002-3005-5796}\inst{\ref{aff106},\ref{aff52}}
\and R.~Cabanac\orcid{0000-0001-6679-2600}\inst{\ref{aff93}}
\and M.~Calabrese\orcid{0000-0002-2637-2422}\inst{\ref{aff107},\ref{aff7}}
\and A.~Cappi\inst{\ref{aff2},\ref{aff108}}
\and D.~Di~Ferdinando\inst{\ref{aff5}}
\and J.~A.~Escartin~Vigo\inst{\ref{aff48}}
\and W.~G.~Hartley\inst{\ref{aff44}}
\and J.~Mart\'{i}n-Fleitas\orcid{0000-0002-8594-569X}\inst{\ref{aff109}}
\and S.~Matthew\orcid{0000-0001-8448-1697}\inst{\ref{aff33}}
\and N.~Mauri\orcid{0000-0001-8196-1548}\inst{\ref{aff31},\ref{aff5}}
\and R.~B.~Metcalf\orcid{0000-0003-3167-2574}\inst{\ref{aff75},\ref{aff2}}
\and A.~Pezzotta\orcid{0000-0003-0726-2268}\inst{\ref{aff12}}
\and M.~P\"ontinen\orcid{0000-0001-5442-2530}\inst{\ref{aff67}}
\and I.~Risso\orcid{0000-0003-2525-7761}\inst{\ref{aff12},\ref{aff20}}
\and V.~Scottez\orcid{0009-0008-3864-940X}\inst{\ref{aff80},\ref{aff110}}
\and M.~Sereno\orcid{0000-0003-0302-0325}\inst{\ref{aff2},\ref{aff5}}
\and M.~Tenti\orcid{0000-0002-4254-5901}\inst{\ref{aff5}}
\and M.~Viel\orcid{0000-0002-2642-5707}\inst{\ref{aff13},\ref{aff14},\ref{aff16},\ref{aff15},\ref{aff111}}
\and M.~Wiesmann\orcid{0009-0000-8199-5860}\inst{\ref{aff55}}
\and Y.~Akrami\orcid{0000-0002-2407-7956}\inst{\ref{aff112},\ref{aff113}}
\and I.~T.~Andika\orcid{0000-0001-6102-9526}\inst{\ref{aff114},\ref{aff115}}
\and S.~Anselmi\orcid{0000-0002-3579-9583}\inst{\ref{aff46},\ref{aff91},\ref{aff116}}
\and M.~Archidiacono\orcid{0000-0003-4952-9012}\inst{\ref{aff4},\ref{aff54}}
\and F.~Atrio-Barandela\orcid{0000-0002-2130-2513}\inst{\ref{aff117}}
\and E.~Aubourg\orcid{0000-0002-5592-023X}\inst{\ref{aff77},\ref{aff118}}
\and D.~Bertacca\orcid{0000-0002-2490-7139}\inst{\ref{aff91},\ref{aff18},\ref{aff46}}
\and M.~Bethermin\orcid{0000-0002-3915-2015}\inst{\ref{aff119}}
\and A.~Blanchard\orcid{0000-0001-8555-9003}\inst{\ref{aff93}}
\and L.~Blot\orcid{0000-0002-9622-7167}\inst{\ref{aff120},\ref{aff70}}
\and H.~B\"ohringer\orcid{0000-0001-8241-4204}\inst{\ref{aff48},\ref{aff53},\ref{aff121}}
\and M.~Bonici\orcid{0000-0002-8430-126X}\inst{\ref{aff86},\ref{aff7}}
\and S.~Borgani\orcid{0000-0001-6151-6439}\inst{\ref{aff122},\ref{aff13},\ref{aff14},\ref{aff15},\ref{aff111}}
\and M.~L.~Brown\orcid{0000-0002-0370-8077}\inst{\ref{aff123}}
\and S.~Bruton\orcid{0000-0002-6503-5218}\inst{\ref{aff124}}
\and A.~Calabro\orcid{0000-0003-2536-1614}\inst{\ref{aff29}}
\and B.~Camacho~Quevedo\orcid{0000-0002-8789-4232}\inst{\ref{aff13},\ref{aff16},\ref{aff14}}
\and F.~Caro\inst{\ref{aff29}}
\and C.~S.~Carvalho\inst{\ref{aff100}}
\and T.~Castro\orcid{0000-0002-6292-3228}\inst{\ref{aff14},\ref{aff15},\ref{aff13},\ref{aff111}}
\and B.~Cl\'ement\orcid{0000-0002-7966-3661}\inst{\ref{aff82},\ref{aff125}}
\and F.~Cogato\orcid{0000-0003-4632-6113}\inst{\ref{aff75},\ref{aff2}}
\and S.~Conseil\orcid{0000-0002-3657-4191}\inst{\ref{aff35}}
\and A.~R.~Cooray\orcid{0000-0002-3892-0190}\inst{\ref{aff126}}
\and O.~Cucciati\orcid{0000-0002-9336-7551}\inst{\ref{aff2}}
\and S.~Davini\orcid{0000-0003-3269-1718}\inst{\ref{aff20}}
\and F.~De~Paolis\orcid{0000-0001-6460-7563}\inst{\ref{aff127},\ref{aff128},\ref{aff129}}
\and G.~Desprez\orcid{0000-0001-8325-1742}\inst{\ref{aff69}}
\and A.~D\'iaz-S\'anchez\orcid{0000-0003-0748-4768}\inst{\ref{aff130}}
\and J.~J.~Diaz\orcid{0000-0003-2101-1078}\inst{\ref{aff32}}
\and S.~Di~Domizio\orcid{0000-0003-2863-5895}\inst{\ref{aff19},\ref{aff20}}
\and J.~M.~Diego\orcid{0000-0001-9065-3926}\inst{\ref{aff6}}
\and P.~Dimauro\orcid{0000-0001-7399-2854}\inst{\ref{aff131},\ref{aff29}}
\and P.-A.~Duc\orcid{0000-0003-3343-6284}\inst{\ref{aff119}}
\and M.~Y.~Elkhashab\orcid{0000-0001-9306-2603}\inst{\ref{aff14},\ref{aff15},\ref{aff122},\ref{aff13}}
\and A.~Enia\orcid{0000-0002-0200-2857}\inst{\ref{aff2},\ref{aff17}}
\and Y.~Fang\orcid{0000-0002-0334-6950}\inst{\ref{aff49}}
\and A.~Finoguenov\orcid{0000-0002-4606-5403}\inst{\ref{aff67}}
\and A.~Fontana\orcid{0000-0003-3820-2823}\inst{\ref{aff29}}
\and A.~Franco\orcid{0000-0002-4761-366X}\inst{\ref{aff128},\ref{aff127},\ref{aff129}}
\and K.~Ganga\orcid{0000-0001-8159-8208}\inst{\ref{aff77}}
\and J.~Garc\'ia-Bellido\orcid{0000-0002-9370-8360}\inst{\ref{aff112}}
\and T.~Gasparetto\orcid{0000-0002-7913-4866}\inst{\ref{aff29}}
\and V.~Gautard\inst{\ref{aff132}}
\and R.~Gavazzi\orcid{0000-0002-5540-6935}\inst{\ref{aff36},\ref{aff81}}
\and E.~Gaztanaga\orcid{0000-0001-9632-0815}\inst{\ref{aff99},\ref{aff97},\ref{aff9}}
\and F.~Giacomini\orcid{0000-0002-3129-2814}\inst{\ref{aff5}}
\and F.~Gianotti\orcid{0000-0003-4666-119X}\inst{\ref{aff2}}
\and A.~H.~Gonzalez\orcid{0000-0002-0933-8601}\inst{\ref{aff133}}
\and G.~Gozaliasl\orcid{0000-0002-0236-919X}\inst{\ref{aff134},\ref{aff67}}
\and M.~Guidi\orcid{0000-0001-9408-1101}\inst{\ref{aff17},\ref{aff2}}
\and C.~M.~Gutierrez\orcid{0000-0001-7854-783X}\inst{\ref{aff135}}
\and S.~Hemmati\orcid{0000-0003-2226-5395}\inst{\ref{aff136}}
\and H.~Hildebrandt\orcid{0000-0002-9814-3338}\inst{\ref{aff137}}
\and J.~Hjorth\orcid{0000-0002-4571-2306}\inst{\ref{aff85}}
\and J.~J.~E.~Kajava\orcid{0000-0002-3010-8333}\inst{\ref{aff138},\ref{aff139}}
\and Y.~Kang\orcid{0009-0000-8588-7250}\inst{\ref{aff44}}
\and V.~Kansal\orcid{0000-0002-4008-6078}\inst{\ref{aff140},\ref{aff141}}
\and D.~Karagiannis\orcid{0000-0002-4927-0816}\inst{\ref{aff1},\ref{aff142}}
\and K.~Kiiveri\inst{\ref{aff64}}
\and J.~Kim\orcid{0000-0003-2776-2761}\inst{\ref{aff143}}
\and C.~C.~Kirkpatrick\inst{\ref{aff64}}
\and S.~Kruk\orcid{0000-0001-8010-8879}\inst{\ref{aff11}}
\and J.~Le~Graet\orcid{0000-0001-6523-7971}\inst{\ref{aff47}}
\and L.~Legrand\orcid{0000-0003-0610-5252}\inst{\ref{aff144},\ref{aff145}}
\and M.~Lembo\orcid{0000-0002-5271-5070}\inst{\ref{aff81},\ref{aff1},\ref{aff105}}
\and F.~Lepori\orcid{0009-0000-5061-7138}\inst{\ref{aff146}}
\and G.~Leroy\orcid{0009-0004-2523-4425}\inst{\ref{aff147},\ref{aff76}}
\and G.~F.~Lesci\orcid{0000-0002-4607-2830}\inst{\ref{aff75},\ref{aff2}}
\and J.~Lesgourgues\orcid{0000-0001-7627-353X}\inst{\ref{aff148}}
\and T.~I.~Liaudat\orcid{0000-0002-9104-314X}\inst{\ref{aff118}}
\and S.~J.~Liu\orcid{0000-0001-7680-2139}\inst{\ref{aff50}}
\and A.~Loureiro\orcid{0000-0002-4371-0876}\inst{\ref{aff149},\ref{aff150}}
\and J.~Macias-Perez\orcid{0000-0002-5385-2763}\inst{\ref{aff151}}
\and M.~Magliocchetti\orcid{0000-0001-9158-4838}\inst{\ref{aff50}}
\and F.~Mannucci\orcid{0000-0002-4803-2381}\inst{\ref{aff152}}
\and R.~Maoli\orcid{0000-0002-6065-3025}\inst{\ref{aff153},\ref{aff29}}
\and C.~J.~A.~P.~Martins\orcid{0000-0002-4886-9261}\inst{\ref{aff154},\ref{aff155}}
\and L.~Maurin\orcid{0000-0002-8406-0857}\inst{\ref{aff45}}
\and C.~J.~R.~McPartland\orcid{0000-0003-0639-025X}\inst{\ref{aff60},\ref{aff102}}
\and M.~Miluzio\inst{\ref{aff11},\ref{aff156}}
\and P.~Monaco\orcid{0000-0003-2083-7564}\inst{\ref{aff122},\ref{aff14},\ref{aff15},\ref{aff13}}
\and C.~Moretti\orcid{0000-0003-3314-8936}\inst{\ref{aff14},\ref{aff13},\ref{aff15},\ref{aff16}}
\and G.~Morgante\inst{\ref{aff2}}
\and C.~Murray\inst{\ref{aff77}}
\and K.~Naidoo\orcid{0000-0002-9182-1802}\inst{\ref{aff9},\ref{aff63}}
\and A.~Navarro-Alsina\orcid{0000-0002-3173-2592}\inst{\ref{aff73}}
\and S.~Nesseris\orcid{0000-0002-0567-0324}\inst{\ref{aff112}}
\and D.~Paoletti\orcid{0000-0003-4761-6147}\inst{\ref{aff2},\ref{aff52}}
\and F.~Passalacqua\orcid{0000-0002-8606-4093}\inst{\ref{aff91},\ref{aff46}}
\and K.~Paterson\orcid{0000-0001-8340-3486}\inst{\ref{aff61}}
\and A.~Pisani\orcid{0000-0002-6146-4437}\inst{\ref{aff47}}
\and D.~Potter\orcid{0000-0002-0757-5195}\inst{\ref{aff146}}
\and S.~Quai\orcid{0000-0002-0449-8163}\inst{\ref{aff75},\ref{aff2}}
\and M.~Radovich\orcid{0000-0002-3585-866X}\inst{\ref{aff18}}
\and P.-F.~Rocci\inst{\ref{aff45}}
\and S.~Sacquegna\orcid{0000-0002-8433-6630}\inst{\ref{aff157},\ref{aff127},\ref{aff128}}
\and M.~Sahl\'en\orcid{0000-0003-0973-4804}\inst{\ref{aff158}}
\and D.~B.~Sanders\orcid{0000-0002-1233-9998}\inst{\ref{aff159}}
\and E.~Sarpa\orcid{0000-0002-1256-655X}\inst{\ref{aff16},\ref{aff111},\ref{aff15}}
\and A.~Schneider\orcid{0000-0001-7055-8104}\inst{\ref{aff146}}
\and D.~Sciotti\orcid{0009-0008-4519-2620}\inst{\ref{aff29},\ref{aff74}}
\and E.~Sellentin\inst{\ref{aff160},\ref{aff26}}
\and L.~C.~Smith\orcid{0000-0002-3259-2771}\inst{\ref{aff161}}
\and J.~G.~Sorce\orcid{0000-0002-2307-2432}\inst{\ref{aff162},\ref{aff45}}
\and K.~Tanidis\orcid{0000-0001-9843-5130}\inst{\ref{aff143}}
\and C.~Tao\orcid{0000-0001-7961-8177}\inst{\ref{aff47}}
\and G.~Testera\inst{\ref{aff20}}
\and R.~Teyssier\orcid{0000-0001-7689-0933}\inst{\ref{aff163}}
\and S.~Tosi\orcid{0000-0002-7275-9193}\inst{\ref{aff19},\ref{aff20},\ref{aff12}}
\and A.~Troja\orcid{0000-0003-0239-4595}\inst{\ref{aff91},\ref{aff46}}
\and M.~Tucci\inst{\ref{aff44}}
\and C.~Valieri\inst{\ref{aff5}}
\and A.~Venhola\orcid{0000-0001-6071-4564}\inst{\ref{aff164}}
\and D.~Vergani\orcid{0000-0003-0898-2216}\inst{\ref{aff2}}
\and G.~Verza\orcid{0000-0002-1886-8348}\inst{\ref{aff165}}
\and P.~Vielzeuf\orcid{0000-0003-2035-9339}\inst{\ref{aff47}}
\and N.~A.~Walton\orcid{0000-0003-3983-8778}\inst{\ref{aff161}}
\and D.~Scott\orcid{0000-0002-6878-9840}\inst{\ref{aff166}}}
										   
\institute{Dipartimento di Fisica e Scienze della Terra, Universit\`a degli Studi di Ferrara, Via Giuseppe Saragat 1, 44122 Ferrara, Italy\label{aff1}
\and
INAF-Osservatorio di Astrofisica e Scienza dello Spazio di Bologna, Via Piero Gobetti 93/3, 40129 Bologna, Italy\label{aff2}
\and
INAF-Osservatorio Astronomico di Capodimonte, Via Moiariello 16, 80131 Napoli, Italy\label{aff3}
\and
Dipartimento di Fisica "Aldo Pontremoli", Universit\`a degli Studi di Milano, Via Celoria 16, 20133 Milano, Italy\label{aff4}
\and
INFN-Sezione di Bologna, Viale Berti Pichat 6/2, 40127 Bologna, Italy\label{aff5}
\and
Instituto de F\'isica de Cantabria, Edificio Juan Jord\'a, Avenida de los Castros, 39005 Santander, Spain\label{aff6}
\and
INAF-IASF Milano, Via Alfonso Corti 12, 20133 Milano, Italy\label{aff7}
\and
Universita di Salerno, Dipartimento di Fisica "E.R. Caianiello", Via Giovanni Paolo II 132, I-84084 Fisciano (SA), Italy\label{aff8}
\and
Institute of Cosmology and Gravitation, University of Portsmouth, Portsmouth PO1 3FX, UK\label{aff9}
\and
INFN -- Gruppo Collegato di Salerno - Sezione di Napoli, Dipartimento di Fisica "E.R. Caianiello", Universita di Salerno, via Giovanni Paolo II, 132 - I-84084 Fisciano (SA), Italy\label{aff10}
\and
ESAC/ESA, Camino Bajo del Castillo, s/n., Urb. Villafranca del Castillo, 28692 Villanueva de la Ca\~nada, Madrid, Spain\label{aff11}
\and
INAF-Osservatorio Astronomico di Brera, Via Brera 28, 20122 Milano, Italy\label{aff12}
\and
IFPU, Institute for Fundamental Physics of the Universe, via Beirut 2, 34151 Trieste, Italy\label{aff13}
\and
INAF-Osservatorio Astronomico di Trieste, Via G. B. Tiepolo 11, 34143 Trieste, Italy\label{aff14}
\and
INFN, Sezione di Trieste, Via Valerio 2, 34127 Trieste TS, Italy\label{aff15}
\and
SISSA, International School for Advanced Studies, Via Bonomea 265, 34136 Trieste TS, Italy\label{aff16}
\and
Dipartimento di Fisica e Astronomia, Universit\`a di Bologna, Via Gobetti 93/2, 40129 Bologna, Italy\label{aff17}
\and
INAF-Osservatorio Astronomico di Padova, Via dell'Osservatorio 5, 35122 Padova, Italy\label{aff18}
\and
Dipartimento di Fisica, Universit\`a di Genova, Via Dodecaneso 33, 16146, Genova, Italy\label{aff19}
\and
INFN-Sezione di Genova, Via Dodecaneso 33, 16146, Genova, Italy\label{aff20}
\and
Department of Physics "E. Pancini", University Federico II, Via Cinthia 6, 80126, Napoli, Italy\label{aff21}
\and
Dipartimento di Fisica, Universit\`a degli Studi di Torino, Via P. Giuria 1, 10125 Torino, Italy\label{aff22}
\and
INFN-Sezione di Torino, Via P. Giuria 1, 10125 Torino, Italy\label{aff23}
\and
INAF-Osservatorio Astrofisico di Torino, Via Osservatorio 20, 10025 Pino Torinese (TO), Italy\label{aff24}
\and
European Space Agency/ESTEC, Keplerlaan 1, 2201 AZ Noordwijk, The Netherlands\label{aff25}
\and
Leiden Observatory, Leiden University, Einsteinweg 55, 2333 CC Leiden, The Netherlands\label{aff26}
\and
Centro de Investigaciones Energ\'eticas, Medioambientales y Tecnol\'ogicas (CIEMAT), Avenida Complutense 40, 28040 Madrid, Spain\label{aff27}
\and
Port d'Informaci\'{o} Cient\'{i}fica, Campus UAB, C. Albareda s/n, 08193 Bellaterra (Barcelona), Spain\label{aff28}
\and
INAF-Osservatorio Astronomico di Roma, Via Frascati 33, 00078 Monteporzio Catone, Italy\label{aff29}
\and
INFN section of Naples, Via Cinthia 6, 80126, Napoli, Italy\label{aff30}
\and
Dipartimento di Fisica e Astronomia "Augusto Righi" - Alma Mater Studiorum Universit\`a di Bologna, Viale Berti Pichat 6/2, 40127 Bologna, Italy\label{aff31}
\and
Instituto de Astrof\'{\i}sica de Canarias, E-38205 La Laguna, Tenerife, Spain\label{aff32}
\and
Institute for Astronomy, University of Edinburgh, Royal Observatory, Blackford Hill, Edinburgh EH9 3HJ, UK\label{aff33}
\and
European Space Agency/ESRIN, Largo Galileo Galilei 1, 00044 Frascati, Roma, Italy\label{aff34}
\and
Universit\'e Claude Bernard Lyon 1, CNRS/IN2P3, IP2I Lyon, UMR 5822, Villeurbanne, F-69100, France\label{aff35}
\and
Aix-Marseille Universit\'e, CNRS, CNES, LAM, Marseille, France\label{aff36}
\and
Institut de Ci\`{e}ncies del Cosmos (ICCUB), Universitat de Barcelona (IEEC-UB), Mart\'{i} i Franqu\`{e}s 1, 08028 Barcelona, Spain\label{aff37}
\and
Instituci\'o Catalana de Recerca i Estudis Avan\c{c}ats (ICREA), Passeig de Llu\'{\i}s Companys 23, 08010 Barcelona, Spain\label{aff38}
\and
Institut de Ciencies de l'Espai (IEEC-CSIC), Campus UAB, Carrer de Can Magrans, s/n Cerdanyola del Vall\'es, 08193 Barcelona, Spain\label{aff39}
\and
UCB Lyon 1, CNRS/IN2P3, IUF, IP2I Lyon, 4 rue Enrico Fermi, 69622 Villeurbanne, France\label{aff40}
\and
Mullard Space Science Laboratory, University College London, Holmbury St Mary, Dorking, Surrey RH5 6NT, UK\label{aff41}
\and
Departamento de F\'isica, Faculdade de Ci\^encias, Universidade de Lisboa, Edif\'icio C8, Campo Grande, PT1749-016 Lisboa, Portugal\label{aff42}
\and
Instituto de Astrof\'isica e Ci\^encias do Espa\c{c}o, Faculdade de Ci\^encias, Universidade de Lisboa, Campo Grande, 1749-016 Lisboa, Portugal\label{aff43}
\and
Department of Astronomy, University of Geneva, ch. d'Ecogia 16, 1290 Versoix, Switzerland\label{aff44}
\and
Universit\'e Paris-Saclay, CNRS, Institut d'astrophysique spatiale, 91405, Orsay, France\label{aff45}
\and
INFN-Padova, Via Marzolo 8, 35131 Padova, Italy\label{aff46}
\and
Aix-Marseille Universit\'e, CNRS/IN2P3, CPPM, Marseille, France\label{aff47}
\and
Max Planck Institute for Extraterrestrial Physics, Giessenbachstr. 1, 85748 Garching, Germany\label{aff48}
\and
Universit\"ats-Sternwarte M\"unchen, Fakult\"at f\"ur Physik, Ludwig-Maximilians-Universit\"at M\"unchen, Scheinerstr.~1, 81679 M\"unchen, Germany\label{aff49}
\and
INAF-Istituto di Astrofisica e Planetologia Spaziali, via del Fosso del Cavaliere, 100, 00100 Roma, Italy\label{aff50}
\and
Space Science Data Center, Italian Space Agency, via del Politecnico snc, 00133 Roma, Italy\label{aff51}
\and
INFN-Bologna, Via Irnerio 46, 40126 Bologna, Italy\label{aff52}
\and
University Observatory, LMU Faculty of Physics, Scheinerstr.~1, 81679 Munich, Germany\label{aff53}
\and
INFN-Sezione di Milano, Via Celoria 16, 20133 Milano, Italy\label{aff54}
\and
Institute of Theoretical Astrophysics, University of Oslo, P.O. Box 1029 Blindern, 0315 Oslo, Norway\label{aff55}
\and
Jet Propulsion Laboratory, California Institute of Technology, 4800 Oak Grove Drive, Pasadena, CA, 91109, USA\label{aff56}
\and
Department of Physics, Lancaster University, Lancaster, LA1 4YB, UK\label{aff57}
\and
Felix Hormuth Engineering, Goethestr. 17, 69181 Leimen, Germany\label{aff58}
\and
Technical University of Denmark, Elektrovej 327, 2800 Kgs. Lyngby, Denmark\label{aff59}
\and
Cosmic Dawn Center (DAWN), Denmark\label{aff60}
\and
Max-Planck-Institut f\"ur Astronomie, K\"onigstuhl 17, 69117 Heidelberg, Germany\label{aff61}
\and
NASA Goddard Space Flight Center, Greenbelt, MD 20771, USA\label{aff62}
\and
Department of Physics and Astronomy, University College London, Gower Street, London WC1E 6BT, UK\label{aff63}
\and
Department of Physics and Helsinki Institute of Physics, Gustaf H\"allstr\"omin katu 2, University of Helsinki, 00014 Helsinki, Finland\label{aff64}
\and
Universit\'e Paris-Saclay, Universit\'e Paris Cit\'e, CEA, CNRS, AIM, 91191, Gif-sur-Yvette, France\label{aff65}
\and
Universit\'e de Gen\`eve, D\'epartement de Physique Th\'eorique and Centre for Astroparticle Physics, 24 quai Ernest-Ansermet, CH-1211 Gen\`eve 4, Switzerland\label{aff66}
\and
Department of Physics, P.O. Box 64, University of Helsinki, 00014 Helsinki, Finland\label{aff67}
\and
Helsinki Institute of Physics, Gustaf H{\"a}llstr{\"o}min katu 2, University of Helsinki, 00014 Helsinki, Finland\label{aff68}
\and
Kapteyn Astronomical Institute, University of Groningen, PO Box 800, 9700 AV Groningen, The Netherlands\label{aff69}
\and
Laboratoire d'etude de l'Univers et des phenomenes eXtremes, Observatoire de Paris, Universit\'e PSL, Sorbonne Universit\'e, CNRS, 92190 Meudon, France\label{aff70}
\and
SKAO, Jodrell Bank, Lower Withington, Macclesfield SK11 9FT, UK\label{aff71}
\and
Centre de Calcul de l'IN2P3/CNRS, 21 avenue Pierre de Coubertin 69627 Villeurbanne Cedex, France\label{aff72}
\and
Universit\"at Bonn, Argelander-Institut f\"ur Astronomie, Auf dem H\"ugel 71, 53121 Bonn, Germany\label{aff73}
\and
INFN-Sezione di Roma, Piazzale Aldo Moro, 2 - c/o Dipartimento di Fisica, Edificio G. Marconi, 00185 Roma, Italy\label{aff74}
\and
Dipartimento di Fisica e Astronomia "Augusto Righi" - Alma Mater Studiorum Universit\`a di Bologna, via Piero Gobetti 93/2, 40129 Bologna, Italy\label{aff75}
\and
Department of Physics, Institute for Computational Cosmology, Durham University, South Road, Durham, DH1 3LE, UK\label{aff76}
\and
Universit\'e Paris Cit\'e, CNRS, Astroparticule et Cosmologie, 75013 Paris, France\label{aff77}
\and
CNRS-UCB International Research Laboratory, Centre Pierre Bin\'etruy, IRL2007, CPB-IN2P3, Berkeley, USA\label{aff78}
\and
University of Applied Sciences and Arts of Northwestern Switzerland, School of Engineering, 5210 Windisch, Switzerland\label{aff79}
\and
Institut d'Astrophysique de Paris, 98bis Boulevard Arago, 75014, Paris, France\label{aff80}
\and
Institut d'Astrophysique de Paris, UMR 7095, CNRS, and Sorbonne Universit\'e, 98 bis boulevard Arago, 75014 Paris, France\label{aff81}
\and
Institute of Physics, Laboratory of Astrophysics, Ecole Polytechnique F\'ed\'erale de Lausanne (EPFL), Observatoire de Sauverny, 1290 Versoix, Switzerland\label{aff82}
\and
Telespazio UK S.L. for European Space Agency (ESA), Camino bajo del Castillo, s/n, Urbanizacion Villafranca del Castillo, Villanueva de la Ca\~nada, 28692 Madrid, Spain\label{aff83}
\and
Institut de F\'{i}sica d'Altes Energies (IFAE), The Barcelona Institute of Science and Technology, Campus UAB, 08193 Bellaterra (Barcelona), Spain\label{aff84}
\and
DARK, Niels Bohr Institute, University of Copenhagen, Jagtvej 155, 2200 Copenhagen, Denmark\label{aff85}
\and
Waterloo Centre for Astrophysics, University of Waterloo, Waterloo, Ontario N2L 3G1, Canada\label{aff86}
\and
Department of Physics and Astronomy, University of Waterloo, Waterloo, Ontario N2L 3G1, Canada\label{aff87}
\and
Perimeter Institute for Theoretical Physics, Waterloo, Ontario N2L 2Y5, Canada\label{aff88}
\and
Centre National d'Etudes Spatiales -- Centre spatial de Toulouse, 18 avenue Edouard Belin, 31401 Toulouse Cedex 9, France\label{aff89}
\and
Institute of Space Science, Str. Atomistilor, nr. 409 M\u{a}gurele, Ilfov, 077125, Romania\label{aff90}
\and
Dipartimento di Fisica e Astronomia "G. Galilei", Universit\`a di Padova, Via Marzolo 8, 35131 Padova, Italy\label{aff91}
\and
Institut f\"ur Theoretische Physik, University of Heidelberg, Philosophenweg 16, 69120 Heidelberg, Germany\label{aff92}
\and
Institut de Recherche en Astrophysique et Plan\'etologie (IRAP), Universit\'e de Toulouse, CNRS, UPS, CNES, 14 Av. Edouard Belin, 31400 Toulouse, France\label{aff93}
\and
Universit\'e St Joseph; Faculty of Sciences, Beirut, Lebanon\label{aff94}
\and
Departamento de F\'isica, FCFM, Universidad de Chile, Blanco Encalada 2008, Santiago, Chile\label{aff95}
\and
Universit\"at Innsbruck, Institut f\"ur Astro- und Teilchenphysik, Technikerstr. 25/8, 6020 Innsbruck, Austria\label{aff96}
\and
Institut d'Estudis Espacials de Catalunya (IEEC),  Edifici RDIT, Campus UPC, 08860 Castelldefels, Barcelona, Spain\label{aff97}
\and
Satlantis, University Science Park, Sede Bld 48940, Leioa-Bilbao, Spain\label{aff98}
\and
Institute of Space Sciences (ICE, CSIC), Campus UAB, Carrer de Can Magrans, s/n, 08193 Barcelona, Spain\label{aff99}
\and
Instituto de Astrof\'isica e Ci\^encias do Espa\c{c}o, Faculdade de Ci\^encias, Universidade de Lisboa, Tapada da Ajuda, 1349-018 Lisboa, Portugal\label{aff100}
\and
Cosmic Dawn Center (DAWN)\label{aff101}
\and
Niels Bohr Institute, University of Copenhagen, Jagtvej 128, 2200 Copenhagen, Denmark\label{aff102}
\and
Universidad Polit\'ecnica de Cartagena, Departamento de Electr\'onica y Tecnolog\'ia de Computadoras,  Plaza del Hospital 1, 30202 Cartagena, Spain\label{aff103}
\and
Infrared Processing and Analysis Center, California Institute of Technology, Pasadena, CA 91125, USA\label{aff104}
\and
Istituto Nazionale di Fisica Nucleare, Sezione di Ferrara, Via Giuseppe Saragat 1, 44122 Ferrara, Italy\label{aff105}
\and
INAF, Istituto di Radioastronomia, Via Piero Gobetti 101, 40129 Bologna, Italy\label{aff106}
\and
Astronomical Observatory of the Autonomous Region of the Aosta Valley (OAVdA), Loc. Lignan 39, I-11020, Nus (Aosta Valley), Italy\label{aff107}
\and
Universit\'e C\^{o}te d'Azur, Observatoire de la C\^{o}te d'Azur, CNRS, Laboratoire Lagrange, Bd de l'Observatoire, CS 34229, 06304 Nice cedex 4, France\label{aff108}
\and
Aurora Technology for European Space Agency (ESA), Camino bajo del Castillo, s/n, Urbanizacion Villafranca del Castillo, Villanueva de la Ca\~nada, 28692 Madrid, Spain\label{aff109}
\and
ICL, Junia, Universit\'e Catholique de Lille, LITL, 59000 Lille, France\label{aff110}
\and
ICSC - Centro Nazionale di Ricerca in High Performance Computing, Big Data e Quantum Computing, Via Magnanelli 2, Bologna, Italy\label{aff111}
\and
Instituto de F\'isica Te\'orica UAM-CSIC, Campus de Cantoblanco, 28049 Madrid, Spain\label{aff112}
\and
CERCA/ISO, Department of Physics, Case Western Reserve University, 10900 Euclid Avenue, Cleveland, OH 44106, USA\label{aff113}
\and
Technical University of Munich, TUM School of Natural Sciences, Physics Department, James-Franck-Str.~1, 85748 Garching, Germany\label{aff114}
\and
Max-Planck-Institut f\"ur Astrophysik, Karl-Schwarzschild-Str.~1, 85748 Garching, Germany\label{aff115}
\and
Laboratoire Univers et Th\'eorie, Observatoire de Paris, Universit\'e PSL, Universit\'e Paris Cit\'e, CNRS, 92190 Meudon, France\label{aff116}
\and
Departamento de F{\'\i}sica Fundamental. Universidad de Salamanca. Plaza de la Merced s/n. 37008 Salamanca, Spain\label{aff117}
\and
IRFU, CEA, Universit\'e Paris-Saclay 91191 Gif-sur-Yvette Cedex, France\label{aff118}
\and
Universit\'e de Strasbourg, CNRS, Observatoire astronomique de Strasbourg, UMR 7550, 67000 Strasbourg, France\label{aff119}
\and
Center for Data-Driven Discovery, Kavli IPMU (WPI), UTIAS, The University of Tokyo, Kashiwa, Chiba 277-8583, Japan\label{aff120}
\and
Max-Planck-Institut f\"ur Physik, Boltzmannstr. 8, 85748 Garching, Germany\label{aff121}
\and
Dipartimento di Fisica - Sezione di Astronomia, Universit\`a di Trieste, Via Tiepolo 11, 34131 Trieste, Italy\label{aff122}
\and
Jodrell Bank Centre for Astrophysics, Department of Physics and Astronomy, University of Manchester, Oxford Road, Manchester M13 9PL, UK\label{aff123}
\and
California Institute of Technology, 1200 E California Blvd, Pasadena, CA 91125, USA\label{aff124}
\and
SCITAS, Ecole Polytechnique F\'ed\'erale de Lausanne (EPFL), 1015 Lausanne, Switzerland\label{aff125}
\and
Department of Physics \& Astronomy, University of California Irvine, Irvine CA 92697, USA\label{aff126}
\and
Department of Mathematics and Physics E. De Giorgi, University of Salento, Via per Arnesano, CP-I93, 73100, Lecce, Italy\label{aff127}
\and
INFN, Sezione di Lecce, Via per Arnesano, CP-193, 73100, Lecce, Italy\label{aff128}
\and
INAF-Sezione di Lecce, c/o Dipartimento Matematica e Fisica, Via per Arnesano, 73100, Lecce, Italy\label{aff129}
\and
Departamento F\'isica Aplicada, Universidad Polit\'ecnica de Cartagena, Campus Muralla del Mar, 30202 Cartagena, Murcia, Spain\label{aff130}
\and
Observatorio Nacional, Rua General Jose Cristino, 77-Bairro Imperial de Sao Cristovao, Rio de Janeiro, 20921-400, Brazil\label{aff131}
\and
CEA Saclay, DFR/IRFU, Service d'Astrophysique, Bat. 709, 91191 Gif-sur-Yvette, France\label{aff132}
\and
Department of Astronomy, University of Florida, Bryant Space Science Center, Gainesville, FL 32611, USA\label{aff133}
\and
Department of Computer Science, Aalto University, PO Box 15400, Espoo, FI-00 076, Finland\label{aff134}
\and
 Instituto de Astrof\'{\i}sica de Canarias, E-38205 La Laguna; Universidad de La Laguna, Dpto. Astrof\'\i sica, E-38206 La Laguna, Tenerife, Spain\label{aff135}
\and
Caltech/IPAC, 1200 E. California Blvd., Pasadena, CA 91125, USA\label{aff136}
\and
Ruhr University Bochum, Faculty of Physics and Astronomy, Astronomical Institute (AIRUB), German Centre for Cosmological Lensing (GCCL), 44780 Bochum, Germany\label{aff137}
\and
Department of Physics and Astronomy, Vesilinnantie 5, University of Turku, 20014 Turku, Finland\label{aff138}
\and
Serco for European Space Agency (ESA), Camino bajo del Castillo, s/n, Urbanizacion Villafranca del Castillo, Villanueva de la Ca\~nada, 28692 Madrid, Spain\label{aff139}
\and
ARC Centre of Excellence for Dark Matter Particle Physics, Melbourne, Australia\label{aff140}
\and
Centre for Astrophysics \& Supercomputing, Swinburne University of Technology,  Hawthorn, Victoria 3122, Australia\label{aff141}
\and
Department of Physics and Astronomy, University of the Western Cape, Bellville, Cape Town, 7535, South Africa\label{aff142}
\and
Department of Physics, Oxford University, Keble Road, Oxford OX1 3RH, UK\label{aff143}
\and
DAMTP, Centre for Mathematical Sciences, Wilberforce Road, Cambridge CB3 0WA, UK\label{aff144}
\and
Kavli Institute for Cosmology Cambridge, Madingley Road, Cambridge, CB3 0HA, UK\label{aff145}
\and
Department of Astrophysics, University of Zurich, Winterthurerstrasse 190, 8057 Zurich, Switzerland\label{aff146}
\and
Department of Physics, Centre for Extragalactic Astronomy, Durham University, South Road, Durham, DH1 3LE, UK\label{aff147}
\and
Institute for Theoretical Particle Physics and Cosmology (TTK), RWTH Aachen University, 52056 Aachen, Germany\label{aff148}
\and
Oskar Klein Centre for Cosmoparticle Physics, Department of Physics, Stockholm University, Stockholm, SE-106 91, Sweden\label{aff149}
\and
Astrophysics Group, Blackett Laboratory, Imperial College London, London SW7 2AZ, UK\label{aff150}
\and
Univ. Grenoble Alpes, CNRS, Grenoble INP, LPSC-IN2P3, 53, Avenue des Martyrs, 38000, Grenoble, France\label{aff151}
\and
INAF-Osservatorio Astrofisico di Arcetri, Largo E. Fermi 5, 50125, Firenze, Italy\label{aff152}
\and
Dipartimento di Fisica, Sapienza Universit\`a di Roma, Piazzale Aldo Moro 2, 00185 Roma, Italy\label{aff153}
\and
Centro de Astrof\'{\i}sica da Universidade do Porto, Rua das Estrelas, 4150-762 Porto, Portugal\label{aff154}
\and
Instituto de Astrof\'isica e Ci\^encias do Espa\c{c}o, Universidade do Porto, CAUP, Rua das Estrelas, PT4150-762 Porto, Portugal\label{aff155}
\and
HE Space for European Space Agency (ESA), Camino bajo del Castillo, s/n, Urbanizacion Villafranca del Castillo, Villanueva de la Ca\~nada, 28692 Madrid, Spain\label{aff156}
\and
INAF - Osservatorio Astronomico d'Abruzzo, Via Maggini, 64100, Teramo, Italy\label{aff157}
\and
Theoretical astrophysics, Department of Physics and Astronomy, Uppsala University, Box 516, 751 37 Uppsala, Sweden\label{aff158}
\and
Institute for Astronomy, University of Hawaii, 2680 Woodlawn Drive, Honolulu, HI 96822, USA\label{aff159}
\and
Mathematical Institute, University of Leiden, Einsteinweg 55, 2333 CA Leiden, The Netherlands\label{aff160}
\and
Institute of Astronomy, University of Cambridge, Madingley Road, Cambridge CB3 0HA, UK\label{aff161}
\and
Univ. Lille, CNRS, Centrale Lille, UMR 9189 CRIStAL, 59000 Lille, France\label{aff162}
\and
Department of Astrophysical Sciences, Peyton Hall, Princeton University, Princeton, NJ 08544, USA\label{aff163}
\and
Space physics and astronomy research unit, University of Oulu, Pentti Kaiteran katu 1, FI-90014 Oulu, Finland\label{aff164}
\and
Center for Computational Astrophysics, Flatiron Institute, 162 5th Avenue, 10010, New York, NY, USA\label{aff165}
\and
Department of Physics and Astronomy, University of British Columbia, Vancouver, BC V6T 1Z1, Canada\label{aff166}}    
    
    \date{\today}

\abstract{
Strong gravitational lensing (SL) by galaxy clusters is a powerful probe of their inner mass distribution and a key test bed for cosmological models. However, the detection of SL events in wide-field surveys such as \Euclid requires robust, automated methods capable of handling the immense data volume generated. In this work, we present an advanced deep learning (DL) framework based on mask region-based convolutional neural networks (Mask R-CNNs), designed to autonomously detect and segment bright, strongly-lensed arcs in \Euclid's multi-band imaging of galaxy clusters. The model is trained on a realistic simulated data set of cluster-scale SL events, constructed by injecting mock background sources into Euclidised \emph{Hubble} Space Telescope images of $10$ massive lensing clusters, exploiting their high-precision mass models constructed with extensive spectroscopic data. The network is trained and validated on over $4500$ simulated images, and tested on an independent set of $500$ simulations, as well as real \Euclid Quick Data Release (Q1) observations. The trained network achieves high performance in identifying gravitational arcs in the test set, with a precision and recall of $76\%$ and $58\%$, respectively, processing $2' \times 2'$ images in a fraction of a second. When applied to a sample of visually confirmed \Euclid Q1 cluster-scale lenses, our model recovers $\approx 66\%$ of gravitational arcs above the area threshold used during training.  While the model shows promising results, limitations include the production of some false positives and challenges in detecting smaller, fainter arcs. Our results demonstrate the potential of advanced DL computer vision techniques for efficient and scalable arc detection, enabling the automated analysis of SL systems in current and future wide-field surveys. The code, \texttt{ARTEMIDE}, is open source and will be available at \url{github.com/LBasz/ARTEMIDE}.
}


\titlerunning{Searching for giant gravitational arcs in galaxy clusters with Mask R-CNNs}
\authorrunning{Euclid Collaboration: L.~Bazzanini et al.}

\maketitle

\section{\label{sec:introduction}Introduction}

Strong gravitational lensing (SL) serves as a valuable tool for probing the total mass distribution in galaxies and galaxy clusters, as well as testing cosmological total models~\citep{meneghetti2021}. Over the past decades, SL has been utilised in various contexts: to study galaxy structures and their evolution~\citep{treu2002, auger2010, sonnenfeld2013}; to estimate the Hubble constant ($H_0$) through time-delay measurements~\citep{suyu2017, grillo2018, millon2020, moresco2022, grillo2024, birrer2025}; to constrain the dark energy (DE) equation of state~\citep{jullo2010, cao2012, collett2014, caminha2022}; and to estimate the dark matter (DM) fraction in massive early-type galaxies~\citep{auger2010, tortora2010, sonnenfeld2015}. On galaxy cluster scales, SL is instrumental in modelling their inner total mass distribution, leveraging multiple images of lensed background sources~\citep{caminha2017, caminha2019, acebron2018, lagattuta2019, bergamini2019, bergamini2021, lagattuta2022, bergamini2023a, bergamini2023b}. Furthermore, the magnification effect of SL allows galaxy clusters to act as cosmic telescopes, enabling the study of faint, high-redshift sources that would otherwise be unobservable~\citep{swinbank2009, richard2011, vanzella2020, vanzella2021}.

Traditionally, visual inspection has been the primary method for confirming lens candidates, typically following an initial pre-selection based on spectroscopic, photometric, or morphological criteria~\citep[e.g.][]{lefevre1988, jackson2008, sygnet2010, pawase2014, Q1-SP057}. However, current and next-generation astronomical surveys, such as \Euclid~\citep{EuclidSkyOverview}, the \emph{Nancy Grace Roman} Space Telescope~\citep{green2012}, and the \emph{Vera Rubin} Observatory Legacy Survey of Space and Time~\citep[][]{LSST2019}, are expected to produce unprecedented volumes of imaging data over the coming decade. For example, \Euclid alone is predicted to yield approximately $10^5$ galaxy-galaxy SL events~\citep{collett2015, AcevedoBarroso24, Q1-SP048}, as well as around $5000$ strongly lensing clusters~\citep{Q1-SP057} out of the $>10^5$ galaxy clusters up to $z=2$ it is expected to detect~\citep{Adam-EP3}.

The visual inspection of \Euclid Q1~\citep{Q1cite} cluster lens candidates carried out in~\citet{Q1-SP057} involved the coordinated effort of around $40$ expert astronomers to assess approximately $1300$ candidate clusters within large $4' \times 4'$ cutouts. This task required several weeks of distributed work, underscoring the resource-intensive nature of such visual inspection. While effective, this approach is not sustainable for future data releases, which will contain orders of magnitude more data; our goal with this work is precisely to address this challenge. This underscores the need for more efficient, scalable, and automated approaches to analyse the vast amounts of data produced by these surveys. In the last decade, several alternative methods have been developed for detecting SL events, particularly on galaxy-scale. These methods range from semi-automated algorithms designed to detect arc and ring-like structures~\citep[e.g.][]{more2012, gavazzi2014, sonnenfeld2018} to crowd-sourced science initiatives~\citep[e.g.][]{marshall2016, sonnenfeld2020}. Several arc-finding algorithms proposed in the literature identify gravitational arcs by detecting elongated structures in the surface brightness distribution, for example by correlating local ellipticities across overlapping image cells~\citep{seidel2007}, or by segmenting objects using local intensity gradients and applying selection criteria based on elongation (e.g.~length-to-width ratio) and morphological properties~\citep{xu2016}. Within this framework, machine learning (ML) and deep learning (DL) techniques have emerged as the most promising tools for identifying SL events~\citep[e.g.][]{metcalf2019}. 

Astronomy has benefited from applying ML and DL to both simulated and real survey data. These applications include, for instance, star detection~\citep{he2023}, quasar and galaxy classification~\citep{bailerjones2019, clarke2020}, photometric redshift estimation~\citep{disanto2018, schuldt2021}, cluster member classification~\citep{angora2020}, optimisation of stochastic models for the generation of gamma-ray burst light curves~\citep{bazzanini2024}; for an in-depth review we refer to~\citet{huertascompany2023}. Due to their efficiency and high performance in image pattern recognition, supervised DL techniques, particularly convolutional neural networks~\citep[CNNs;][]{lecun1998, lecun2015}, are becoming increasingly crucial for analysing astrophysical data sets.
CNNs indeed have emerged as a powerful tool for identifying SL systems in imaging surveys~\citep[e.g.][]{petrillo2017, canameras2020, huang2020, jia2023, angora2023, Q1-SP048}. Once trained, these networks can process single images in a fraction of a second.

The field of computer vision has recently greatly advanced in areas such as object detection, classification, and semantic segmentation. The mask region-based convolutional neural network framework~\citep[Mask R-CNN;][]{he2017} has set the benchmark for DL models in instance segmentation. Mask R-CNNs have been applied in astrophysics for detection, deblending, and classification of astronomical sources~\citep{burke2019, merz2024}, for detection and morphological classification of galaxies~\citep{farias2020}, for detecting `ghosting artefacts'~\citep{tanoglidis2022}, for detecting magnetic bright points in the solar photosphere~\citep{yang2019, yang2023}, and for the detection and morphological classification of extended radio sources~\citep{lao2023}.

Deep learning methods, however, require training on suitable simulated data sets due to the limited number of confirmed SL systems, especially in galaxy clusters. To this end, considerable efforts have been made in recent years to simulate SL events akin to those observed by both ongoing and upcoming surveys. These simulations typically consist of generating mock images of SL events by superimposing simulated lensed sources onto foreground galaxies using various techniques~\citep{meneghetti2008, meneghetti2010, metcalf2019, Leuzzi-TBD}. These sources are then combined with real or synthetic images using ray-tracing techniques, e.g.~\texttt{GLAMER}~\citep{metcalf2014, petkova2014} and \texttt{GRAVLENS}~\citep{keeton2001}.

In this work, we focus on detecting strongly lensed bright arcs on galaxy cluster scales, rather than galaxy-scale lensing events. Compared to galaxy-scale SL, typically searched for in small cutouts (e.g.~$10'' \times 10''$, \citealt{Q1-SP048}), cluster-scale arc detection presents additional challenges. The larger field of view used in our case ($2' \times 2'$) contains hundreds of sources per image, increasing the likelihood of source confusion and false positives. Furthermore, the cluster environment is more complex, featuring numerous bright and extended foreground galaxies and diverse arc morphologies caused by the complex multi-component mass distribution. These factors make automated detection particularly demanding.

The challenges of gravitational arc detection in galaxy clusters are well-suited for the Mask R-CNN framework; this advanced neural network (NN) indeed offers key advantages over standard CNNs. In particular, it can process full-size images of galaxy clusters thanks to its ability to handle inputs of variable size, without requiring resizing to a fixed input dimension, thereby preserving morphological features and spatial information. We adopt Mask R-CNN not for pixel-level arc measurements, but because its instance-segmentation framework naturally enables simultaneous detection, classification, and segmentation of multiple objects in a single image at different scales, allowing it to distinguish individual arc instances within large, crowded fields. Our primary goal in this work is therefore lens finding on cluster scales, not obtaining high-precision masks for e.g.~downstream automated lens modelling. These capabilities make Mask R-CNNs an effective choice for addressing the complexity of the cluster-scale lensing regime, positioning this work as an innovative application of advanced computer vision techniques to SL in wide-field surveys.

We train, validate, and test the performance of a Mask R-CNN using simulated cluster-scale SL events in \Euclid simulated images~\citep{hst2euclid}. The mock \Euclid imaging data is generated from \emph{Hubble} Space Telescope (HST) observations. To build the training set, we simulate thousands of galaxy cluster strong lensing (GCSL) arcs by taking advantage of high-precision cluster lens models constructed by~\citet{caminha2017, caminha2019, bergamini2019, bergamini2021, bergamini2023a, bergamini2023b} using the public software {\textsc{lenstool}}~\citep{kneib1996, jullo2007, jullo2009}. We then further test the NN by applying it to real \Euclid images of lensing clusters found in~\citet{Q1-SP057}, comparing the results of the NN inference with the lensing events found with a visual inspection. After training, this method is extremely efficient, detecting gravitational arcs in \Euclid in a $3\times1199\times1199\,\mathrm{pixels}$ image in a fraction of a second using a single NVIDIA Quadro RTX 6000 GPU. Our code, \texttt{ARTEMIDE} (ARcs in clusTErs using Mask r-cnn IDEntifier), is free and open source, and available in our GitHub repository\footnote{\url{https://github.com/LBasz/ARTEMIDE}}.

This paper is organised as follows. In Sect.~\ref{sec:dataset} we describe how the training set is produced via SL simulations. In Sect.~\ref{sec:network}, we introduce the Mask R-CNN framework, describing in detail the architecture of our implementation, and the training setup. In Sect.~\ref{sec:results}, we present the results of the trained network, we evaluate its performance on the test set, and we further test it on real \Euclid Q1 images. In Sect.~\ref{sec:discussion}, we discuss the results. Finally, in Sect.~\ref{sec:conclusions} we summarise and conclude.

Throughout this paper, we assume a flat $\Lambda\mathrm{CDM}$ cosmology with $\Omega_\Lambda = 0.7$, $\Omega_{\mathrm{m}} = 0.3$, and $H_0 = 70\,\mathrm{km}\,\mathrm{s}^{-1}\,\mathrm{Mpc}^{-1}$. Magnitudes are reported in the AB system~\citep{oke1983}, unless otherwise stated.

\section{\label{sec:dataset}Data set}

In this section, we outline the procedure for generating the training set. We support multi-band Flexible Image Transport System files~\citep[FITS;][]{pence2010} as image input for the Mask R-CNN. In order to train a NN with a supervised learning approach, we typically need a large training set of lensing and non-lensing examples. However, the number of real cluster-scale gravitational arcs is limited to a small sample; therefore, to produce the `positive' class examples we have to resort to SL simulations. 

Since the most accurate lens models of galaxy clusters are available for objects observed by the HST, to generate the training set we had to degrade the HST images, in order to reproduce the observational parameters of the \Euclid survey. We perform this task starting from the \texttt{Python} code \texttt{HST2EUCLID}~\citep{hst2euclid}, a tool for converting HST observations into \Euclid-like imaging data. With this software, we created the simulated data set of mock \Euclid images, starting from $10$ HST observations of galaxy clusters (see Table~\ref{table:clusters}) observed in the Cluster Lensing and Supernova Survey with Hubble (CLASH\footnote{\url{https://archive.stsci.edu/prepds/clash/}}; \citealt{postman2012}) and the Hubble Frontier Fields (HFF\footnote{\url{https://archive.stsci.edu/prepds/frontier/}}; \citealt{lotz2014, lotz2017}) programmes. Since the mock images are based on real observations, they inherently capture all the complexities present in the observed galaxy clusters.
\begingroup
\setlength{\tabcolsep}{6pt}       
\renewcommand{\arraystretch}{1.3} 
\begin{table}
\caption{Description of the cluster sample included in the GCSL set of simulations.} 
\centering
\label{table:clusters} 
    \begin{tabular}{l l}
    \hline\hline
    Galaxy clusters          & $z_{\mathrm{cl}}$ \\
    \hline\hline
    \textbf{Training + Validation Set} ($4000+500$ images) &\\
    \hline
    RX J2129$+$0005   (R2129)& $0.234$ \\
    Abell 2744        (A2744)& $0.308$ \\
    MACS J1931$-$2635 (M1931)& $0.352$ \\
    MACS 1115$+$0129  (M1115)& $0.352$ \\
    MACS J0416$-$2403 (M0416)& $0.397$ \\
    MACS J1206$-$0847 (M1206)& $0.439$ \\
    MACS J0329$-$0211 (M0329)& $0.450$ \\
    RX J1347$-$1145   (R1347)& $0.451$ \\
    MACS J2129$-$0741 (M2129)& $0.587$ \\
    \hline\hline
    \textbf{Test Set} ($500$ images) & \\
    \hline
    Abell S1063       (A1063)& $0.348$ \\
    \hline
    \hline
    \end{tabular}
\end{table}
\endgroup

\subsection{\Euclid}

The \Euclid mission is a space-based survey by the European Space Agency that uses a telescope equipped with a $1.2\,\mathrm{m}$ mirror~\citep{EuclidSkyOverview}. The Euclid Wide Survey~\citep[EWS;][]{Scaramella-EP1} will map approximately $14\,000\,\deg^2$ of extragalactic sky with low zodiacal background and low Galactic extinction. The Euclid Deep Survey, instead, will cover about $ 50\,\deg^2$, achieving depths two magnitudes deeper than the wide survey. The first \Euclid Quick Data Release~\citep[Q1;][]{Q1cite} comprises $63.1\,\deg^2$ of the Euclid Deep Fields to nominal wide-survey depth, and contains about 30 million objects~\citep{Q1-TP001}.

The \Euclid telescope is equipped with two instruments: a visible imager~\citep[VIS;][]{EuclidSkyVIS}, and a Near-Infrared Spectrometer and Photometer~\citep[NISP;][]{EuclidSkyNISP}. VIS is a large-format imager with a field of view (FoV) of $0.54\,\deg^2$ sampled at $\ang{;;0.1}\,\mathrm{pixel}^{-1}$, operating in a single red passband~\citep{EuclidSkyOverview}. NISP is a near-infrared imager and slitless spectrometer, and provides multiband photometry and slitless grism spectroscopy in the wavelength range $920\text{--}2020\,\mathrm{nm}$, using the light transmitted by the dichroic beamsplitter~\citep{EuclidSkyOverview}. With a pixel scale of $\ang{;;0.3}\,\mathrm{pixel}^{-1}$, the NISP FoV covers a nearly square-shaped $0.57\,\deg^2$~\citep{EuclidSkyOverview}. The VIS photometric channel offers one passband, $\IE$ ($550\text{--}920\,\mathrm{nm}$), while the NISP photometric channel offers three of them: $\YE$ ($949.6\text{--}1212.3\,\mathrm{nm}$); $\JE$ ($1167.6\text{--}1567.0\,\mathrm{nm}$); and $\HE$ ($1521.5\text{--}2021.4\,\mathrm{nm}$). The requirements on the VIS point-spread function (PSF) is a FWHM smaller than $\ang{;;0.18}$ at $800\,\mathrm{nm}$, while for the NISP PSFs the FWHM is $1.10$ pixel in $\YE$, 1.17 pixel in $\JE$, and $1.19$ pixel in $\HE$, when fitting a Moffat profile~\citep{EuclidSkyOverview}.

\subsection{\texttt{HST2EUCLID}}

\texttt{HST2EUCLID} is a code developed by~\citet{hst2euclid} to create simulated \Euclid images in the $\IE$, $\YE$, $\JE$, and $\HE$, bands starting from real HST observations. Although originally designed to produce mock \Euclid observations of galaxy clusters, this tool can also be applied to any sufficiently deep HST image to create the corresponding \Euclid-like observations, provided that the relevant HST filters overlap with the \Euclid photometric bands that we want to simulate. Specifically, the HST/F606W and HST/F814W filters are combined to simulate the \Euclid $\IE$ band, while from the HST/F105W, HST/F125W, and HST/F160W filters we simulate the $\YE$, $\JE$, and $\HE$ bands. For further details on the methodology and implementation, we refer the reader to the original publication.

The \texttt{HST2EUCLID} simulation pipeline was validated through a series of three tests on the resulting Euclidised images. First, a check that the simulated images reach the expected depth of the EWS observations. Then, the distribution of galaxy sizes were measured in blank fields. Finally, the galaxy number counts detected in the Euclidised images were estimated to ensure consistency with expectations.

In our work, we exploited the Euclidised images in the four \Euclid photometric bands of $10$ massive galaxy clusters (see Table~\ref{table:clusters}), obtained with \texttt{HST2EUCLID}. We then simulated the GCSL events, and co-added them to the generated Euclidised image.

\subsection{\label{subsec:simulations}Lensing simulations}

For the GCSL event simulations, we utilise deflection angle maps derived from high-precision SL models of $10$ galaxy clusters as presented by~\citet{caminha2019} and~\citet[see Table~\ref{table:clusters}]{bergamini2019, bergamini2021, bergamini2023a, bergamini2023b}\footnote{These lens models are publicly available at \url{https://www.fe.infn.it/astro/lensing/}.} with the \textsc{lenstool} software~\citep{kneib1996, jullo2007, jullo2009}. These models utilise extensive spectroscopic data of multiple images to accurately represent both the large-scale mass distribution of the cluster and the sub-halo mass distribution (i.e.~the cluster member galaxies). This detailed modelling captures the impact on the morphology, brightness, and occurrence of GCSL events.

Each cluster's total mass distribution is represented by a parametric model of the overall lensing potential. This model incorporates the following components:
\begin{itemize}
    \item A cluster-scale component, consisting of DM halos and, when available, the smooth intra-cluster hot-gas mass from \emph{Chandra} X-ray observations~\citep{bonamigo2017, bonamigo2018}.
    \item A sub-halo mass component associated with cluster galaxies. The mass density profile of each sub-halo, including both dark and baryonic matter, is typically characterised by either a spherical or an elliptical singular dual-pseudo isothermal profile~\citep{limousin2005, eliasdottir2007}. This profile is further refined using measured stellar velocity dispersions from extensive samples of cluster member galaxies~\citep{bergamini2021}.
    \item Any other massive objects in the galaxy cluster outskirts or along the line of sight.
\end{itemize}
These lens models effectively reconstruct the observed positions of numerous multiple images (ranging from approximately $20$ to $200$), with a typical accuracy of $\lesssim \ang{;;0.5}$.

\textsc{lenstool}~\citep{kneib1996, jullo2007, jullo2009} reconstructs the cluster potential by minimising the offset between the observed positions of multiple images and those predicted by the model, given a specific set of model parameters. The reduced deflection angle, $\vec{\alpha}$, characterises the relation between the true position of the source $\vec{\beta}$ and its observed location $\vec{\theta}$ using the lens equation~\citep{meneghetti2021}
\begin{equation}
    \vec{\beta} = \vec{\theta} - \vec{\alpha} (\vec{\theta}) \;.
    \label{eq:lensequation}
\end{equation}

\begin{figure*}
    \centering
    \includegraphics[origin=c,width=1.0\linewidth]{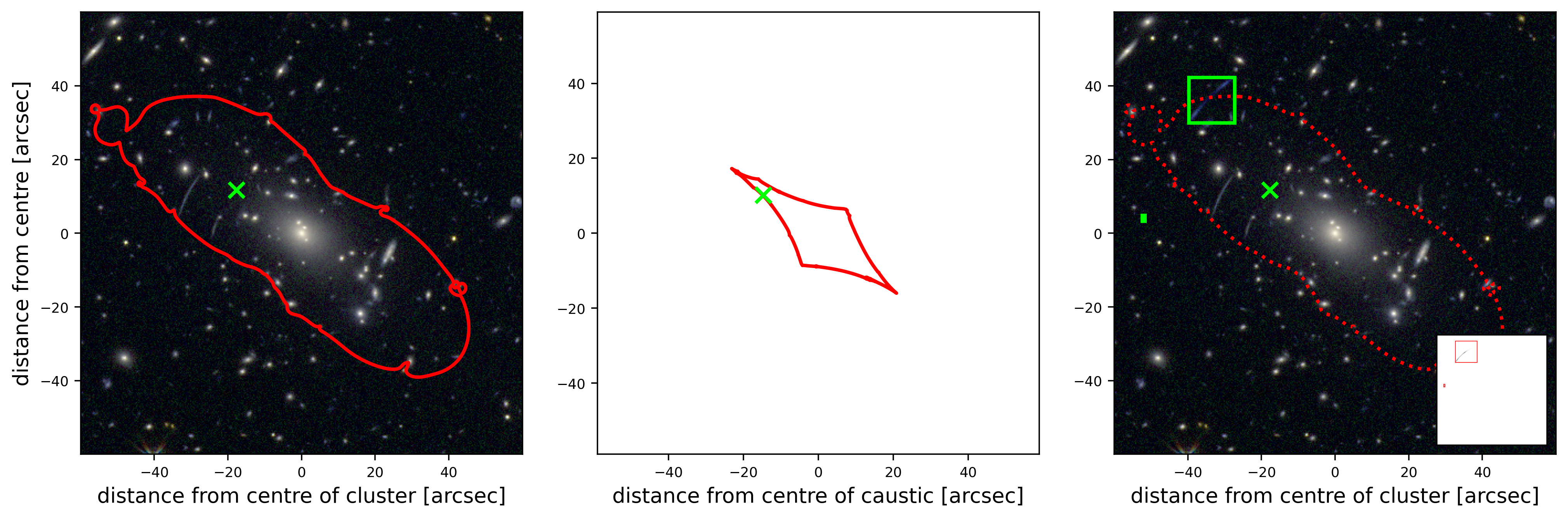}
    \caption{Steps of a \Euclid GCSL simulation. \emph{Left}: \texttt{HST2EUCLID} $2'\times2'$ RGB image of the galaxy cluster Abell~S1063 ($z_\mathrm{cl}=0.348$), with the main critical line (in red) for a source at $z_{\mathrm{s}}=2.92$, based on the lens model by~\citet{bergamini2019}. The critical line has a circularised Einstein radius of $\theta_\mathrm{E} \simeq 33''$. The green cross marks the position of the injected source to be lensed. \emph{Middle}: Source plane at $z_{\mathrm{s}}=2.92$, displaying the caustic (in red) related to the main critical line. The injected source (green cross) features a \sersic profile (index $n=1.22$, $r_{\mathrm{eff}}=\ang{;;0.11}$), with $\YE=24.8$ and the SED of a star-forming galaxy; these parameters are sampled according to the procedure described in Sect.~\ref{subsec:simulations}. \emph{Right}: Colour-composite image of the simulated GCSL system, including the critical line (red dotted line). Green boxes enclose the gravitational arcs resulting from the lensing simulation, which are also shown in the bottom inset.}
    \label{fig:HST2EUCLID_A1063_sersic_0_000009}
\end{figure*}

The simulation process is implemented using the \texttt{Python} library \texttt{PyLensLib}~\citep{meneghetti2021}, and the key steps can be summarised as follows.
\begin{enumerate}
    \item Given the mass distribution and redshift of the lens galaxy cluster and the redshift of the source, we numerically compute the deflection angle maps. Then, we derive the convergence and shear maps, these being the elements of the Jacobian matrix describing the image distortion, whose inverse matrix is known as the `magnification tensor'. Critical curves are then identified as the set of points where magnification diverges to infinity. An example of a tangential critical curve for the galaxy cluster Abell~S1063 ($z_\mathrm{cl}=0.348$) and a source at redshift $z_\mathrm{s}=2.92$ is shown in red in Fig.~\ref{fig:HST2EUCLID_A1063_sersic_0_000009} (left panel).
     \item The primary critical line is mapped into the corresponding caustic line on the source plane (central panel in Fig.~\ref{fig:HST2EUCLID_A1063_sersic_0_000009}) using the lens equation (Eq.~\ref{eq:lensequation}).
     \item We simulate the lensing event by injecting a \sersic surface brightness profile~\citep{sersic1963, sersic1968}, $I_\mathrm{s}(\vec{\beta})$, near the primary caustic line, within a buffer of width $\ang{;;0.5}$. We choose only the points within the buffer with a magnification factor $\mu$ greater than $50$. Since lens mapping conserves surface brightness, we can relate the observed and intrinsic surface brightnesses, $I(\vec{\theta})$ and $I_\mathrm{s}(\vec{\beta})$, using the lens equation $I(\vec{\theta}) = I_\mathrm{s}(\vec{\beta})$; this allows us to directly reconstruct the source image in the lens plane through ray tracing~\citep{meneghetti2021}.
     \item Finally, the GCSL event is produced by convolving the simulated arcs with the \Euclid PSF for each band, and then co-adding it to the \texttt{HST2EUCLID} base image in each filter (right panel in Fig.~\ref{fig:HST2EUCLID_A1063_sersic_0_000009}).
\end{enumerate}

For the injected sources, we use a source spectral energy distribution (SED) based on a star-forming galaxy template from~\citet{kinney1996}. Table~\ref{tab:sersic_params} lists the \sersic parameters along with their adopted value ranges. The \sersic index, $n$, is sampled from a uniform distribution within $[1.0,2.0]$, which corresponds to typical starburst profiles of late-type galaxies. The axis ratio $q$ and position angle $\varphi$ are randomly drawn from uniform distributions in the ranges $[0.2, 1.0]$ and $[0, \pi]$, respectively. To closely reproduce the observed properties of the \sersic sources, we follow the approach of~\citet{angora2023} and adopt a non-uniform sampling of the remaining parameters. For the source intrinsic magnitudes and redshifts we estimated the number counts in the $i$-band (that is, the number of galaxies per square degree per magnitude bin) from the COSMOS~2015 catalogue~\citep{scoville2007, laigle2016}. Moreover, we complemented this with HST Deep Field North and South observations~\citep{williams1996, metcalfe2001} in F814W~\citep{capak2007}, which extends the galaxy counts to the faint end, down to $\mathrm{F814W}=29$. Then, we used the COSMOS photometric redshift catalogue to fit a redshift probability density function (PDF), i.e.~$p(z\, |\Delta i)$, in six magnitude bins, using a function of the form $p(z\, |\Delta i) = A z^2 \mathrm{e}^{-z/z_0}$ for $i \in [22,24)$ and $p(z\, |\Delta i) = A z^2 \mathrm{e}^{-(z/z_0)^{1/2}}$ for the other magnitude bins~\citep{lombardi1999, lombardi2005}; see also Fig.~3 in~\citealt{angora2023}. We then use these PDFs to assign a source $\IE$ magnitude to each background galaxy and, given this magnitude, a corresponding redshift.

For the injected sources, we retained only those with sampled magnitudes up to one magnitude fainter than the \Euclid $\IE$-band limiting magnitude, i.e.~up to $\IE = 25.5$~\citep{Scaramella-EP1}. Following~\cite{meneghetti2022}, we also enforce a lower limit for the source redshift of $z_\mathrm{s} = z_{\mathrm{cl}}+0.4$. Indeed, their study, which examined the lensing cross-section of the galaxy clusters used in this work, showed that the lensing cross-section increases substantially beyond zero for source redshifts roughly greater than $z_\mathrm{cl}+0.4$. 

Finally, to assign an effective radius $r_\mathrm{eff}$ to the injected background galaxies, we use an empirical relation from~\citet{shibuya2015}, describing how galaxy sizes evolve with redshift. This relation takes the form $r_\mathrm{eff} = B(1+z)^\beta$, based on galaxy size estimates in both optical and ultraviolet bands. However, since a comparison with the effective radii measured by~\cite{tortorelli2018} for low-redshift galaxies shows a significant overestimation, we apply this relation only for $z>1$, adopting a constant $r_\mathrm{eff}$ for $z \le 1$ (see left panel of Fig.~4 in~\citealt{angora2023}).

\begingroup
\setlength{\tabcolsep}{6pt}       
\renewcommand{\arraystretch}{1.2} 
\begin{table}
    \caption{List of \sersic parameters and their adopted value ranges for the injected sources.} \label{tab:sersic_params}%
    \center%
    \begin{tabular}{ll}
    \hline\hline%
    Parameter                                                    & Sampling strategy \\
    \hline
    \multirow{2}{*}{Coordinate ($y_{\mathrm{s}}$, source plane)} & Sampled within a buffer\\
                                                                 & around the main caustic \\
    \multirow{2}{*}{Source magnitude ($i$)}        & Sampled from PDF, $p(i)$\\
                                                                 & COSMOS + HST fields \\
    \multirow{2}{*}{Source redshift ($z_\mathrm{s}$)}            & Sampled from PDF, $p(z|\Delta i)$\\
                                                                 & COSMOS \\
    \multirow{3}{*}{Effective radius ($r_{\mathrm{eff}}$)}       & $r_{\mathrm{eff}}(z) = 2.54$\,kpc, $z\le1$\\
                                                                 & $r_{\mathrm{eff}}(z) = B(1+z)^\beta$, $z>1$\\
                                                                 & \citep{shibuya2015} \\
    \sersic\ index ($n$)                                         & Sampled from $\mathcal{U}[1.0,2.0]$ \\
    Axis ratio ($q$)                                             & Sampled from $\mathcal{U}[0.2,1.0]$ \\
    Position angle ($\varphi$)                                   & Sampled from $\mathcal{U}[0,\pi]$   \\
    \hline\hline
    \end{tabular}
\end{table}
\endgroup

\subsection{Building the knowledge base}

The described method enables the simulation of a large number of realistic GCSL events involving bright gravitational arcs. The cluster's lens model is required to produce the knowledge base consisting of these lensing simulations. In addition, all galaxy clusters used in the training set already exhibit real cluster-scale bright gravitational arcs. The training set was generated as follows
\begin{enumerate}
    \item Use the $10$ available highly-accurate lens models of the HST clusters, as done in~\citet[][see Table~\ref{table:clusters}]{angora2023}.
    \item Generate \Euclid-like images of these galaxy clusters using \texttt{HST2EUCLID}~\citep{hst2euclid}. 
    \item Simulate $500$ SL events for each galaxy cluster using \texttt{pyLensLib}~\citep{meneghetti2021}. The simulations in the NISP bands are generated at pixelscale $\ang{;;0.3}\,\mathrm{pixel}^{-1}$, and then upsampled at the VIS one ($\ang{;;0.1}\,\mathrm{pixel}^{-1}$).
\end{enumerate}

The combined training and validation set consists of $4500$ \Euclid-like galaxy cluster images, each with dimensions of $120''\times 120''$. On these images, simulated gravitational lensed arcs (in addition to the existing ones) were added by randomly injecting \sersic sources in the background, near the regions of higher magnification ($\mu \geq 50$). We kept the injected arcs only if their areas were larger than $400$\,pixels; this area cut was implemented to focus the training specifically on the detection of giant and bright arcs, which are the primary targets of this work. This approach avoids the ambiguity inherent in smaller, fainter lensed arcs that could be confused by the NN with other objects, e.g.~edge-on galaxies or stellar spikes, which could potentially degrade the network's performance and increase the rate of false positives. This simulation process enables the direct extraction of bounding boxes and binary masks of the injected arcs. Conversely, the masks of pre-existing gravitational arcs were manually extracted for each cluster.

All the $500$ simulations corresponding to the last galaxy cluster in Table~\ref{table:clusters}, namely A1063, were excluded from the training set. Instead, they were used exclusively as a test set for evaluating the performance of the NN.

Although the training set is based on only $10$ clusters, each of them provides a very rich lensing environment: they contain numerous real arcs and highly structured mass distributions, and by injecting simulated sources we can generate thousands of independent realisations that preserve realistic cluster morphologies.

\section{\label{sec:network}Network and training}

In this section, we describe our implementation of the Mask R-CNN for detecting bright gravitational lensing arcs in galaxy clusters. Our training procedure follows a supervised learning approach, utilising simulated SL events (and corresponding masks) on top of Euclidised images.

\subsection{Implementation}

\begin{figure}
    \centering
    \includegraphics[width=0.99\linewidth]{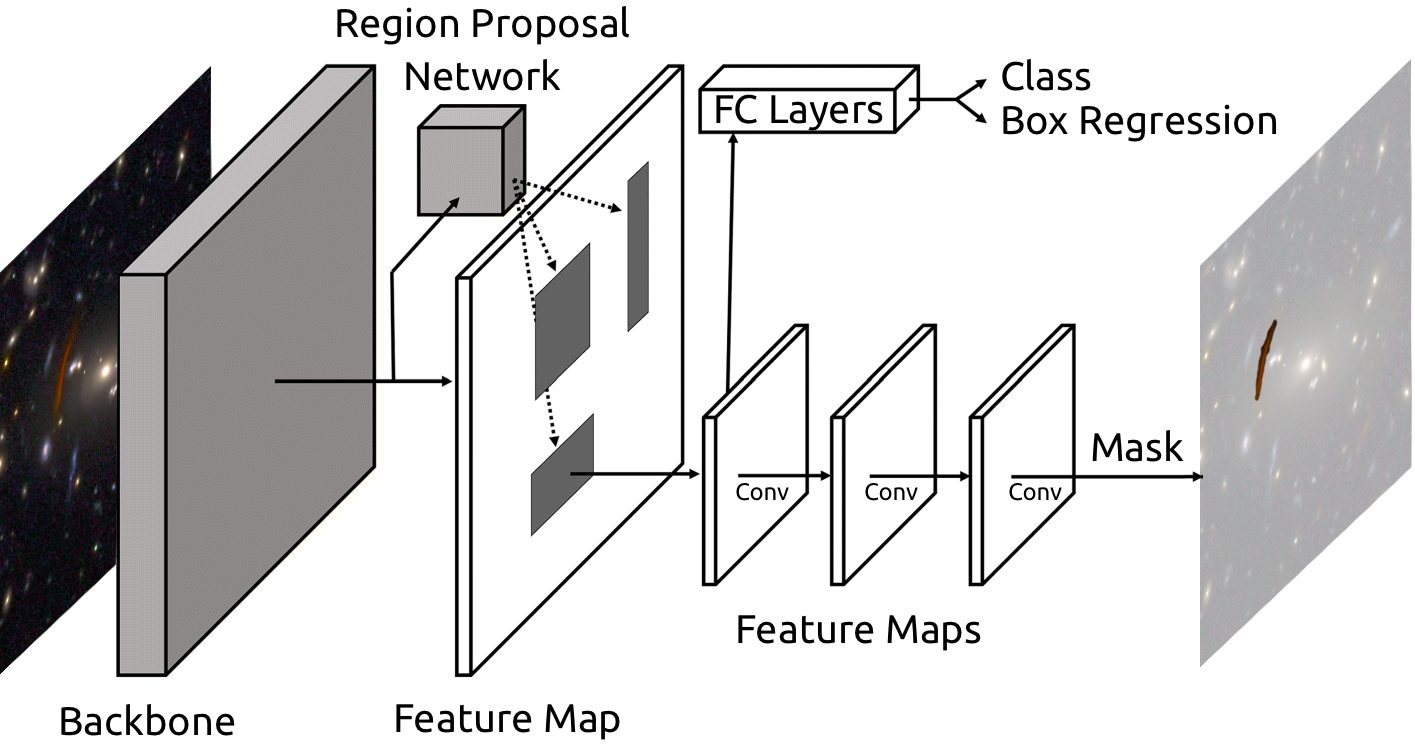}
    \caption{Mask R-CNN architecture. Adapted from~\citet{jung2019}.}
    \label{fig:astromaskrcnn}
\end{figure}

Object detection and classification involve identifying objects within an image and assigning them to predefined classes. Semantic segmentation, on the other hand, labels each pixel of an image according to its class, without distinguishing between individual object instances. Instance segmentation combines these tasks by both detecting objects and generating a separate segmentation mask for each individual instance~\citep{hafiz2020}. \citet{he2017} proposed the Mask R-CNN, an advanced framework for instance segmentation, extending the Faster R-CNN architecture~\citep{girshick2015, ren2015}. Mask R-CNN adds a branch for pixel-level segmentation and operates in two stages: first, generating object proposals and classifications, and then applying bounding boxes and segmentation masks to the image (see Fig.~\ref{fig:astromaskrcnn}).

The Mask R-CNN architecture offers several key advantages over traditional CNNs when applied to lens detection tasks.
\begin{itemize}
    \item Instance segmentation. Unlike conventional CNNs that primarily perform image classification or basic object detection, Mask R-CNN generates detailed pixel-level segmentation masks for each detected object. It seamlessly integrates classification, bounding box regression, and segmentation tasks, whereas standard CNNs typically focus on a single task, such as whole-image classification.
    \item Multi-object detection. Mask R-CNN is capable of detecting and segmenting multiple objects within a single image, even at different scales, while effectively distinguishing individual instances. In contrast, traditional CNNs often face challenges in identifying and separating overlapping objects.  
    \item Flexible input handling. Mask R-CNN can process images of varying sizes without the need to resize them to a fixed input dimension, unlike many standard CNN implementations. This flexibility helps preserve fine spatial details and prevents the loss of morphological information that can occur by interpolation during image resizing--an important advantage when working with high-resolution astrophysical data.    
    \item No need to produce the `negatives'. In the Mask R-CNN architecture  every object that is not flagged as belonging to the `positive' class (i.e.~gravitational arcs) automatically pertains to the `negative' one, removing in this way the need to create a fraction of the training set containing no arcs.
\end{itemize}

The code used in this work is based on the \texttt{Python}\footnote{\url{https://www.python.org/}} implementation of the Mask R-CNN from \texttt{torchvision}~\citep{torchvision2016}, built on the \texttt{pytorch} framework~\citep{pytorch, Ansel_PyTorch_2_Faster_2024}. The initial stage of the Mask R-CNN utilises a pre-trained CNN, known as the `backbone', to extract feature maps from input images. The standard Mask R-CNN model~\citep[\texttt{maskrcnn\_resnet50\_fpn}\footnote{\url{https://pytorch.org/vision/stable/models/generated/torchvision.models.detection.maskrcnn_resnet50_fpn.html}.} in \texttt{torchvision}]{he2017} utilises as backbone the ResNet-50 residual neural network~\citep{he2015}. ResNets function as a feature extractor, with their initial layers capturing low-level features, such as edges and corners, and their deeper layers capturing high-level features via residual learning. ResNets are CNNs that employ `skip connections', which allow the creation of deep architectures with many layers while avoiding the accuracy degradation issues commonly affecting deep NNs~\citep{he2015}. This accuracy degradation happens because, as network depth increases, accuracy initially saturates and then swiftly declines~\citep{bengio1994, he2017}. The root cause of this problem lies in the backpropagation process, where continual multiplications by small weights diminish gradient sizes to the point of ineffectiveness; this issue is commonly known as the `vanishing gradient problem'~\citep{pascanu2012}.

To handle objects of varying scales, features are extracted at different backbone stages and combined using a feature pyramid network~\citep[FPN;][]{lin2016}, which integrates multiscale features by sharing information across hierarchical layers~\citep{lakshmanan2021}. The extracted feature maps are processed by the region proposal network~\citep[RPN;][]{ren2015}, which generates object proposals using a sliding window approach that defines anchor boxes of varying sizes and aspect ratios at each spatial location on the feature map~\citep{ren2015, elgendy2020}. The RPN is trained to classify anchors as containing an object or not, while refining their bounding boxes, using a dedicated loss function to address both classification and regression tasks. In our implementation, we employ a set of anchors of sizes $\{32, 64, 128, 256, 512\}$ pixels, with aspect ratios $\{0.5, 1, 2\}$. The subsampling factor of ResNet-50, i.e.~the ratio by which the spatial dimensions of an image are reduced as it passes through the NN, is $32$.

Non-maximum suppression~\citep[NMS;][]{girshick2015} is subsequently applied to the proposed anchor boxes. This procedure begins by sorting all generated boxes according to their objectness scores, i.e.~the confidence estimated by the network that a given region contains an object of interest. The algorithm then iteratively selects the box with the highest score and compares it against all remaining boxes using the intersection over union (IoU) metric,
\begin{equation}
    \mathrm{IoU} = \frac{|A \cap B|}{|A \cup B|} \;,
    \label{eq:IoU}
\end{equation}
where $A$ and $B$ denote two bounding boxes, the numerator represents the area of their intersection, while the denominator corresponds to the area of their union. To filter the RPN proposals, we adopt the default threshold of $\mathrm{IoU}_{\mathrm{thr}}^{\mathrm{RPN}}=0.7$~\citep{he2017}. Boxes that exhibit high IoU overlap with the selected box are suppressed, effectively removing redundant proposals. This process is repeated with the next highest-scoring unsuppressed box until all boxes have been processed or a predefined maximum number of proposals is reached. NMS thus helps to reduce computational load and enhances detection accuracy by retaining only the most promising, non-overlapping candidate regions.

The highest-ranked boxes, determined by their likelihood of containing an `object', are selected as the region of interest (RoI) for the following stage of the pipeline. In our work, we categorise RoIs into two classes, `gravitational arcs' and `background'; the latter, being trivial, will not be considered.

In the final stage of the Mask R-CNN, three tasks are performed simultaneously: a softmax classifier assigns probabilities for $K+1$ categories ($K=1$ in our case since we have only the `gravitational arcs' class, while the $+1$ is for the background), a regression module refines bounding box coordinates, and a fully convolutional network generates pixel-wise binary masks within the RoI for semantic segmentation. The overall loss function of the Mask R-CNN is computed as the sum of these three contributions.

Transfer learning is a technique that enables NNs to leverage knowledge acquired from one task to enhance performance on a related but distinct task (refer to~\citealt{tan2018} for an overview). This methodology capitalises on pre-existing weight values derived from one data set, employing them as the foundation for network initialisation for training the network on another data set. The advantages of this technique are twofold: it accelerates the training process, and it mitigates network overfitting. 

In our work, we utilise as a starting point the Mask R-CNN weights pre-trained on the Microsoft Common Objects in Context~\citep[MS COCO;][]{lin2014} data set. The MS COCO is a large-scale object detection, segmentation, key-point detection, and captioning dataset. It consists of $328\,000$ RGB $3$-channel images of common objects, categorised into $91$ classes. By utilizing transfer learning, we leverage the ``universal'' low-level geometric features learned from the images in MS COCO (e.g.~edges, curves, and corners), allowing our model to generalize effectively despite the relatively small number of instances in our training set. This allows the subsequent training on our specialized dataset to focus exclusively on learning the high-level morphology unique to gravitational arcs. After initialising the pre-trained NN parameters, we made trainable all its $\approx$\,44 million parameters, including also the backbone ones.

\subsection{Data preparation}

Since the \texttt{pytorch} implementation of Mask R-CNN is designed for standard $3$-channel RGB images, an additional pre-processing step was required to retain as much information as possible. The third channel was then created by averaging the $\JE$ and $\HE$ \Euclid bands (henceforth called the $JH_\sfont{E}$ band). The three RGB channels are thus $JH_\sfont{E}$, $\YE$, and $\IE$, respectively.

Before feeding each image to the NN, we standardised the intensity of the pixels of each image. First, we clipped each band of the images independently at the $98{\mathrm{th}}$-percentile. Then, for each band of each image in the training set, we rescaled the pixel values using the $z$-score normalisation~\citep{bishop2023}, in order for the input image values to span a similar range of values, as
\begin{align}
        R&=\left(JH_\sfont{E}-\langle JH_\sfont{E}\rangle\right)/\sigma_{JH_\sfont{E}} \;,\\
        G&=\left(\YE-\langle \YE\rangle\right)/\sigma_{\YE} \;,\\
        B&=\left(\IE-\langle \IE\rangle\right)/\sigma_{\IE} \;,
\end{align}
where $\langle \IE \rangle$ is the average value of the mean of the images in the $\IE$-band, while $\sigma_{\IE}$ is the average standard deviation, both computed over the whole training set (and similarly for the $\YE$- and $JH_\sfont{E}$-bands). This allows a faster convergence of the NN, and prevents vanishing or exploding gradients~\citep{lecun2012}. 
This standardisation also ensures that the NN is not affected by the exposure time, detector gain, or the final normalisation of the pipeline-reduced images. The same data standardisation is then applied to real images during the inference phase. 

We performed data augmentation on the training set, which makes the NN invariant with respect to the included transformations applied~\citep{mikolajczyk2018}. In particular, we augmented the data by randomly applying for each image the following transformations, preserving classes, bounding boxes, and object masks:
\begin{itemize}
    \item horizontal flip, with $50\%$ probability;
    \item vertical flip, with $50\%$ probability.
\end{itemize}
These basic augmentations replicate additional observational setups and conditions with minimal computational overhead, aiding the NN in better generalising its results.

\subsection{Loss function}

During training, each sampled RoI has a corresponding multitask loss, representing the three assignments of the Mask R-CNN (classification, location, and segmentation), expressed as $L_{\mathrm{RoI}} = L_{\mathrm{cls}} + L_{\mathrm{box}} + L_{\mathrm{mask}}$~\citep{he2017}, where:
\begin{itemize}
    \item the classification loss is $L_{\mathrm{cls}}=-\ln p_u$, where $p_u$ is the probability estimate from the NN for the true class $u$ and a discrete probability distribution $p = (p_0, \dots, p_K)$ is computed by a softmax on the output of a fully connected layer over $K + 1$ classes for each RoI~\citep{girshick2015};
    \item the bounding box loss $L_{\mathrm{box}}$ again follows the definition in~\citet{girshick2015}, using an $L1$ smooth loss of the offsets between predicted and real bounding boxes for each of the $K$ object classes;
    \item for the mask branch, which contains $K$ binary masks of resolution $m \times m$ for each class ($m$ being the size of the RoI), a per-pixel sigmoid function is applied, leading to $L_{\mathrm{mask}}$ being the average binary cross-entropy loss (refer to~\citealt{he2017} for the details).
\end{itemize}
The final training loss $L$ is the combination of the classification, box, and mask losses from the final branches of the Mask R-CNN, along with the classification and box losses from the RPN~\citep{lakshmanan2021}.

\subsection{Training}

To train the NN, we used an improved version of the stochastic gradient descent algorithm~\citep[SGD;][]{goodfellow2016}. SGD updates the model weights $\vec{\Theta}$ by minimising the loss function $L(\vec{\Theta})$, as described by
\begin{equation}
    \vec{\Theta}_{j+1} = \vec{\Theta}_j - \eta \frac{\partial}{\partial \vec{\Theta}_j} L(\vec{\Theta}_j) \; , 
    \label{eq:sgd}
\end{equation}
where $\eta$ represents the learning rate, a hyperparameter that is fine-tuned to avoid local minima and ensure convergence. 

In all epochs, we train all the weights in all layers, using a learning rate scheduler, starting with a value of $\eta = 10^{-3}$ and gradually decreasing it by a factor of $0.5$ every $20$ epochs. This approach facilitates deeper learning and more precise tuning of the weights, while at the same time reducing the risk of overfitting~\citep{goodfellow2016}. The SGD optimiser was initialised with a momentum value of $0.9$, and a weight decay coefficient of $5\times 10^{-4}$. Furthermore, to prevent overfitting we included the early stopping regularisation~\citep{prechelt1997, raskutti2011}. 

We trained the NN for $100$ epochs in total, processing $4\,000$ training and $500$ validation $3\times1199\times1199\,\mathrm{pixel}$ images per epoch, in batches of $16$. We used a single NVIDIA Quadro RTX 6000 GPU, with $3840$ cores and $24\,\mathrm{GB}$ GDDR5 memory, to train on $4000$ simulated $3$-band FITS images. The training took approximately $10$ hours to complete (wall time). After this initial cost to train the NN, detection and inference on images of the same size can be performed in a fraction of a second.

\section{\label{sec:results}Results}

In this section, we present the results of the training, the performance of the NN, and the results of the application to real \Euclid imaging data. We begin by testing the model's performance using traditional ML metrics for classification tasks. Following this, we examine the outcomes of applying the trained Mask R-CNN to real \Euclid Q1 observations, in order to assess its performance in discovering bright gravitational arcs in real galaxy cluster images.

\subsection{Performance metrics}

Two classes are present in our data set, the gravitational arcs (`positive' class) and everything else, i.e.~the background and all other astronomical objects (`negative' class). We use the MS COCO Evaluation Metrics~\citep{lin2014} to quantitatively evaluate the performance of the NN. In particular, to measure the classification performance on the test data set, we compute precision (or purity, $P$), recall (or completeness, $R$), and F1 score (their harmonic mean). Precision tells us how many of the positive predictions of the model are actually correct. Recall, on the other hand, is the ratio of correctly identified positive samples to the total number of actual positive samples in the data set. These metrics are defined as
\begin{align}
    P &= \frac{\mathrm{TP}}{\mathrm{TP} + \mathrm{FP}}  \;,\\
    R &= \frac{\mathrm{TP}}{\mathrm{TP} + \mathrm{FN}} \;,\\ 
    \mathrm{F1} &= 2\frac{P \, R}{P+R} \;, 
\end{align}
where TP stands for true positive, the number of objects correctly classified as the positive class, TN for true negative, the number of objects correctly classified as the negative class, FP for false positive, the number of objects incorrectly labelled as belonging to the positive class, and FN for false negative, the number of objects incorrectly labelled as belonging to the negative class. 

A detection is considered `positive' if its classification confidence score exceeds a specified minimum threshold $p_{\mathrm{thr}}$, and such that its predicted bounding box overlaps with the ground truth one, giving an IoU larger than a threshold ($\mathrm{IoU}_{\mathrm{thr}}$). Therefore, TPs are identified when a detection has a confidence score above $p_{\mathrm{thr}}$ and can be matched to a ground truth object with $\mathrm{IoU}>\mathrm{IoU}_{\mathrm{thr}}$, FNs refer to ground truth objects that lack a corresponding detection, and FPs are detections with a high confidence score that do not correspond to any ground truth object.

Precision and recall are not particularly informative when considered separately; an object detector is considered effective only if its precision remains high as recall increases~\citep{ivezic2020}. Therefore, to measure the performance of the NN, we use the average precision (AP) score~\citep{everingham2010}, a widely adopted metric in the DL and computer vision communities, which has superseded the area under the receiver operating characteristic curve. The AP (for a specific class) is calculated as the area under the precision-recall ($P$--$R$) curve, representing the precision averaged across uniformly distributed recall values from $0$ to $1$. Generally, a higher precision at a given recall value indicates superior model detection performance. This is evidenced by a larger area under the $P$--$R$ curve; hence, higher AP values indicate better detection capabilities of the model. In this work, AP is computed for each image by averaging precision values at $101$ equally spaced recall levels, \( R \in [0, 0.01, \dots, 1.0] \), namely as~\citep{everingham2012}
\begin{equation}
    \mathrm{AP} = \frac{1}{101} \sum_{R \in [0, 0.01, \dots, 1.0]} P(R) \; , 
    \label{eq:ap}
\end{equation}
where $P(R)$ is the maximum precision within the interval $\Delta R$. We then calculate the average of the $P$--$R$ curves and mean AP scores across all test set images. 

In the MS COCO evaluation metrics, the network’s ability to distinguish and classify objects is assessed by computing AP in different ways. We use the standard AP score variants commonly used in the computer vision field. First, we repeat the computation of AP varying $\mathrm{IoU}_{\mathrm{thr}}$ from $0.50$ to $0.95$ in steps of $0.05$, resulting in $10$ sets of $P$--$R$ curves. The $10$ AP values derived from these $10$ sets are then averaged and denoted as $\mathrm{AP}_{@50:5:95}$, the primary COCO challenge metric. Then, $\mathrm{AP}_{50}$~\citep[the Pascal VOC metric;][]{everingham2010} and $\mathrm{AP}_{75}$ are the AP metrics computed with $\mathrm{IoU}_{\mathrm{thr}}$ fixed at $0.5$ and $0.75$, respectively. Additionally, the AP for objects of different pixel areas is computed to evaluate the network’s detection performance across various object sizes. For gravitational arcs detection, objects with an area smaller than $16^2$ pixels are denoted as $\mathrm{AP}_{\mathrm{S}}$, those with an area between $16^2$ and $32^2$ pixels are denoted as $\mathrm{AP}_{\mathrm{M}}$, and those larger than $32^2$ pixels are denoted as $\mathrm{AP}_{\mathrm{L}}$, calculated using the same IoU threshold range as that of $\mathrm{AP}_{@50:5:95}$\footnote{Notice that the values of $\mathrm{AP}_{\mathrm{S}}$, $\mathrm{AP}_{\mathrm{M}}$, and $\mathrm{AP}_{\mathrm{L}}$ used in our work are smaller with respect to those in the COCO Evaluation Metrics, since the pixel areas of gravitational arcs are generally smaller than typical COCO images.}. The average recall (AR) instead is the maximum recall given a fixed number of detections per image ($100$ in our case), averaged over all the categories and IoU threshold values ranging from $0.50$ to $0.95$ in steps of $0.05$.

\subsection{Performance on the test set}

\begin{figure}
    \centering
    \includegraphics[width=1.0\hsize]{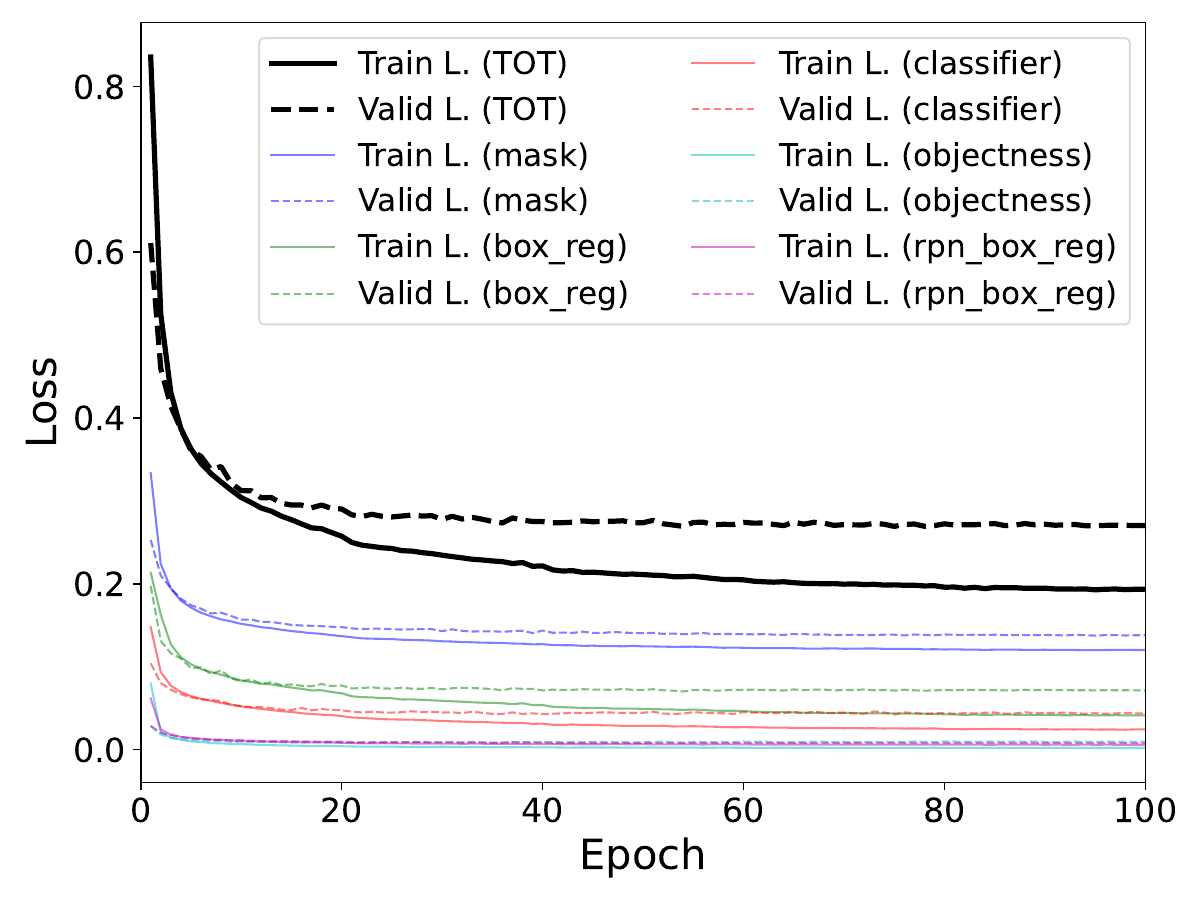}
    \caption{Training history of the total loss and all its individual components as a function of the training epoch, for the training (solid lines) and validation (dashed lines) data sets. The total loss (black lines) is the sum of the classification, box, and mask losses from the Mask R-CNN, along with the classification and box losses from the RPN.}
    \label{fig:losses}
\end{figure}

In Fig.~\ref{fig:losses} we show the training history of the total loss and its individual components as a function of the training epoch for both training and validation sets. The validation loss value stabilised after approximately $20$ iterations.

\begin{figure*}
    \centering
    \includegraphics[angle=0,width=1.0\hsize]{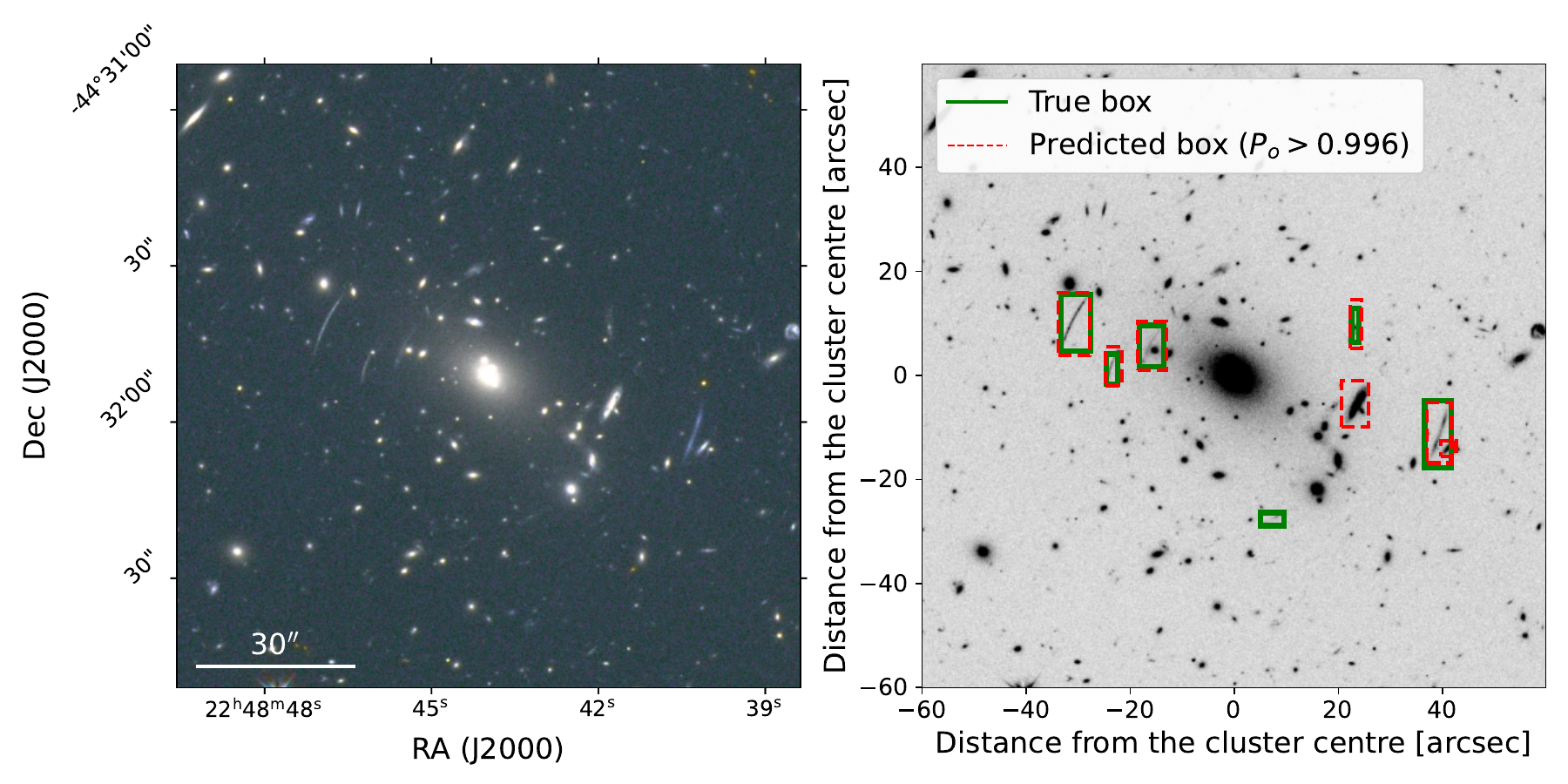} 
    \caption{\emph{Left}: Euclidised $2' \times 2'$ RGB image ($R=JH_\sfont{E}$, $G=\YE$, $B=\IE$) of the galaxy cluster Abell 1063, belonging to the test set; the right-most arc is the one injected via SL simulation, while the others are real. \emph{Right}: single channel $2' \times 2'$ Euclidised $\IE$ image of the same cluster. The green boxes enclose the gravitational arcs (both real and simulated) present in the field, i.e.~the ground truth, while the red dashed boxes are the `gravitational arcs' found by the NN, having an object confidence score greater than $0.996$.}
    \label{fig:plot_3_0.0.29}
\end{figure*}

As an example of the inference, we show in Fig.~\ref{fig:plot_3_0.0.29} the result of the application of the trained Mask R-CNN on a galaxy cluster belonging to the test set, and thus, not seen during the training phase. The green bounding boxes correspond to the ground truth, i.e.~they enclose the gravitational arcs present in the galaxy cluster (both the real and simulated ones), while the red dashed boxes are the final proposed `positives' by the NN. We can see from this example that the NN correctly recovers all the largest gravitational arcs, missing only the smaller ones and producing a few false positives.

In Table~\ref{table:AP}, we present the COCO AP and AR metrics used to assess the performance of the NN on the test set, showing the model’s accuracy across various IoU thresholds and object sizes. These metrics offer insight into the strengths and limitations of the network. 
We observe that both AP and AR increase with arc size, with the best performance achieved in the large object category. This is expected, since larger arcs are more easily distinguishable from background structures and provide more pixels for the network to identify and segment. In contrast, smaller arcs are more affected by noise, and are often harder to resolve, given the resolution and depth of the images. This trend is also a direct consequence of the fact that our training set included only simulated arcs with an area larger than $400$\,pixels, biasing the network toward high completeness and purity in the detection of large arcs. Furthermore, the comparison between $\mathrm{AP}_{50}$ ($54.3\%$) and $\mathrm{AP}_{75}$ ($20.1\%$) highlights that while the NN is relatively good at localising arcs with moderate overlap, it struggles to precisely match their shape and boundaries -- a known challenge in segmenting faint, irregular sources.

\begingroup
\setlength{\tabcolsep}{6pt}       
\renewcommand{\arraystretch}{1.2}
\begin{table*}
\centering
\caption{Summary of AP and AR bounding box results of the NN on the test set.}
\label{table:AP}
\begin{tabular}{ccc|ccc|ccc}
    \hline\hline
    $\mathrm{AP}_{@50:5:95}$ & $\mathrm{AP}_{50}$ & $\mathrm{AP}_{75}$ & $\mathrm{AP}_\mathrm{S}$ & $\mathrm{AP}_\mathrm{M}$ & $\mathrm{AP}_\mathrm{L}$ & $\mathrm{AR}_\mathrm{S}$ & $\mathrm{AR}_\mathrm{M}$ & $\mathrm{AR}_\mathrm{L}$\\ 
    \hline
    $21.8\%$ & $54.3\%$ & $20.1\%$ & $-1$ & $15.8\%$ & $41.9\%$ & $-1$ & $18.1\%$ & $57.9\%$\\
    \hline\hline
\end{tabular}
\tablefoot{$\mathrm{AP}_{@50:5:95}$ represents the AP score averaged over IoU thresholds of $\mathrm{IoU}_{\mathrm{thr}} \in \{0.50, 0.55, \dots, 0.95\}$. $\mathrm{AP}_\mathrm{50}$ and $\mathrm{AP}_{75}$ refer to the AP score at IoU threshold of $0.5$ and $0.75$, respectively. $\mathrm{AP}_\mathrm{S}$, $\mathrm{AP}_\mathrm{M}$, and $\mathrm{AP}_\mathrm{L}$ denote the AP values for sources with small ($\mathrm{area}<16^2 \, \mathrm{pixel}$), medium ($16^2<\mathrm{area}<32^2 \, \mathrm{pixel}$), and large ($\mathrm{area}>32^2 \, \mathrm{pixel}$) bounding box area, respectively, averaged over IoU threshold values ranging from $0.50$ to $0.95$ in steps of $0.05$. Same for the three $\mathrm{AR}_{\mathrm{S,M,L}}$. The $-1$ means that there were no objects belonging to that particular class.}
\end{table*}
\endgroup

\begin{figure*}
    \centering
    \begin{minipage}{0.33\textwidth}
        \centering
        \includegraphics[width=\textwidth]{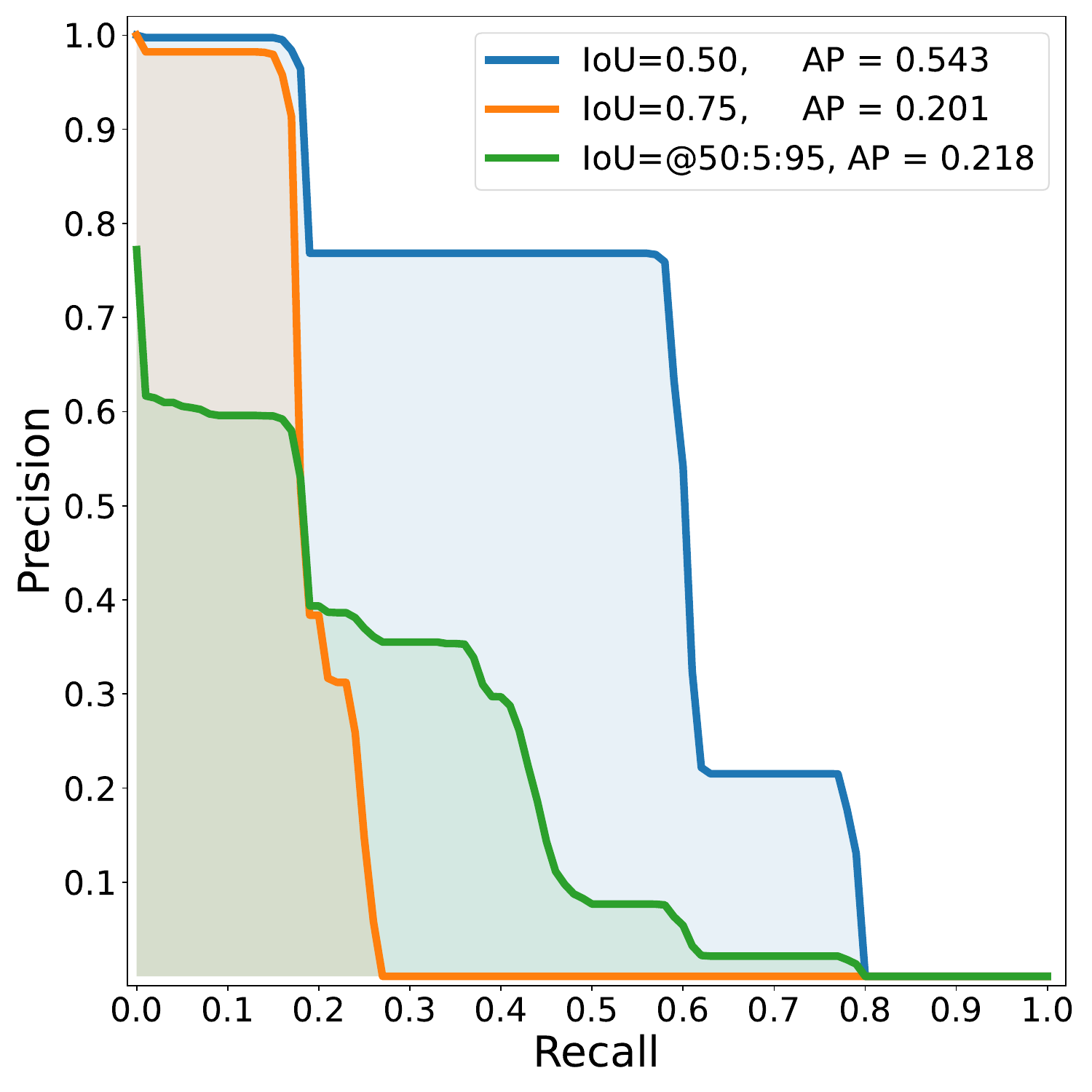}
        \vspace{-0.5cm} %
        \\\textbf{(a)}
    \end{minipage}
    \hfill
    \begin{minipage}{0.33\textwidth}
        \centering
        \includegraphics[width=\textwidth]{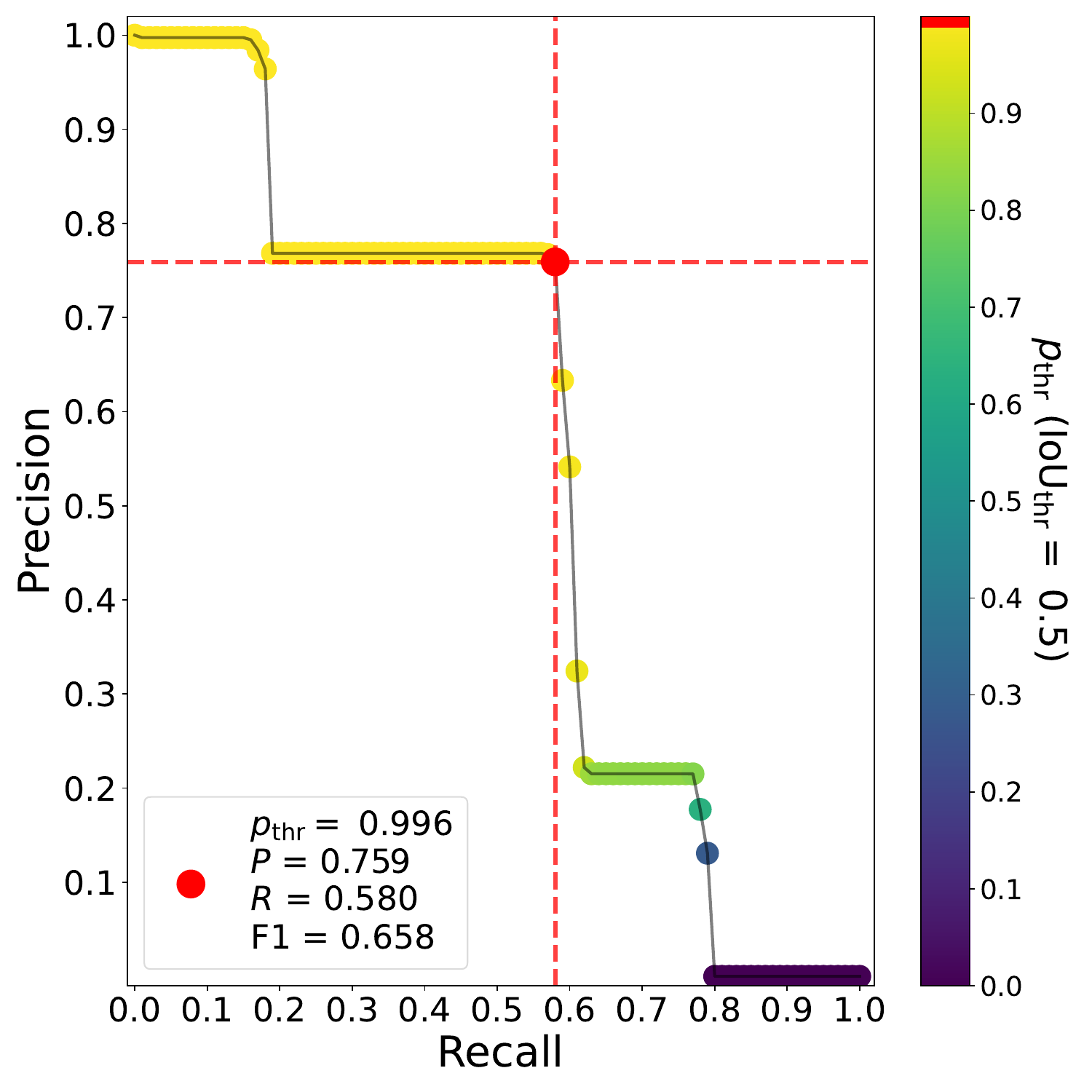}
        \vspace{-0.5cm}
        \\\textbf{(b)}
    \end{minipage}
    \hfill
    \begin{minipage}{0.33\textwidth}
        \centering
        \includegraphics[width=\textwidth]{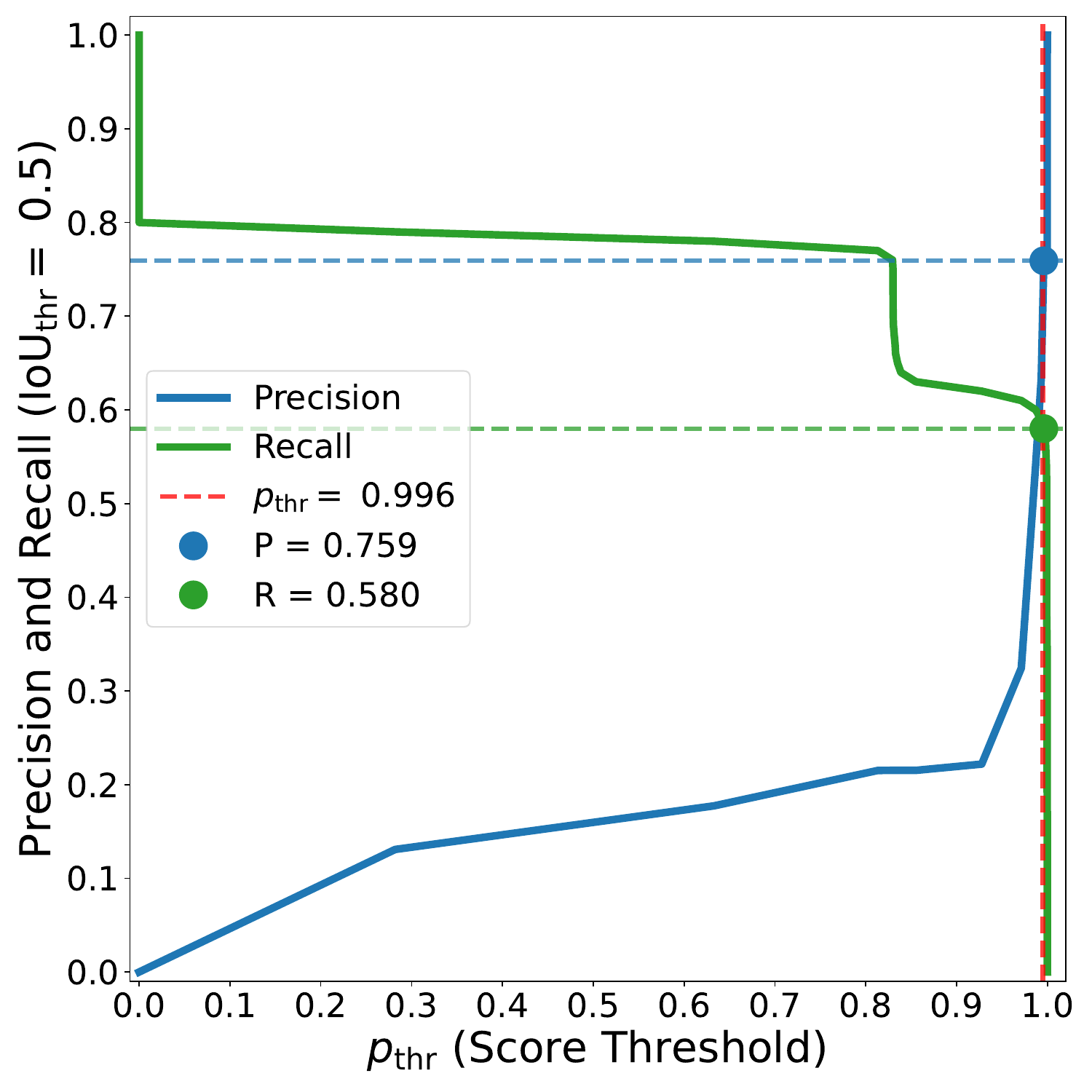}
        \vspace{-0.5cm}
        \\\textbf{(c)}
    \end{minipage}
    \caption{In panel~(a) we show two sets of three $P$--$R$ curves for $\mathrm{IoU}_{\mathrm{thr}} = 0.5, 0.75$, @$50$:$5$:$95$, each obtained by varying the object score threshold $p_{\mathrm{thr}}$. In panel~(b) we show the $P$--$R$ curve for $\mathrm{IoU}_{\mathrm{thr}} = 0.5$, where the dots are colour-coded by the score threshold $p_{\mathrm{thr}}$ used to evaluate them. In particular, we emphasise the $P$--$R$ pair associated to our choice of score threshold $p_{\mathrm{thr}}=0.996$. Note that the recall remains relatively high even at moderate precision, reflecting the network's ability to recover most bright arcs while keeping the number of false positives manageable. Increasing the threshold shifts the operating point toward higher precision at the expense of recall. The extended horizontal segments are an artifact of the test set construction: all $500$ test images are simulations based on the same cluster (Abell 1063). As a consequence, a subset of detections (and their associated confidence scores) remains identical across the test set, producing clustered score values and thus plateau-like features in the PR curve. In panel~(c) we show two precision and the recall curves as a function of the score threshold for $\mathrm{IoU}_{\mathrm{thr}} = 0.5$.}
    \label{fig:pr-curves}
\end{figure*}

In Fig.~\ref{fig:pr-curves}~(a) we present the three $P$--$R$ curves for $\mathrm{IoU}_{\mathrm{thr}} = 0.5, 0.75$, and @$50$:$5$:$95$\footnote{@$50$:$5$:$95$ means the value of the reference metric obtained by averaging it over IoU thresholds of $\mathrm{IoU}_{\mathrm{thr}} \in \{0.50, 0.55, \dots, 0.95\}$.}, each obtained by varying the object score classification threshold $p_{\mathrm{thr}}$. In Fig.~\ref{fig:pr-curves}~(b) we show the $P$--$R$ curve for $\mathrm{IoU}_{\mathrm{thr}} = 0.5$, where the dots are colour coded by the score threshold $p_{\mathrm{thr}}$ used to evaluate them. Finally, in Fig.~\ref{fig:pr-curves}~(c) we show the precision and the recall as a function of the score threshold for $\mathrm{IoU}_{\mathrm{thr}} = 0.5$. By choosing a fixed value of the object probability threshold, $p_{\mathrm{thr}}=0.996$, which can be tuned depending if we want to maximise precision over completeness, or vice versa, we can then compute the precision, recall, and F1 score of the NN on the test set, obtaining $P = 75.9\%$, $R = 58.0\%$, and $\mathrm{F1} = 65.8\%$.

We note that the test set is composed of images containing several positive examples by construction. In a realistic scenario, where most clusters do not host giant arcs, the effective precision would decrease due to the lower prevalence of true positives relative to false positives.

\subsection{Inference on \Euclid Q1 lens clusters}

As a further test, we wanted to assess the performance of the NN on the $20$ galaxy clusters with lens probability $\mathcal{P}_{\mathrm{lens}} > 0.90$ obtained from the~\citet{Q1-SP057} visual inspection. 
\citet{Q1-SP057} presented the first catalogue of SL galaxy clusters identified in the \Euclid Q1 observations, based on a systematic visual inspection of $1260$ richness-selected galaxy clusters from~\citet{wen2024}, over an effective area of $4.4\,\deg^2$. They identified $83$ gravitational lenses with $\mathcal{P}_{\mathrm{lens}} > 0.5$, including $14$ systems with $\mathcal{P}_{\mathrm{lens}} = 1$ that exhibit secure SL features such as giant tangential and radial arcs, as well as multiple images. 

\begin{figure}
    \centering
    \includegraphics[width=1.0\hsize]{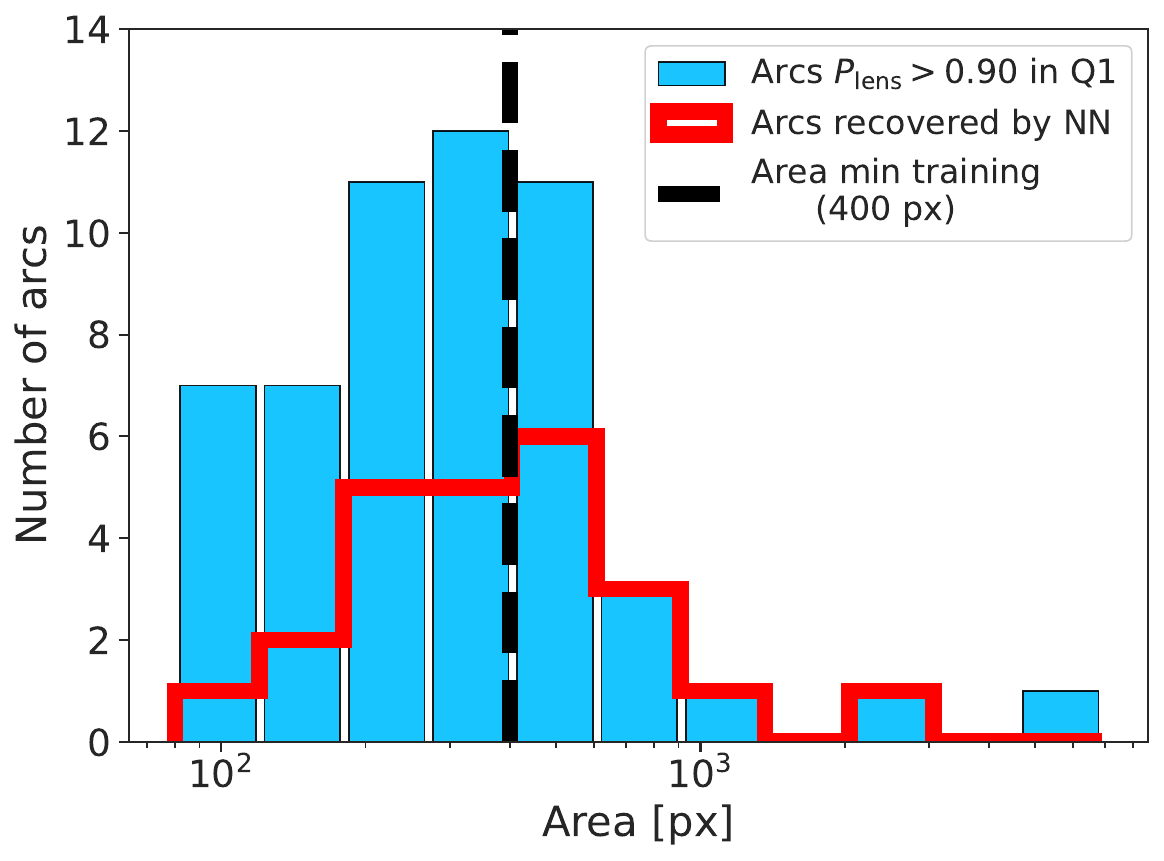}
    \caption{Distribution of area in pixels of the gravitational arcs found by expert astronomers in the 20 \Euclid Q1 clusters with $\mathcal{P}_{\mathrm{lens}}>90\%$~\citep{Q1-SP057}. The red dashed line shows the minimum value of the area of the arcs used as the ground truth for the training (400\,pixels).}
    \label{fig:area_hist_euclid_q1}
\end{figure}

In Fig.~\ref{fig:area_hist_euclid_q1}, we show in blue the distribution of the areas of the visually selected gravitational arcs\footnote{These binary masks were manually drawn based on the bounding boxes surrounding the lensing events provided by the expert astronomers in~\citet{Q1-SP057}.}; in particular, we highlight the fact that the majority of them ($\approx65\%$) have an area which is smaller than the minimum area used in the training set for the simulated arcs (i.e.~$400$\,pixels, indicated in the plot by the black dashed line). The red line instead shows the area distribution of the gravitational arcs correctly recovered by the NN. As expected, the majority of arcs with area larger than the training threshold are found ($12/18 \approx 66\,\%$). Overall, by considering all the arcs found during the visual inspection in the $20$ $\mathcal{P}_{\mathrm{lens}} > 0.90$ galaxy clusters as the ground truth for a further test set, the NN achieves a precision value of $P=19\%$, and recall of $R=42\%$. In Appendix~\ref{apdx:A} we show the predictions of the NN as bounding boxes over-plotted on the images of these 20 \Euclid Q1 clusters, alongside the real arcs found during the visual inspection.

\section{\label{sec:discussion}Discussion}

Despite the promising results, certain limitations remain. Bright stars and elongated shapes occasionally mimicking arcs, produce false positives, highlighting the need for improved pre-processing of the images and more diversity in the training set. Increasing the prevalence of such challenging objects in the training set could help the network better distinguish true arcs from lens-like false positives. While the simulated data set effectively mimics real observations, the reliance on simulations may limit generalisability. As \Euclid acquires more observational data, the performance of the Mask R-CNN is expected to improve for several reasons. With more real \Euclid observations available, we can generate larger and more representative training data sets. This will help mitigate domain shift issues between simulated and real data, improving the generalization of the network. Moreover, new mock data sets can be generated with more realistic arc properties, taking into account improved lens models, observational systematics, and the full range of image conditions and artefacts in \Euclid data. For instance, using real \Euclid cluster images as the background onto which mock lensed sources are injected ensures that the simulations accurately reflect the properties of the actual observations, including realistic noise, PSF, photometric depth, and the presence of instrumental features such as diffraction spikes or saturated stars. Once a sufficiently large real data set is available, fine-tuning the existing Mask R-CNN model with actual \Euclid images, rather than only relying on simulated arcs, will lead to better performance.

While the average precision and average recall test set results for large arcs ($\mathrm{AP}_\mathrm{L}$ and $\mathrm{AR}_\mathrm{L}$) are promising, the situation for medium- and small-sized ones reflected the inherent challenges in detecting these structures, which are often faint, blended with cluster galaxies, and affected by noise. In our work, we aimed particularly at detecting large bright gravitational arcs; therefore, the arcs injected in the training set reflected this goal. With continued improvements in training data sets, preprocessing, and model architecture, the detection of medium-sized arcs is expected to improve. However, small arcs will remain particularly challenging to detect due to their intrinsically low signal-to-noise ratios, and the higher abundance of small objects in the field that can mimic arc-like morphologies, compared to the relatively rarer large arc-like features that are not genuine arcs.

\section{\label{sec:conclusions}Summary and conclusions}

In this paper, we have developed a DL methodology to autonomously identify cluster-scale SL gravitational arcs within \Euclid multi-band imaging of galaxy clusters. The challenges of gravitational arc detection in wide-field surveys are well-suited for the Mask R-CNN architecture, an advanced computer vision framework offering several important advantages over traditional CNNs. First, it performs instance segmentation, generating detailed pixel-level masks for each detected object by integrating classification, bounding box regression, and segmentation into a unified framework. Standard CNNs, instead, are typically limited to single-task operations such as whole-image classification. In addition, Mask R-CNN can detect and segment multiple objects within a single image, even when they appear at different scales or overlap, effectively distinguishing individual instances, something conventional CNNs often struggle with. Another important advantage is interpretability: the pixel-level segmentation masks produced by the Mask R-CNN framework make it possible to visualise exactly which image regions the network has identified as arcs. This helps in understanding the origin of false positives, a task that is far more challenging with conventional CNNs that provide only image-level classifications. Finally, it can handle input images of varying sizes without requiring resizing to a fixed dimension, preserving structural details. In this context, our work represents one of the first applications of instance segmentation networks to the automated detection of cluster-scale gravitational arcs in real survey data, paving the way for future large-scale lensing searches in \Euclid and other wide-field surveys.

The effectiveness of the network in detecting and classifying GCSLs within \Euclid images is evaluated through several standard DL performance metrics. We also assess the capability of the NN to recover bright gravitational arcs in real \Euclid Q1 observations, comparing the results of the inference with those obtained by human experts in a visual inspection~\citep{Q1-SP057}. Our analysis on the test set revealed that the Mask R-CNN framework performs well in arc detection, achieving a good level of precision-recall trade-off. Notably, the architecture’s ability to simultaneously localise, classify, and segment arcs streamlined the detection process and facilitated accurate identification in crowded fields such as those of galaxy clusters. The successful application of this framework to real \Euclid Q1 images has validated its practical utility, with the majority of large arcs identified aligning with expert classifications.

The Mask R-CNN framework was selected for this study due to its strong performance in instance segmentation and its ability to simultaneously classify, localise, and segment gravitational arcs. It is not necessarily the definitive or optimal solution, and there are several alternative architectures that could be explored in future work to further improve arc detection and segmentation, like U-Nets~\citep{ronneberger2015} and YOLOs~\citep{redmon2016}; such networks, however, generally perform poorer on typical computer vision benchmarks~\citep{he2017}.

Deploying the NN blindly across all \Euclid tiles would inevitably lead to a significant number of false positives, which, while expected, would require extensive post-processing. The best strategy would be then to apply the network primarily to a selected subset of the richest galaxy clusters rather than across the entire survey tiles. This approach aims to balance minimising false positives with ensuring that genuine SL events are not missed. Moreover, while the DL models provide a fairly high level of accuracy in arc detection, a human validation step remains crucial to minimise false positives and ensure high purity in the final candidate sample. This step can be performed by expert astronomers visually inspecting the top-ranked candidates. A key advantage of our approach lies in its ability to significantly reduce the visual inspection effort required in large-scale surveys. In the~\citet{Q1-SP057} analysis of \Euclid Q1 data, around $40$ experts visually examined approximately $1300$ candidate clusters, selected with a cut on richness from the~\citet{wen2024} catalogue. This process, while effective, is not scalable to future releases; indeed, \citet{Q1-SP057} estimate that, with the same number of human experts, it would take more than $15$ years to visually inspect the entire EWS area. 

This study further illustrates the potential of DL techniques in astrophysics, enabling the efficient and automated analysis of gravitational lensing phenomena in large-scale surveys. As current and next-generation surveys will produce unprecedentedly large data sets, advanced computer vision methodologies like this will be crucial in unlocking the full potential of these observations.

\begin{acknowledgements}
We thank the anonymous referee for the helpful and insightful comments that improved the paper. 
We acknowledge financial support through grant 2020SKSTHZ.
LB is indebted to the communities behind the multiple free, libre, and open-source software packages on which we all depend. MM was supported by INAF Grants “The Big-Data era of
cluster lensing" and "Probing Dark Matter and Galaxy Formation in Galaxy Clusters through Strong Gravitational Lensing", and ASI Grant n.
2024-10-HH.0 "Attività scientifiche per la missione Euclid – fase E". The research activities described in this paper have been
co-funded by the European Union – NextGeneration EU within PRIN 2022
project no. 20229YBSAN – Globular clusters in cosmological simulations and in lensed fields: from their birth to the present epoch. 
  \AckEC 
  \AckQone 
  This work used the following software packages:
    \href{https://www.python.org/}{\texttt{Python}}~\citep{python},
    \href{https://pytorch.org/}{\texttt{PyTorch}}~\citep{pytorch, Ansel_PyTorch_2_Faster_2024},
    \href{https://github.com/pytorch/vision}{\texttt{torchvision}}~\citep{torchvision2016},
    \href{https://github.com/numpy/numpy}{\texttt{NumPy}}~\citep{numpy1, numpy2},
    \href{https://github.com/scipy/scipy}{\texttt{SciPy}}~\citep{scipy}
    \href{https://scikit-learn.org/stable/index.html}{\texttt{scikit-learn}}~\citep{scikit-learn}
    \href{https://github.com/astropy/astropy}{\texttt{Astropy}}~\citep{astropy1, astropy2},
    \href{https://github.com/matplotlib/matplotlib}{\texttt{matplotlib}}~\citep{matplotlib},
    \href{https://git-cral.univ-lyon1.fr/lenstool/lenstool}{\textsc{lenstool}}~\citep{kneib1996, jullo2007, jullo2009},
    \href{https://sites.google.com/cfa.harvard.edu/saoimageds9/home}{\texttt{ds9}}~\citep{ds9},
    \href{https://aladin.cds.unistra.fr/}{\texttt{Aladin}}~\citep{aladin},
    \href{https://www.star.bris.ac.uk/~mbt/topcat/}{\texttt{topcat}}~\citep{topcat},
    \href{https://git-scm.com/}{\texttt{git}},
    \href{https://www.gnu.org/software/bash/}{\texttt{bash}}~\citep{gnu2007free}.
\end{acknowledgements}

\bibliography{bibliography_full}

\begin{appendix}
\section{Predictions on \Euclid Q1 clusters\label{apdx:A}}
In this Appendix, we show the inference results of the trained Mask R-CNN on the $20$ $\mathcal{P}_{\mathrm{lens}} > 0.90$ galaxy clusters from~\citet{Q1-SP057}. The NN has been applied on a $2' \times 2'$ MER cutout on the nominal centre of the cluster, according to the catalogue. The red masks enclose the real arcs identified by expert astronomers; the dashed rectangles instead show the predicted `gravitational arcs' by the NN, using the object score threshold $p_{\mathrm{thr}}=0.996$ employed for the test set (cf.~Fig.~\ref{fig:plot_3_0.0.29}). We note the presence of false positives, often associated with bright elongated galaxies, edge-on discs, or image artefacts that mimic arc-like morphologies, highlighting the need for further refinement of the training data or post-processing steps.

\begin{figure*}[htbp!]
\centering
\begin{minipage}{0.43\textwidth}
    \centering
    \includegraphics[width=\textwidth]{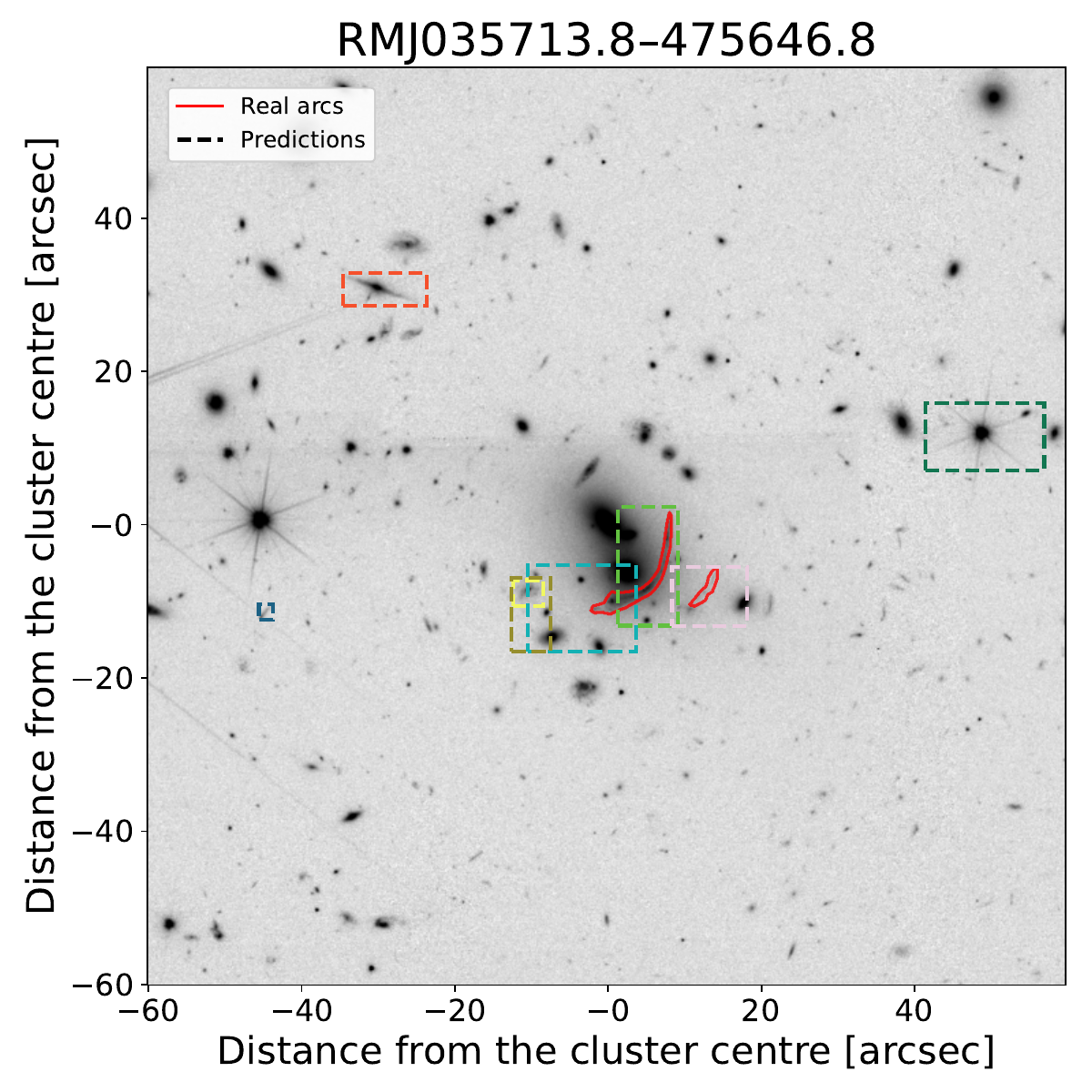}
\end{minipage}
\hfill
\begin{minipage}{0.43\textwidth}
    \centering
    \includegraphics[width=\textwidth]{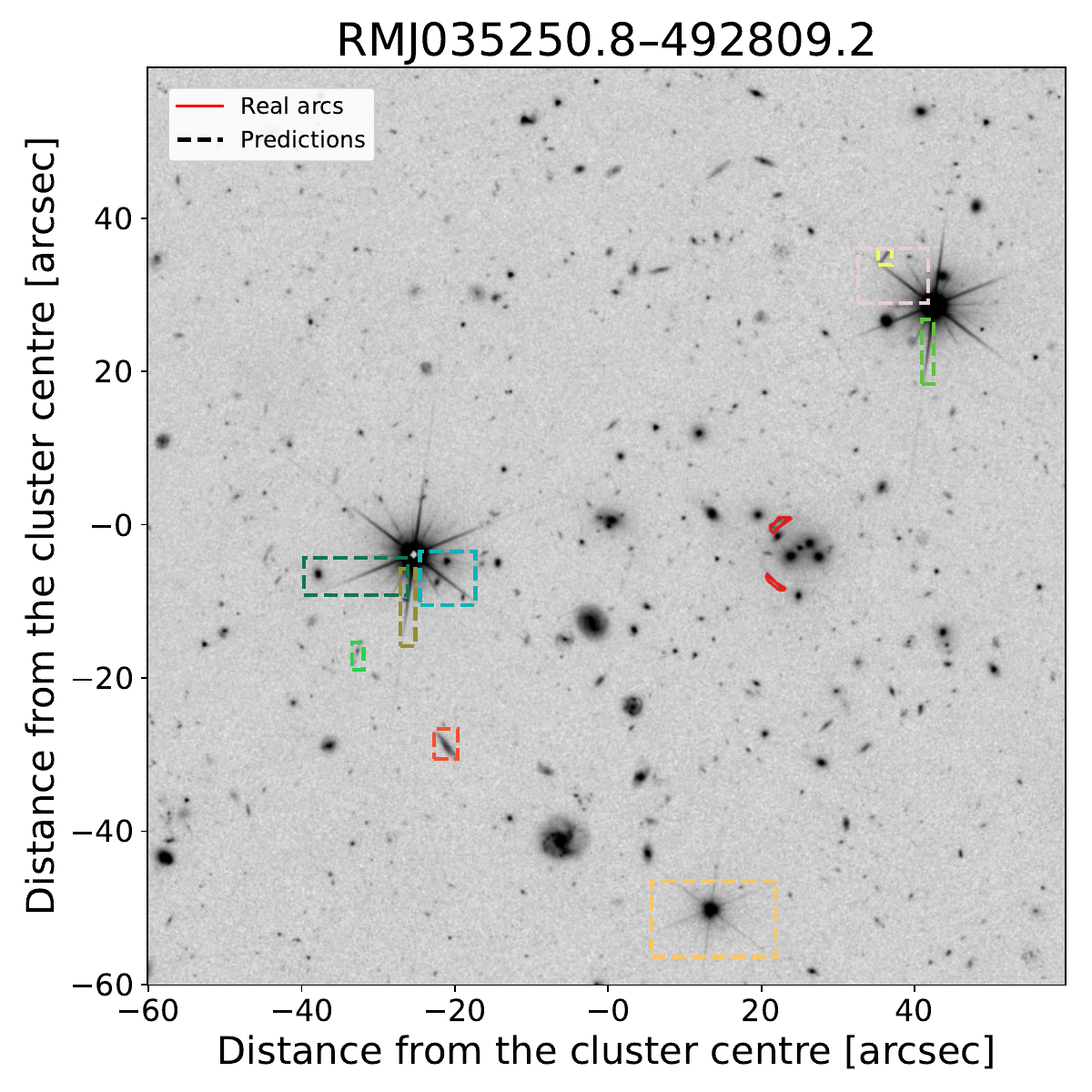}
\end{minipage}

\vspace{0.5cm}

\begin{minipage}{0.43\textwidth}
    \centering
    \includegraphics[width=\textwidth]{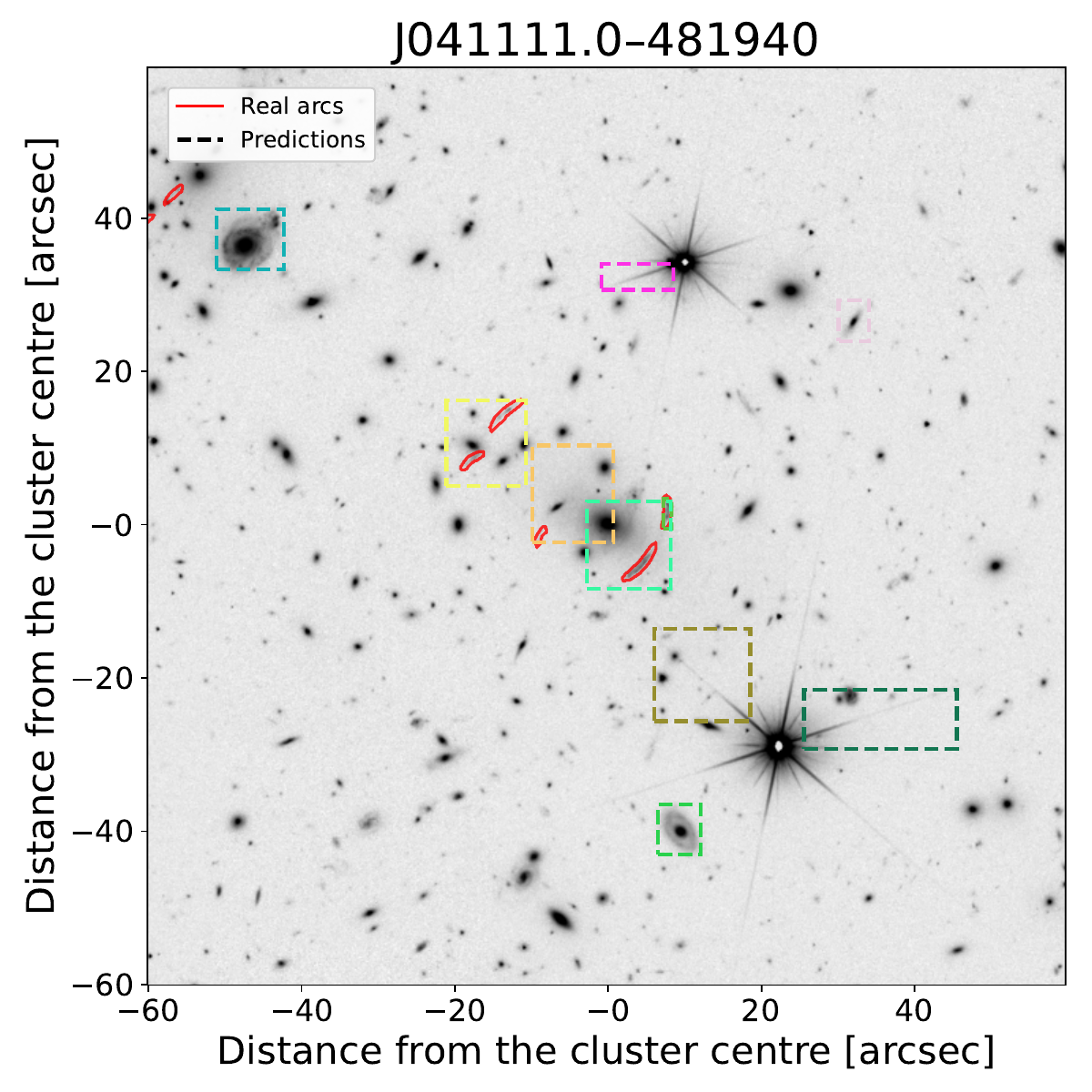}
\end{minipage}
\hfill
\begin{minipage}{0.43\textwidth}
    \centering
    \includegraphics[width=\textwidth]{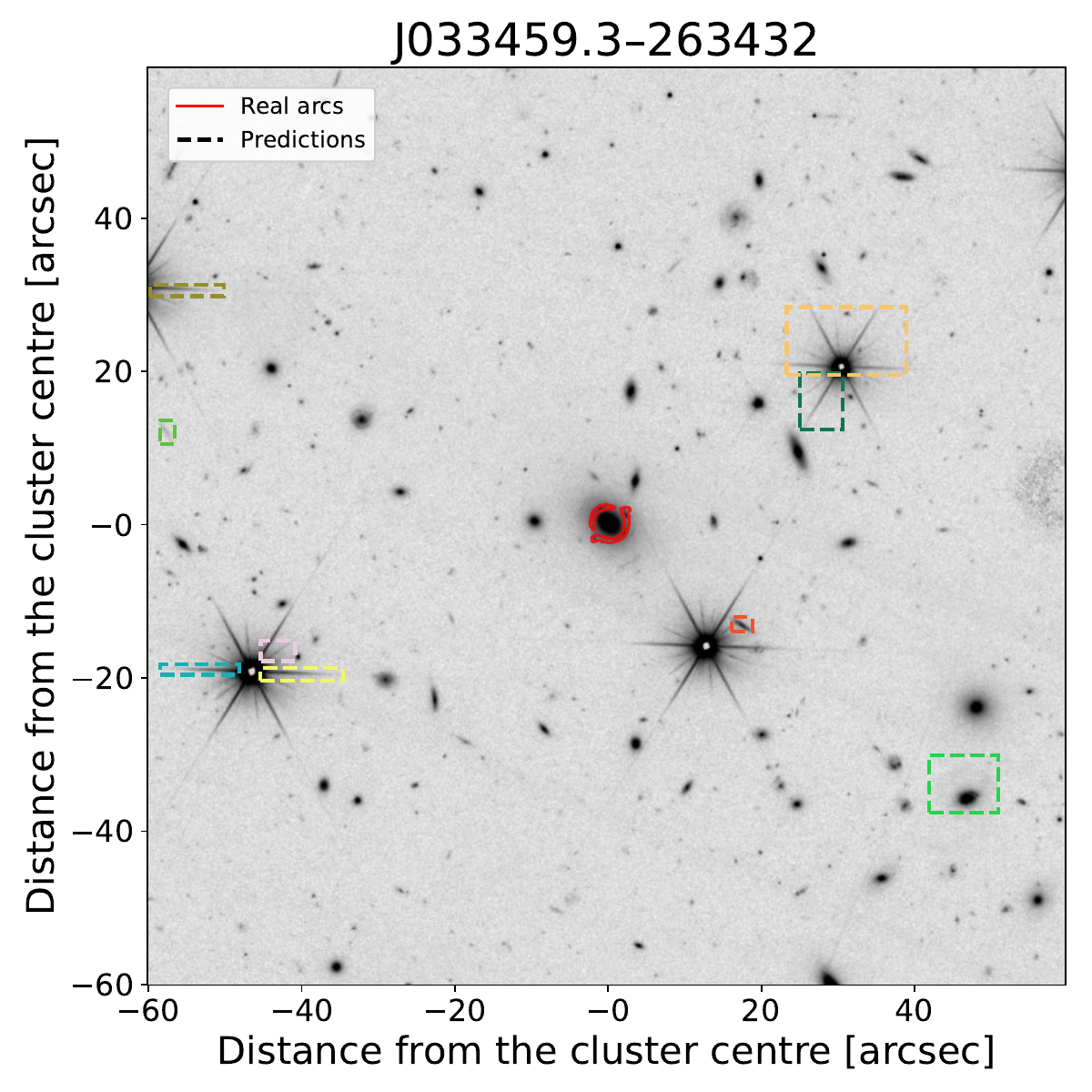}
\end{minipage}

\vspace{0.5cm}

\begin{minipage}{0.43\textwidth}
    \centering
    \includegraphics[width=\textwidth]{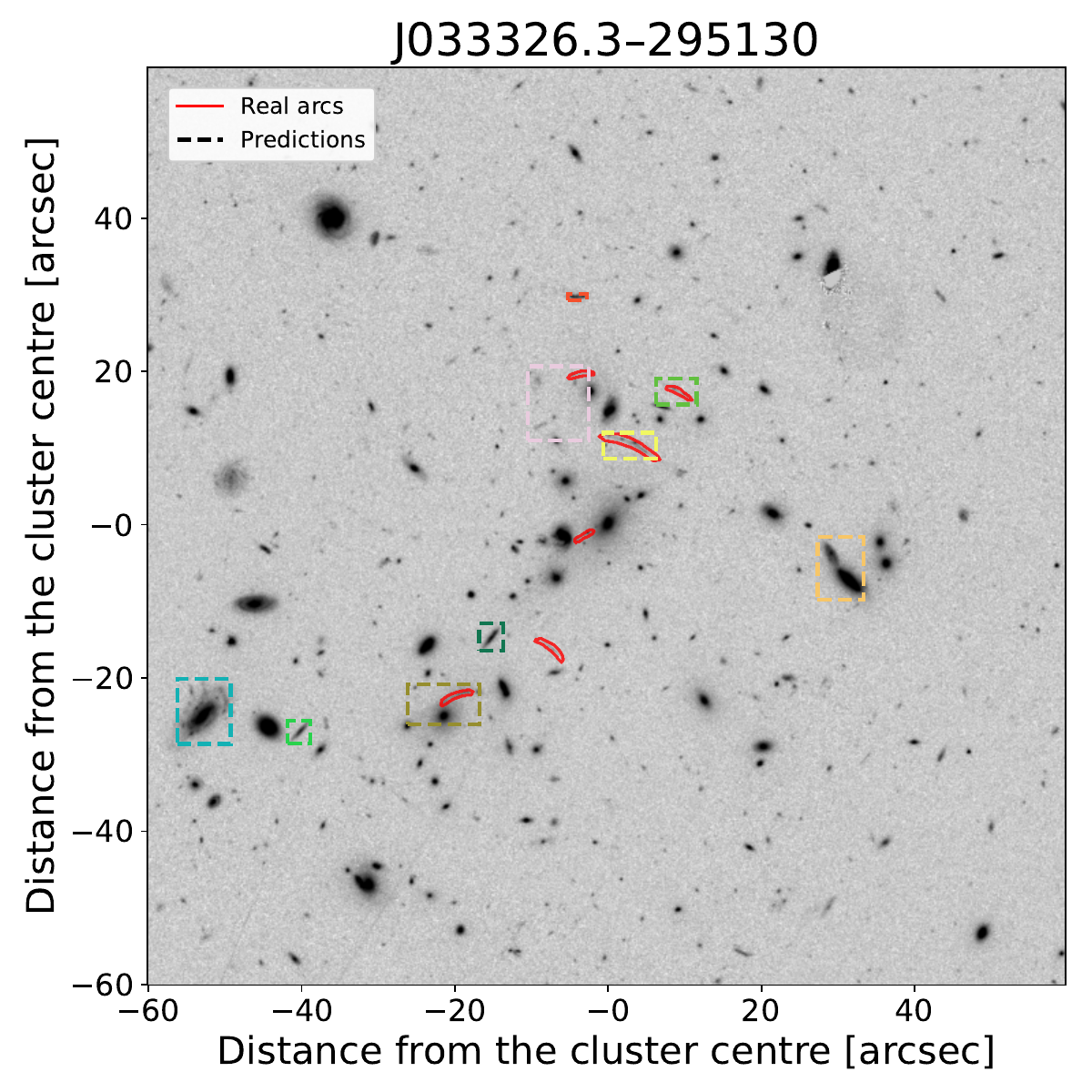}
\end{minipage}
\hfill
\begin{minipage}{0.43\textwidth}
    \centering
    \includegraphics[width=\textwidth]{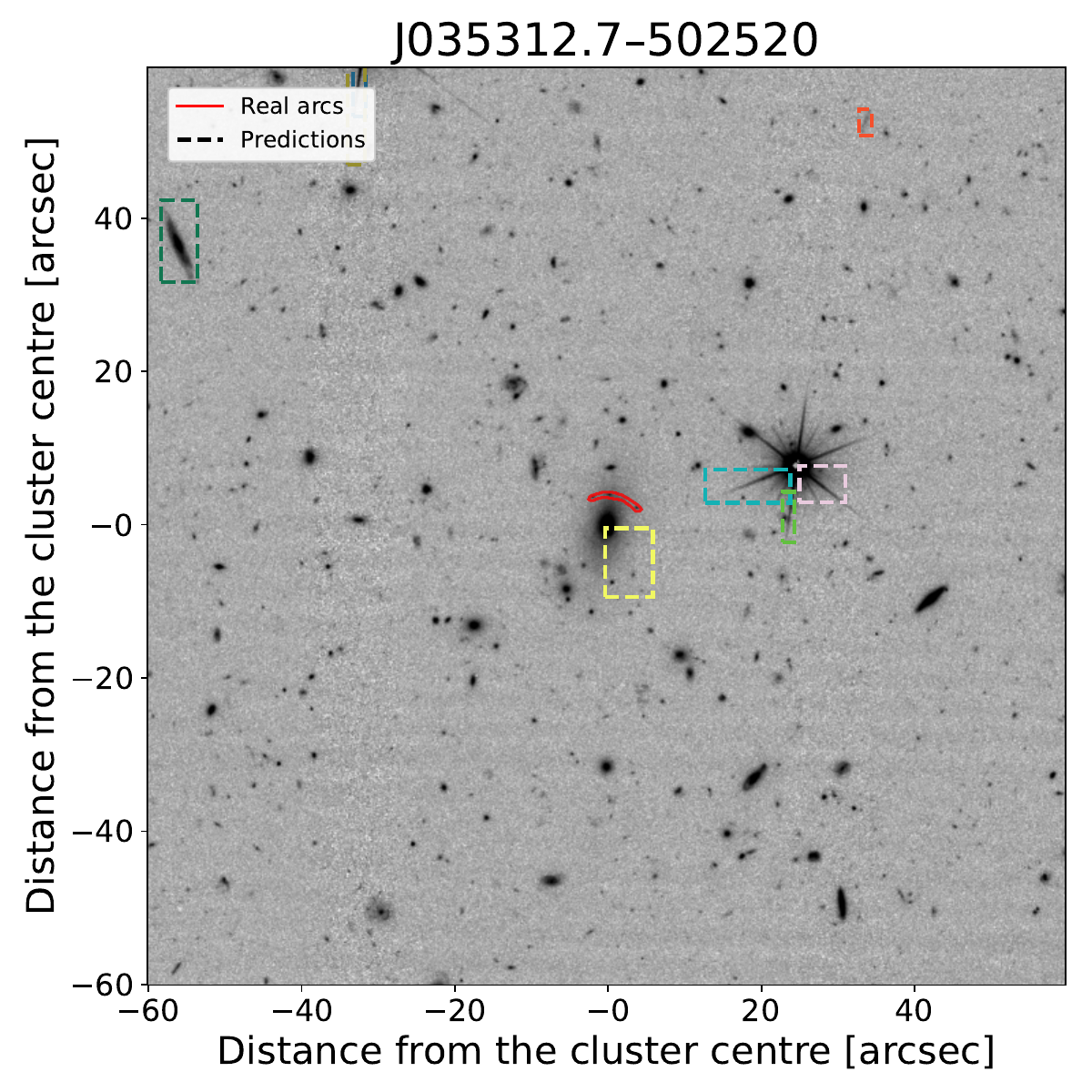}
\end{minipage}
\end{figure*}

\begin{figure*}[htbp!]
\centering
\begin{minipage}{0.43\textwidth}
    \centering
    \includegraphics[width=\textwidth]{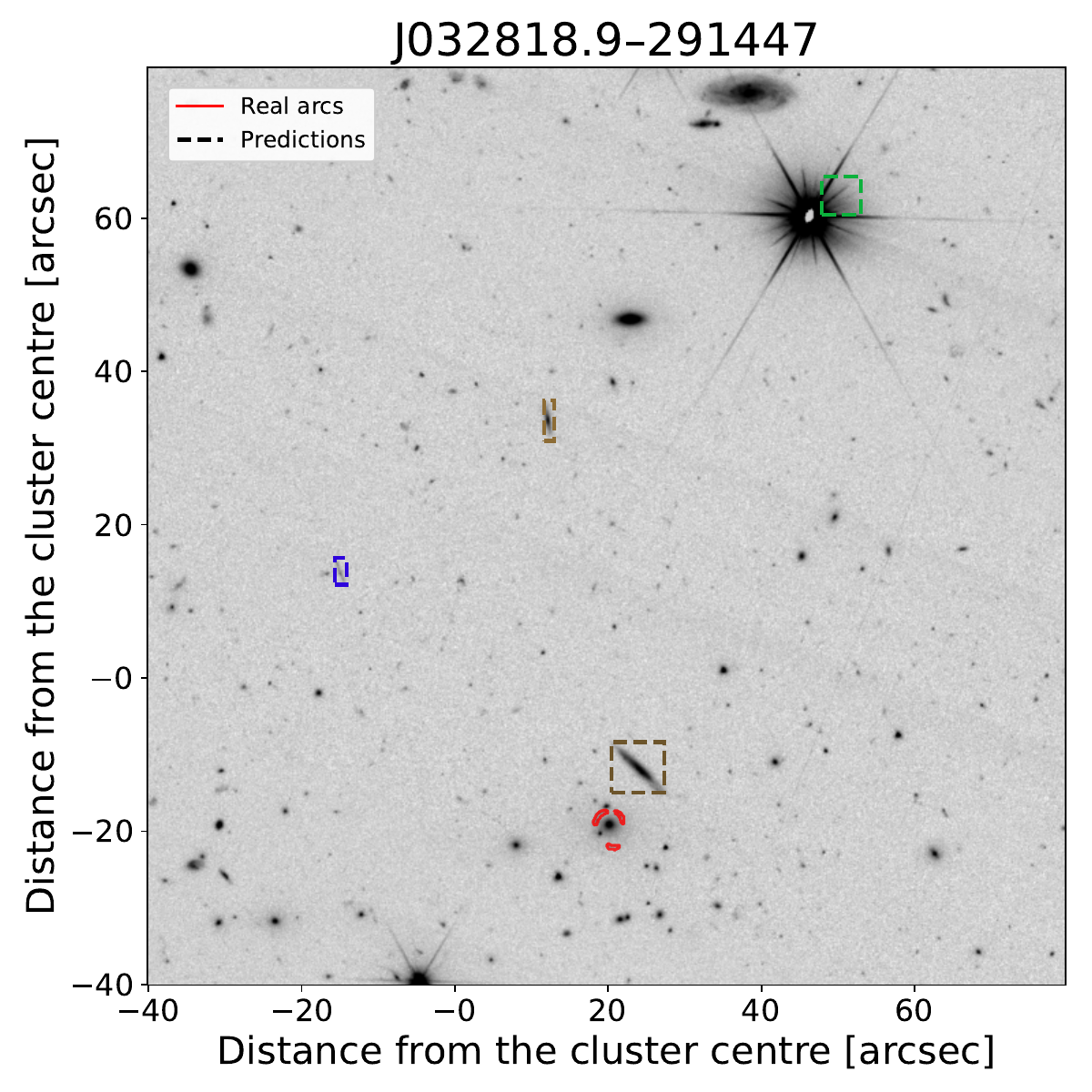}
\end{minipage}
\hfill
\begin{minipage}{0.43\textwidth}
    \centering
    \includegraphics[width=\textwidth]{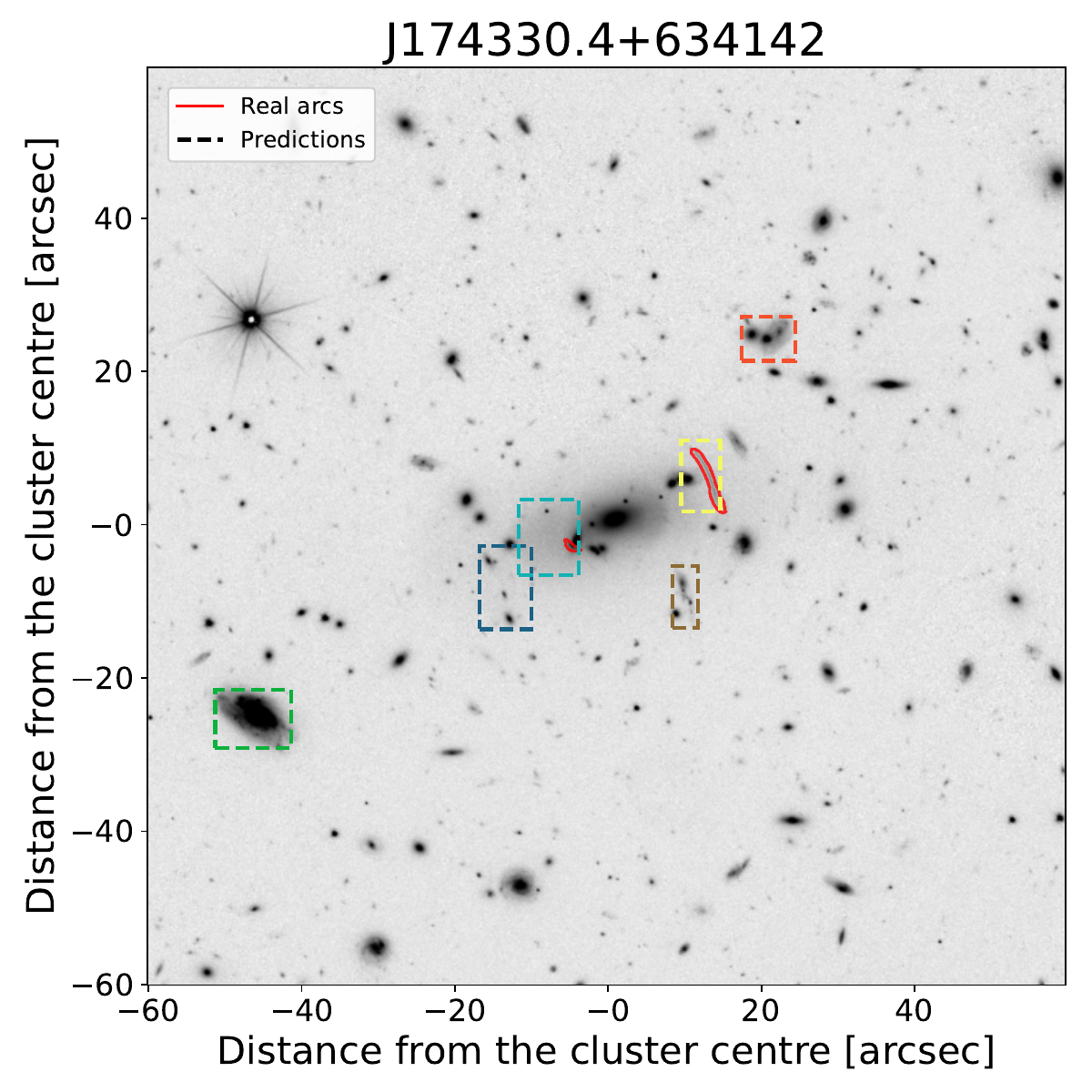}
\end{minipage}

\vspace{0.5cm}

\begin{minipage}{0.43\textwidth}
    \centering
    \includegraphics[width=\textwidth]{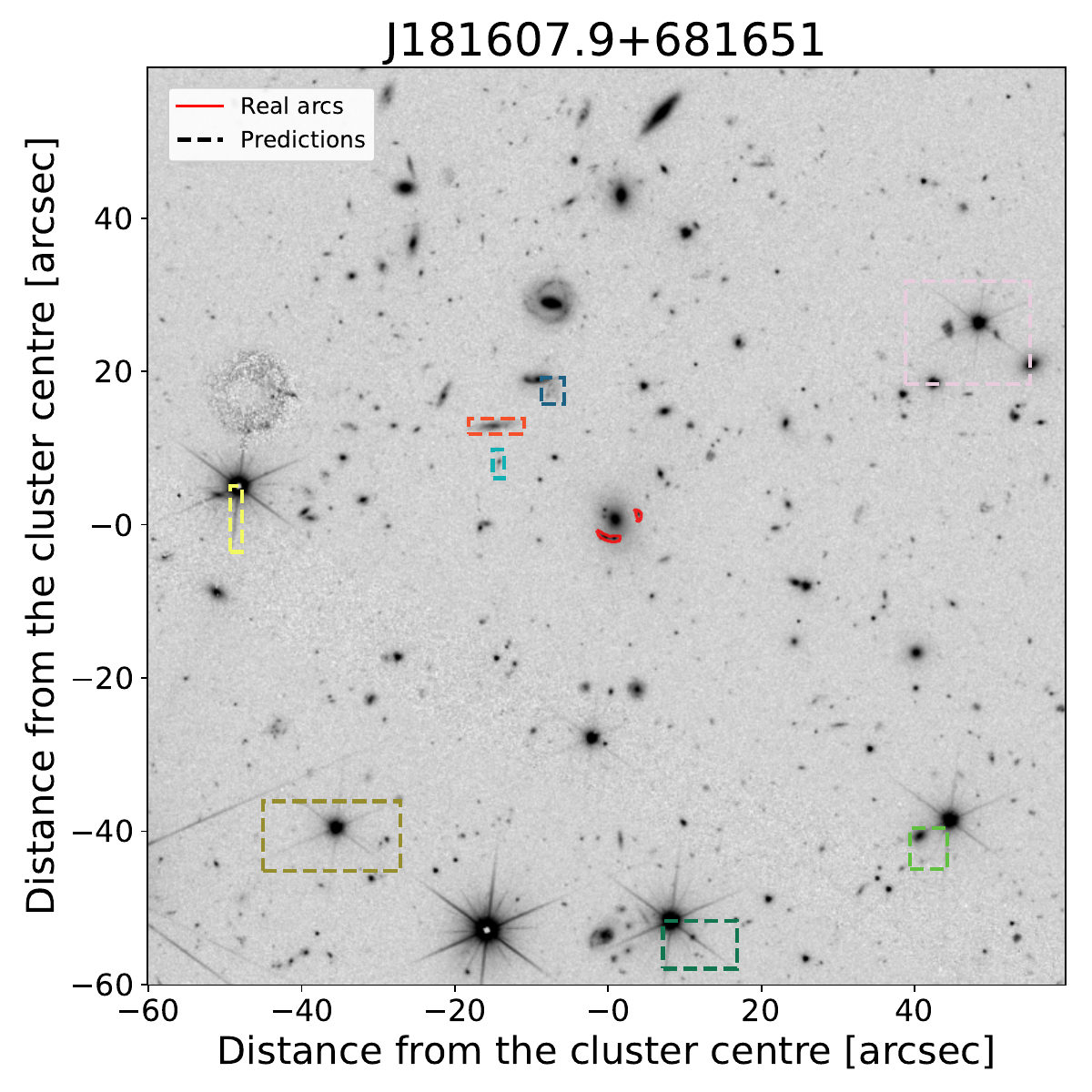}
\end{minipage}
\hfill
\begin{minipage}{0.43\textwidth}
    \centering
    \includegraphics[width=\textwidth]{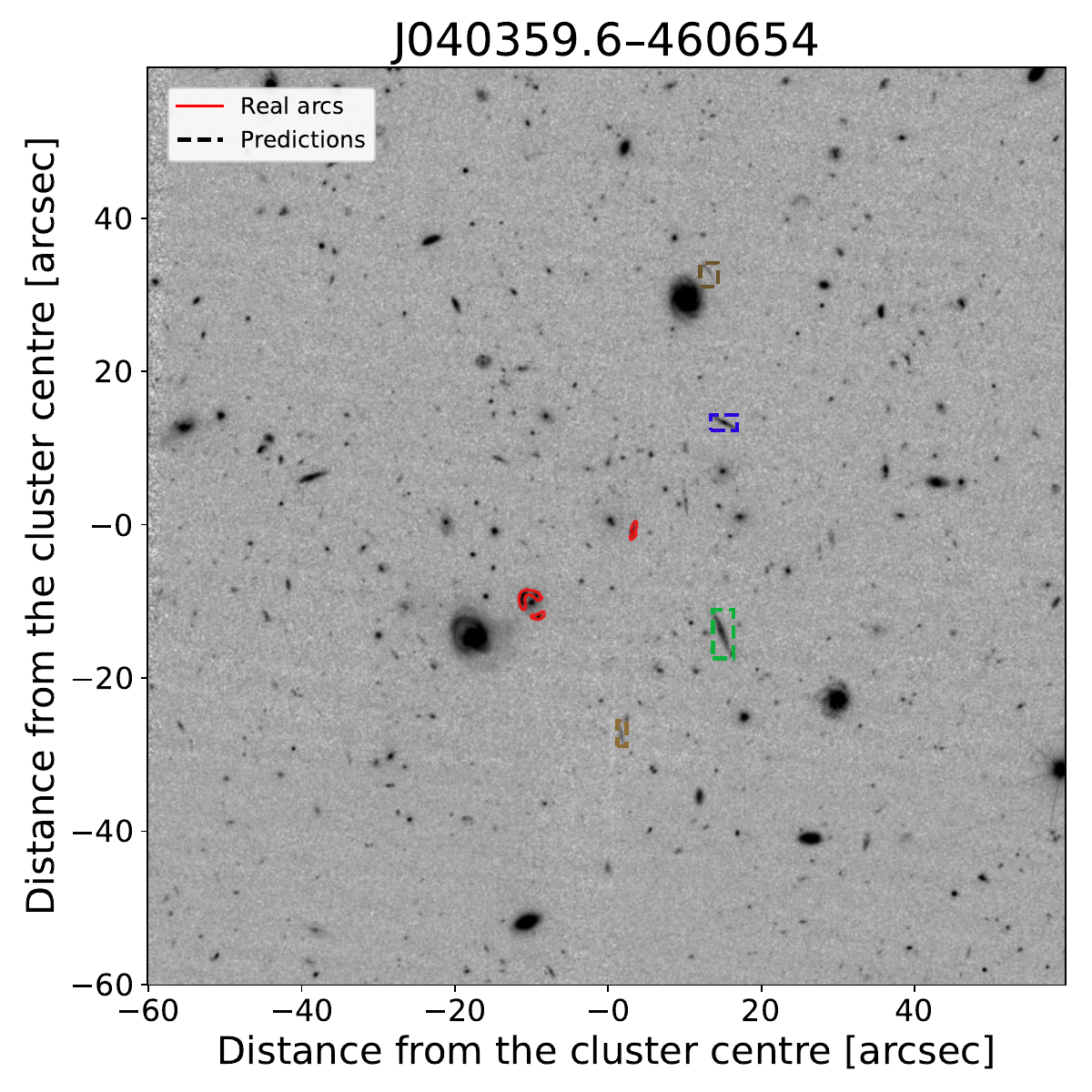}
\end{minipage}

\vspace{0.5cm}

\begin{minipage}{0.43\textwidth}
    \centering
    \includegraphics[width=\textwidth]{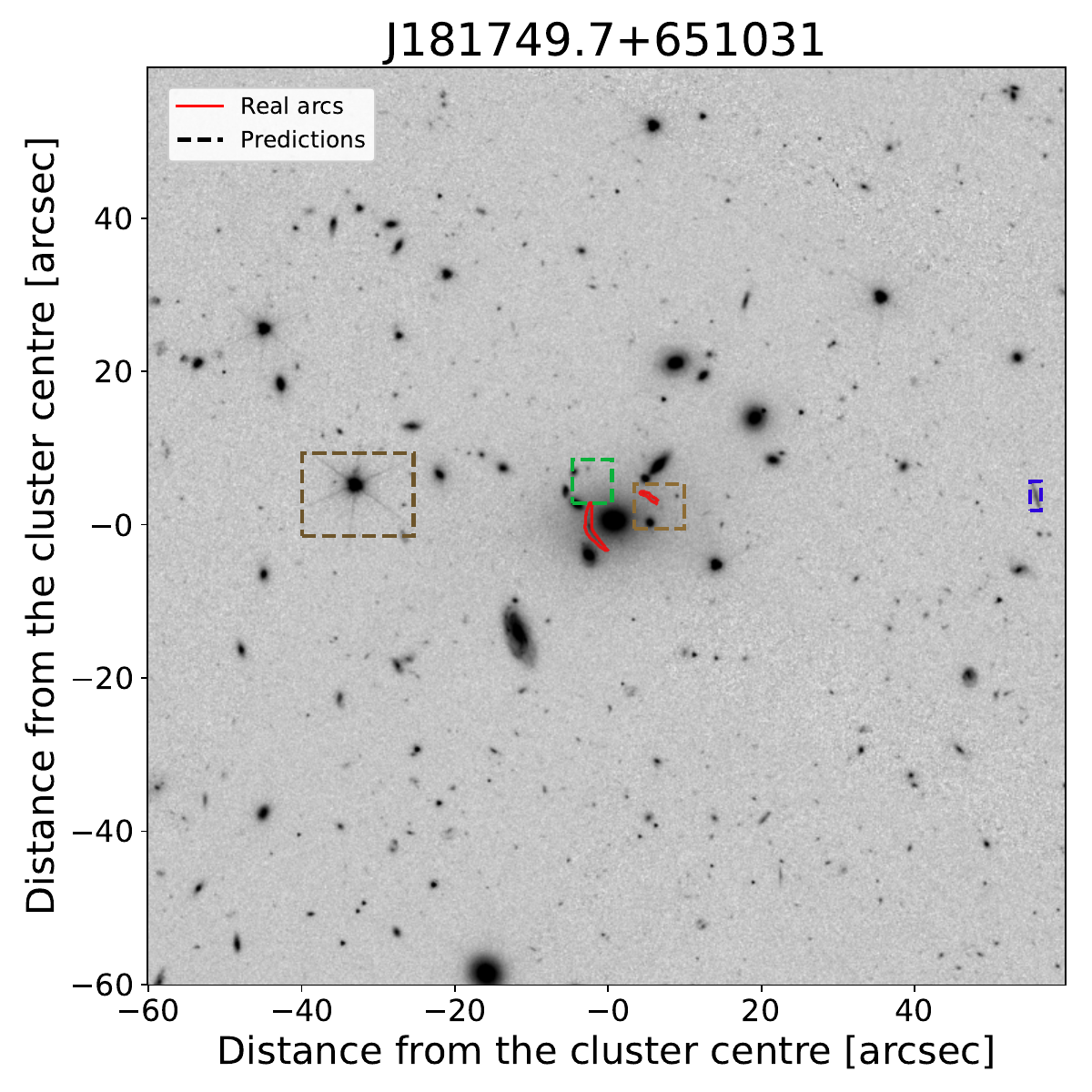}
\end{minipage}
\hfill
\begin{minipage}{0.43\textwidth}
    \centering
    \includegraphics[width=\textwidth]{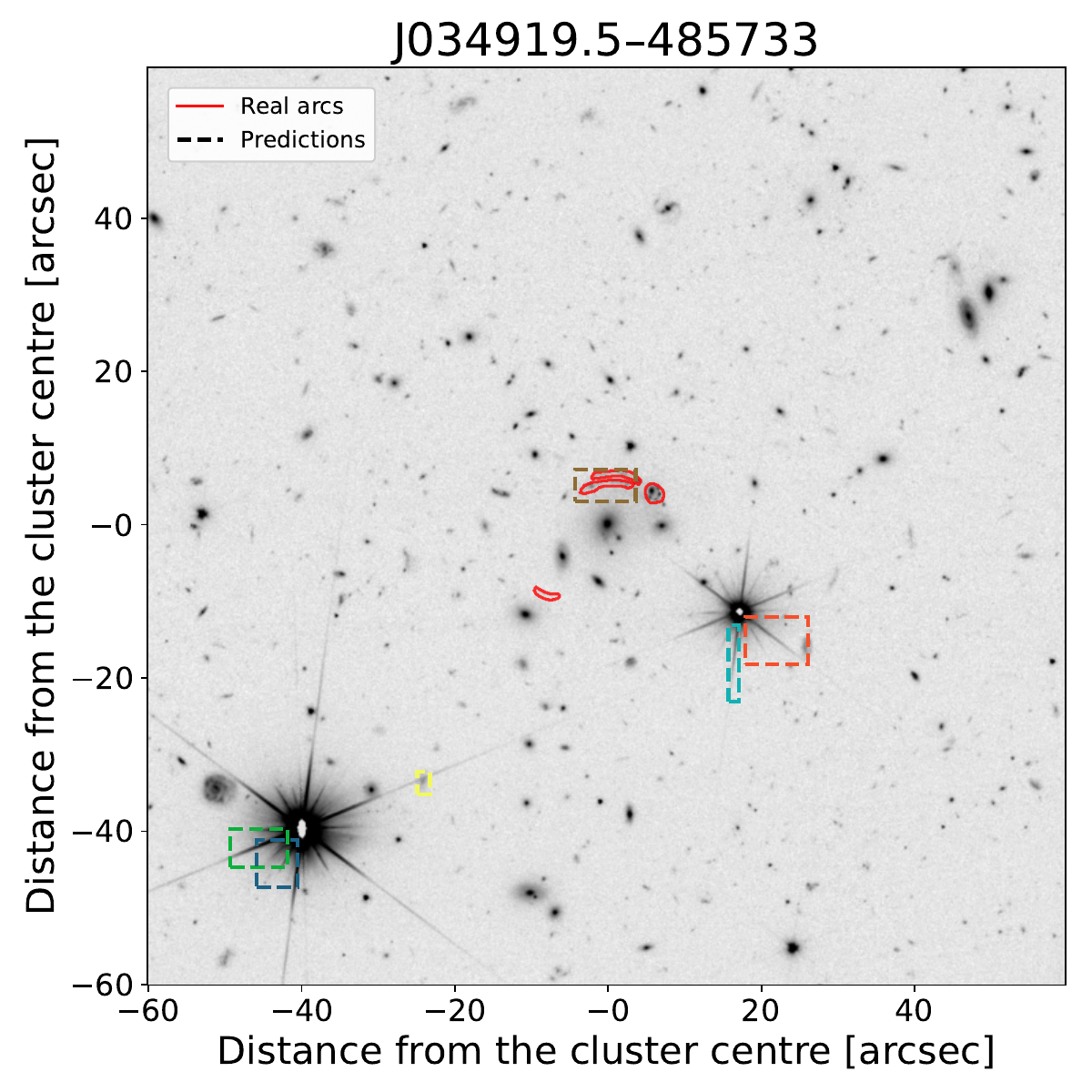}
\end{minipage}
\end{figure*}

\begin{figure*}[htbp!]
\centering
\begin{minipage}{0.43\textwidth}
    \centering
    \includegraphics[width=\textwidth]{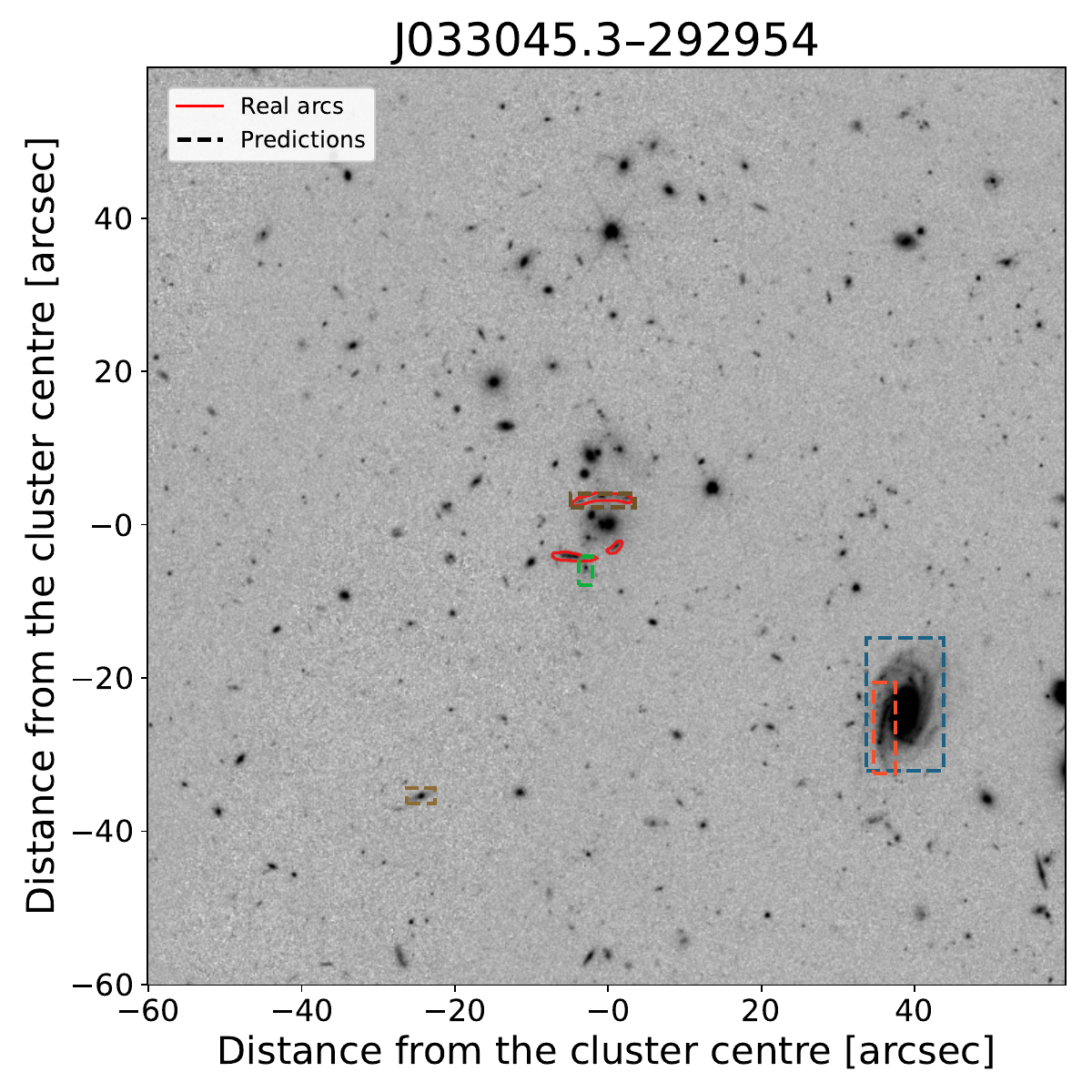}
\end{minipage}
\hfill
\begin{minipage}{0.43\textwidth}
    \centering
    \includegraphics[width=\textwidth]{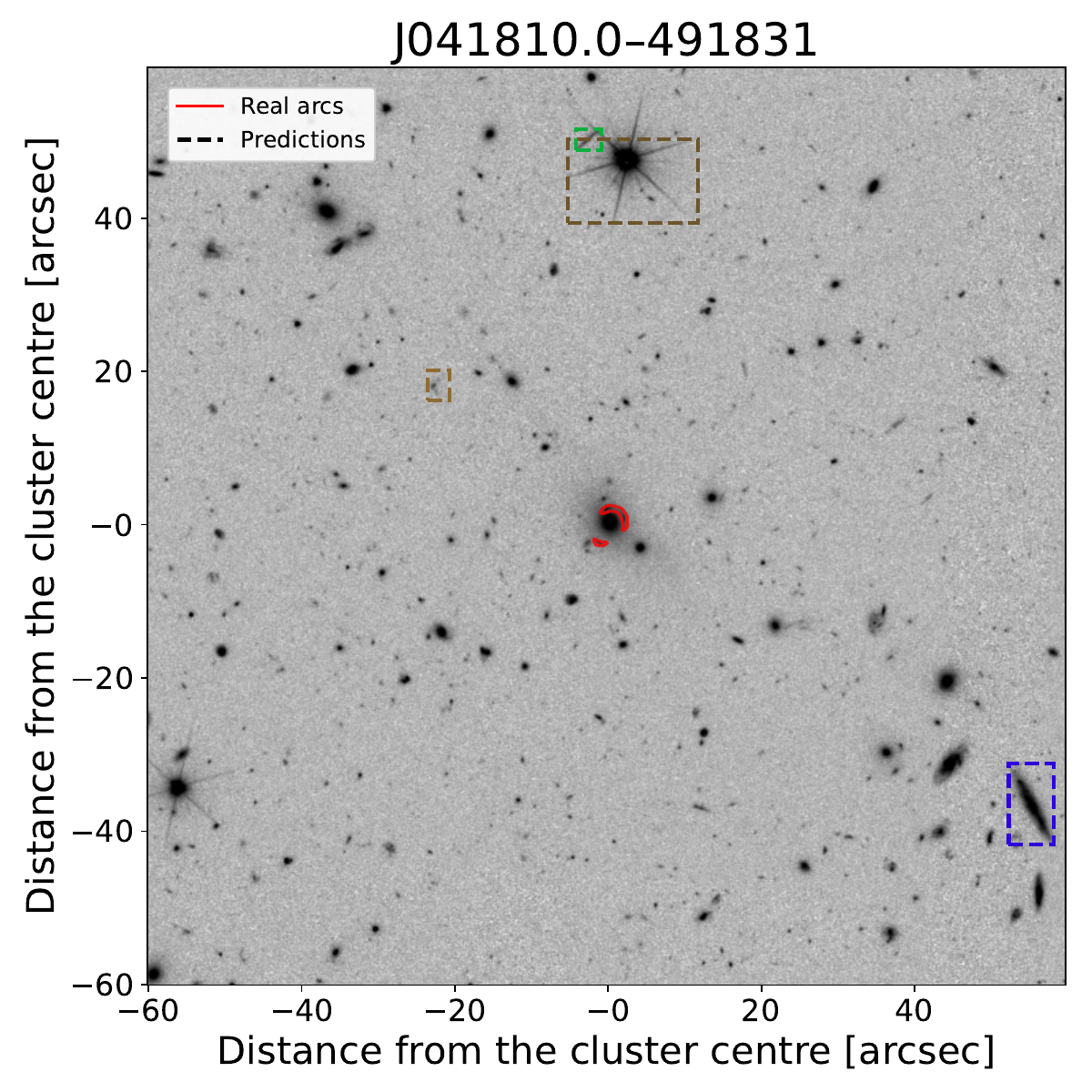}
\end{minipage}

\vspace{0.5cm}

\begin{minipage}{0.43\textwidth}
    \centering
    \includegraphics[width=\textwidth]{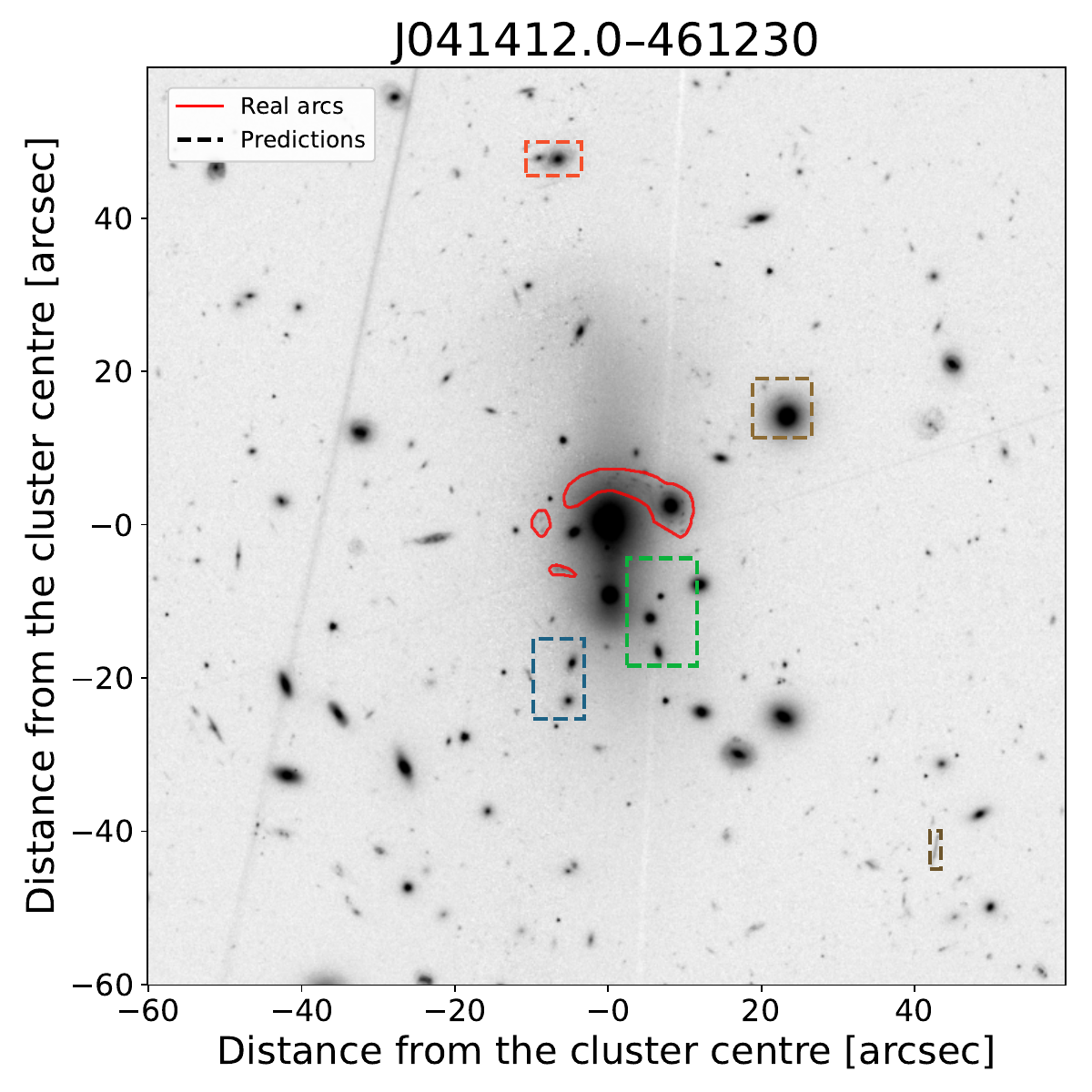}
\end{minipage}
\hfill
\begin{minipage}{0.43\textwidth}
    \centering
    \includegraphics[width=\textwidth]{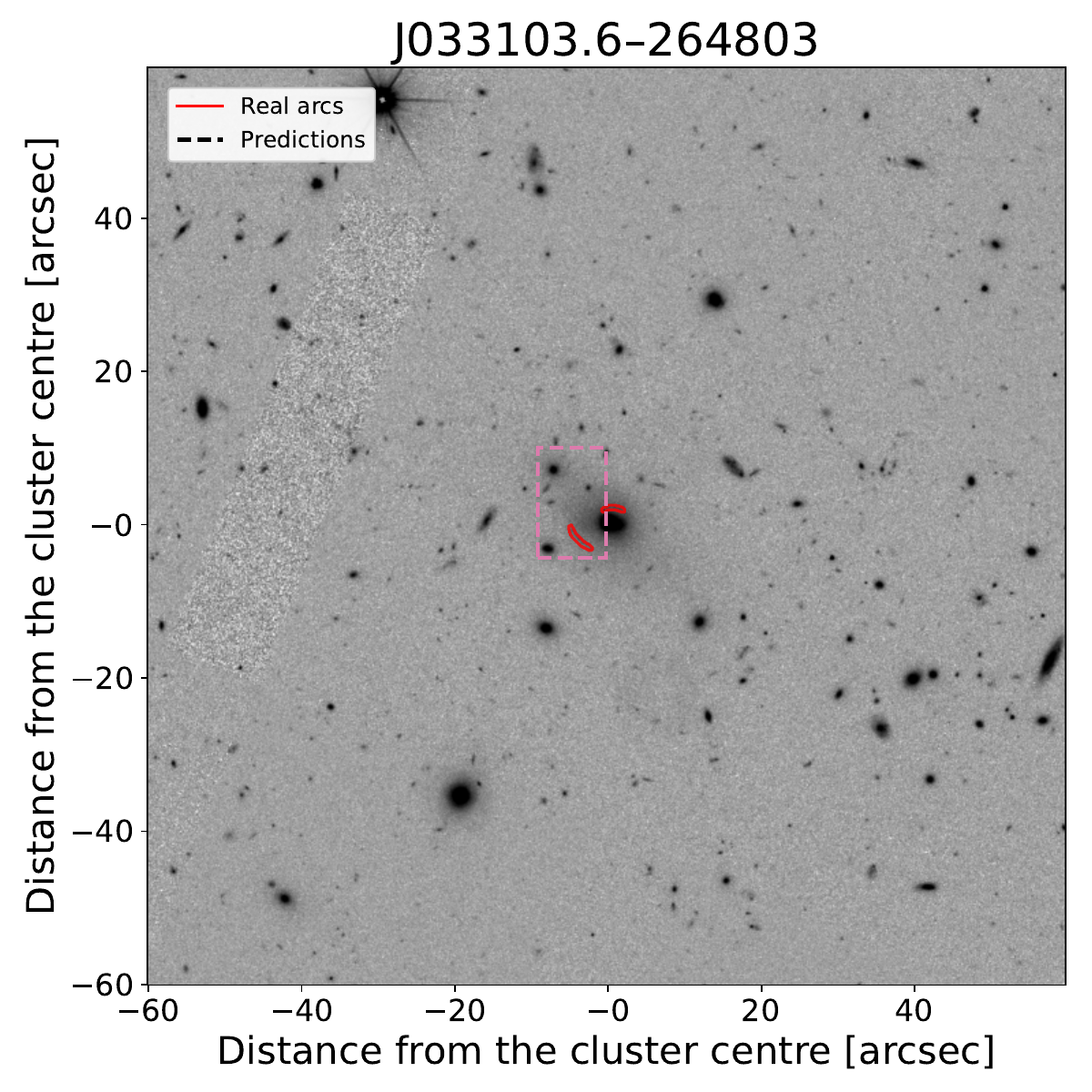}
\end{minipage}

\vspace{0.5cm}

\begin{minipage}{0.43\textwidth}
    \centering
    \includegraphics[width=\textwidth]{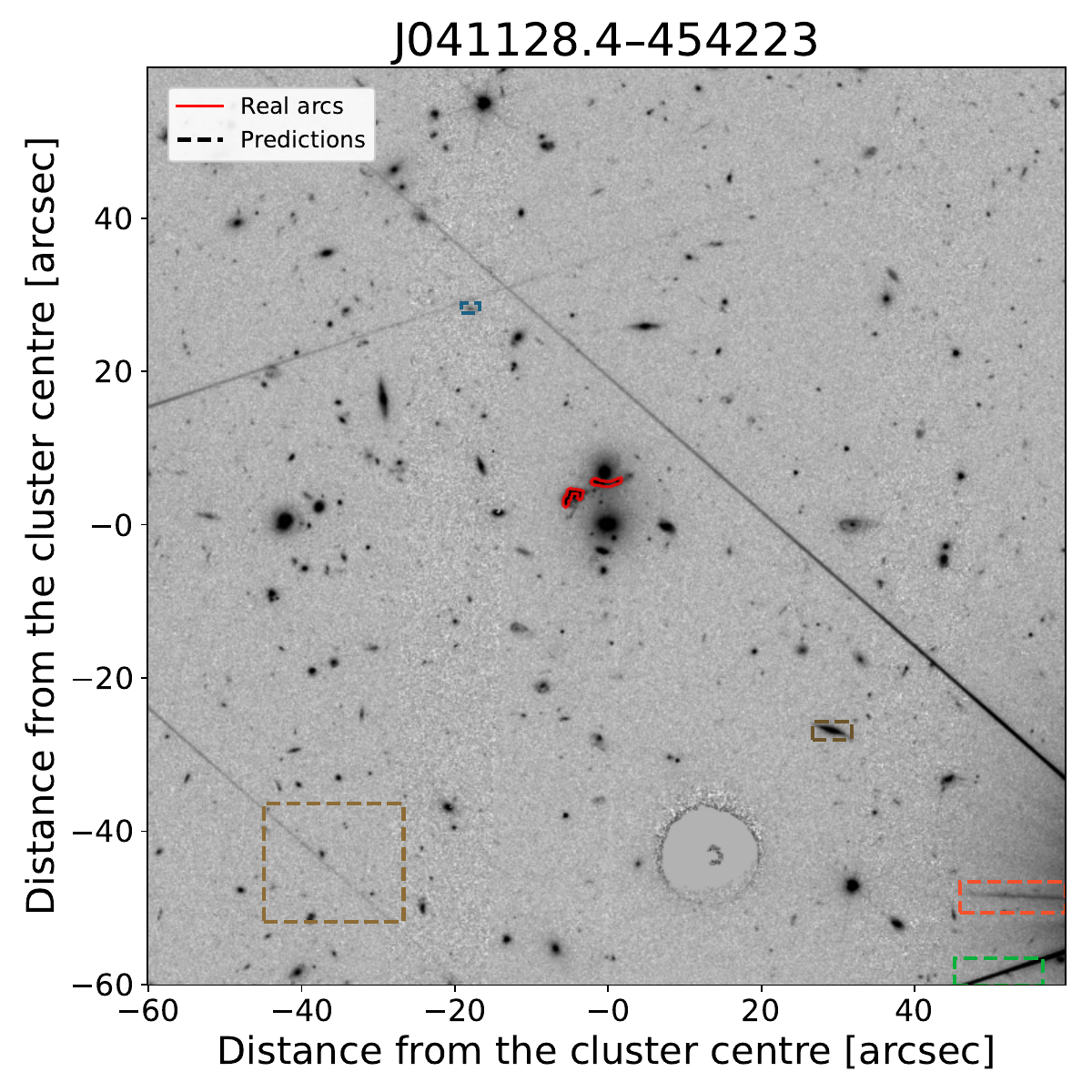}
\end{minipage}
\hfill
\begin{minipage}{0.43\textwidth}
    \centering
    \includegraphics[width=\textwidth]{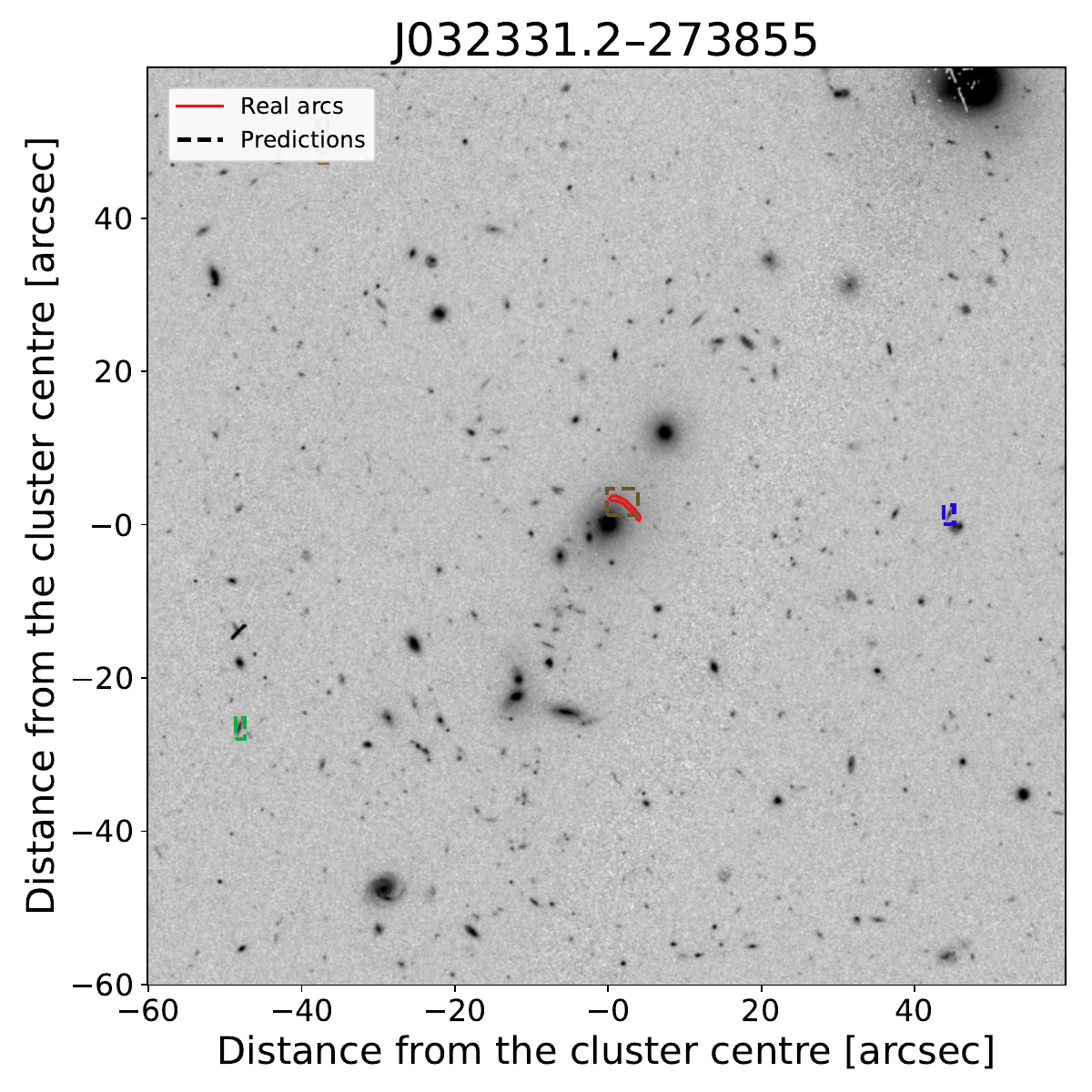}
\end{minipage}
\end{figure*}

\begin{figure*}[htbp!]
\centering
\begin{minipage}{0.43\textwidth}
    \centering
    \includegraphics[width=\textwidth]{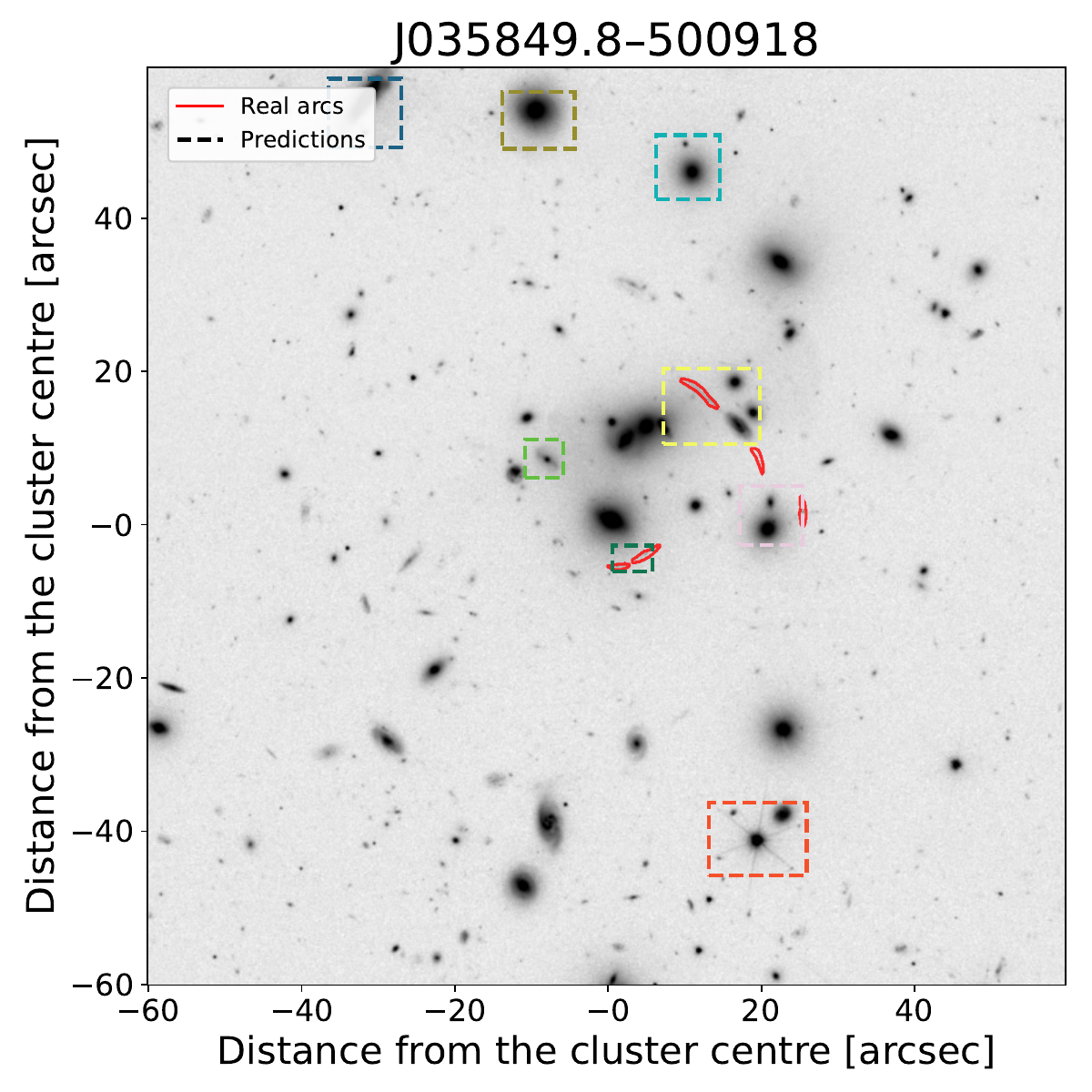}
\end{minipage}
\hfill
\begin{minipage}{0.43\textwidth}
    \centering
    \includegraphics[width=\textwidth]{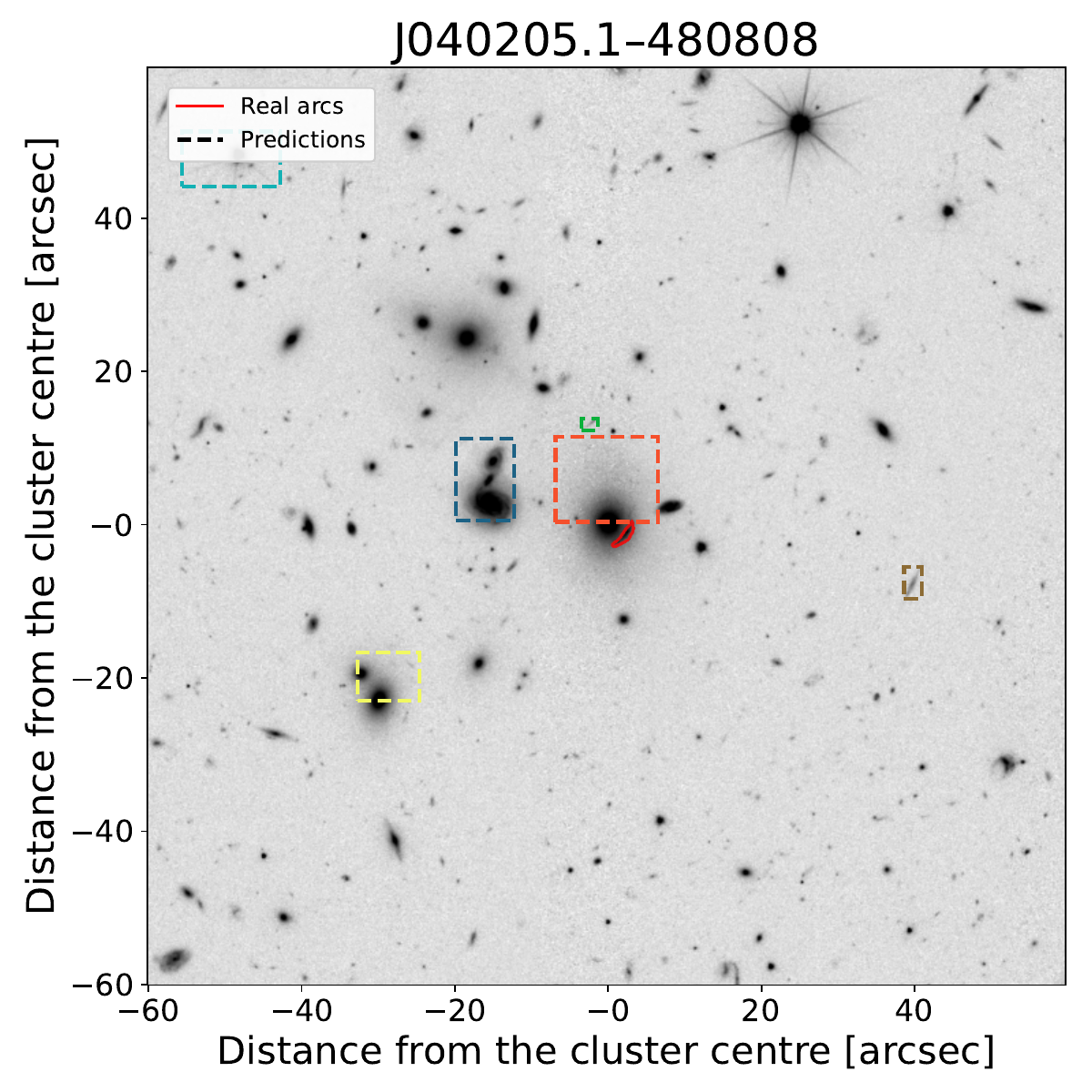}
\end{minipage}
\caption{$2' \times 2'$ VIS cutouts of the 20 $\mathcal{P}_{\mathrm{lens}} > 0.90$ galaxy clusters from~\citet{Q1-SP057}. The red masks enclose the real arcs identified by expert astronomers; the dashed rectangles show the predictions of the NN.}
\label{LastPage}
\end{figure*}

\end{appendix}

\end{document}